\documentclass[twocolumn]{aastex631}

\newcommand{\kblue}{K_\mathrm{blue}}
\newcommand{\kred}{K_\mathrm{red}}
\newcommand{\signmad}{\sigma_\textsubscript{NMAD}}

\shorttitle{Introducing the FLAMINGOS-2 Split-K Medium Band Filters}
\shortauthors{Esdaile et al.}
\graphicspath{{./}{figures/}}

\begin{document}

\title{Introducing the FLAMINGOS-2 Split-K Medium Band Filters: The Impact on Photometric Selection of High-z Galaxies in the FENIKS-pilot survey}

\correspondingauthor{James Esdaile}
\email{jesdaile@swin.edu.au}

\author[0000-0001-6941-7662]{James Esdaile} 
\affiliation{Centre for Astrophysics and Supercomputing, Swinburne University of Technology, Melbourne, VIC 3122, AUSTRALIA} 
\affiliation{ARC Centre for Excellence in All-Sky Astrophysics in 3D (ASTRO 3D)}

\author[0000-0002-2057-5376]{Ivo Labb\'e} 
\affiliation{Centre for Astrophysics and Supercomputing, Swinburne University of Technology, Melbourne, VIC 3122, AUSTRALIA}

\author[0000-0002-3254-9044]{Karl Glazebrook} 
\affiliation{Centre for Astrophysics and Supercomputing, Swinburne University of Technology, Melbourne, VIC 3122, AUSTRALIA} 
\affiliation{ARC Centre for Excellence in All-Sky Astrophysics in 3D (ASTRO 3D)}

\author[0000-0002-0243-6575]{Jacqueline Antwi-Danso}
\affiliation{Department of Physics and Astronomy, Texas A\&M University, College Station, TX, 77843-4242 USA} \affiliation{George P.\ and Cynthia Woods Mitchell Institute for Fundamental Physics and Astronomy, Texas A$\&$M University, College Station, TX, 77843-4242 USA}

\author[0000-0001-7503-8482]{Casey Papovich}
\affiliation{Department of Physics and Astronomy, Texas A\&M University, College Station, TX, 77843-4242 USA} \affiliation{George P.\ and Cynthia Woods Mitchell Institute for Fundamental Physics and Astronomy, Texas A$\&$M University, College Station, TX, 77843-4242 USA}

\author[0000-0002-3958-0343]{Edward Taylor}
\affiliation{Centre for Astrophysics and Supercomputing, Swinburne University of Technology, Melbourne, VIC 3122, AUSTRALIA}

\author[0000-0002-7248-1566]{Z. Cemile Marsan}
\affiliation{Department of Physics and Astronomy, York University, Toronto, Ontario, Canada}

\author[0000-0002-9330-9108]{Adam Muzzin}
\affiliation{Department of Physics and Astronomy, York University, Toronto, Ontario, Canada}

\author[0000-0001-5937-4590]{Caroline M. S. Straatman} 
\affiliation{Department of Physics and Astronomy, Ghent University, Krijgslaan 281 S9, B-9000 Gent, Belgium}

\author[0000-0001-9002-3502]{Danilo Marchesini}
\affiliation{Physics and Astronomy Department, Tufts University, Medford, MA, USA}

\author[0000-0001-9716-5335]{Ruben Diaz}
\affiliation{Gemini Observatory, NSF’s NOIRLab, 950 N. Cherry Ave., Tucson, AZ 85719, USA}

\author[0000-0001-5185-9876]{Lee Spitler} 
\affiliation{Department of Physics and Astronomy, Macquarie University, Sydney, NSW 2109, Australia} \affiliation{Research Centre in Astronomy, Astrophysics \& Astrophotonics, Macquarie University, Sydney, NSW 2109, Australia}

\author[0000-0001-9208-2143]{Kim-Vy H. Tran}
\affiliation{ARC Centre for Excellence in All-Sky Astrophysics in 3D (ASTRO 3D)}
\affiliation{School of Physics, University of New South Wales, Kensington, Australia}  \affiliation{Department of Astronomy, University of Washington}

\author[0000-0002-4144-5116]{Stephen Goodsell}
\affiliation{Gemini Observatory, NSF’s NOIRLab, 950 N. Cherry Ave., Tucson, AZ 85719, USA}
\affiliation{Physics Department, Durham University, Stockton Road, Durham, DH1 3LE, UK}



\begin{abstract}

Deep near-infrared photometric surveys are efficient in identifying high-redshift galaxies, however they can be prone to systematic errors in photometric redshift. This is particularly salient when there is limited sampling of key spectral features of a galaxy's spectral energy distribution (SED), such as for quiescent galaxies where the expected age-sensitive Balmer/4000~\AA\ break enter the $K$-band at $z>4$. With single filter sampling of this spectral feature, degeneracies between SED models and redshift emerge. A potential solution to this comes from splitting the $K$-band into multiple filters.  We use simulations to show an optimal solution is to add two medium-band filters, $\kblue$  ($\lambda\textsubscript{cen}$=2.06 $\mu$m, $\Delta\lambda$=0.25 $\mu$m) and $\kred$ ($\lambda\textsubscript{cen}$=2.31 $\mu$m, $\Delta\lambda$=0.27 $\mu$m), that are complementary to the existing $K_\mathrm{s}$ filter. We test the impact of the $K$-band filters with simulated catalogues comprised of galaxies with varying ages and signal-to-noise.  The results suggest that the $K$-band filters do improve photometric redshift constraints on $z>4$ quiescent galaxies, increasing precision and reducing outliers by up to 90$\%$.  We find that the impact from the $K$-band filters depends on the signal-to-noise, the redshift and the SED of the galaxy. The filters we designed were built and used to conduct a pilot of the FLAMINGOS-2 Extra-galactic Near-Infrared $K$-band Split (FENIKS) survey.  While no new $z>4$ quiescent galaxies are identified in the limited area pilot, the $\kblue$ and $\kred$ filters indicate strong Balmer/4000~\AA\ breaks in existing candidates.  Additionally we identify galaxies with strong nebular emission lines, for which the $K$-band filters increase photometric redshift precision and in some cases indicate extreme star-formation.

\end{abstract}

\keywords{galaxy: evolution, galaxies: high-redshift, galaxies: photometry}


\section{Introduction}


There is now substantial evidence of massive (Milky Way stellar mass equivalent and above) galaxies at 3 $< z <$ 4 from deep Near-Infrared (NIR) surveys \citep{Marchesini2010, Muzzin2013b, Spitler2014, Straatman2014, Stefanon2015, Laigle2016}.  These surveys have been instrumental in the recent discovery and spectroscopic confirmation of quenched and massive galaxies at 3 $< z <$ 4 \citep{ Marsan2015, Glazebrook2017, Marsan2017, Schreiber2018, Valentino2019, Forrest2020}.  Massive quiescent galaxies at $z \gtrsim 3$ are very different compared to local early-type galaxies (ETG) despite sharing similar characteristics, e.g. red rest-frame colors \citep[see the introduction in][for a detailed summary of this population and the differences from local galaxies with similar attributes]{Schreiber2018}.  Most notably, they have very compact sizes of 1 kpc half-light radius \citep{Marsan2015, Straatman2015}, which are very rare in the local Universe \citep{Trujillo2009,Taylor2010,Valentino2019}, and heavily suppressed (not necessarily zero) star-formation compared to the star-forming main-sequence at similar epochs.  Finally, massive quiescent galaxies are exceedingly rare at $3 < z < 4$ with a number density 1--$2\times 10^{-5}$ Mpc$^{-3}$ \citep{Straatman2014,Schreiber2018, Merlin2019}, a factor of 80x less than similar mass ETGs at $z \sim~ 0.1$.  While all massive galaxies are less abundant at higher redshift \citep{Conselice2016}, the quiescent fraction is also significantly smaller by a factor of 4-10 over this same redshift range \citep{Martis2016}.

The discovery of massive quiescent galaxies at these epochs proved to be an important test for galaxy formation models as the simulations at the time were unable to reproduce them in sufficient number \citep[][and references therein]{Glazebrook2017,Schreiber2018, Merlin2019, Forrest2020}.  The tension existed partly due to limited simulation box sizes, where the relatively small volumes allow for only a few massive galaxies at high redshifts, and the treatment of sub-grid physics such as AGN feedback, which is a potential mechanism to quench galaxies at these epochs.  With the latest generation of simulations, both semi-analytic models \citep{Henriques2015,Qin2017} and hydro-dynamical simulations \citep{Montero2019} appear to be able to reproduce number densities of massive quiescent galaxies at these epochs.  However, these agreements rely heavily on how the comparisons between observations and simulations are made, such as the use of emission line templates in SED fitting for observations or the selection of the radii within which the star-formation is calculated in simulations \citep[see][for a detailed discussion]{Merlin2019}.

Discrepancies remain however when considering star formation histories (SFH) and the evolution of these galaxies beyond $z \sim 4$.  The spectroscopically confirmed galaxy ZF-COSMOS-20115 at z=3.71 in \citet{Glazebrook2017} implies a rapid formation timescale $<250$ Myr and abrupt end to star formation prior to $z = 5$.  Similarly, spectroscopic confirmation of the most massive $z > 3$ quiescent galaxy, has a SFH that indicates this galaxy is quiescent at $z > 4$ \citep{Forrest2020}.  Despite now achieving sufficient cosmic volumes and more detailed treatment of AGN feedback, massive quiescent galaxies are predicted not to be present in simulations at $z \sim 5$.  


\begin{figure*}
\begin{tabular}{cc}
\begin{minipage}{\linewidth}
\centering
    \includegraphics[width=0.7\linewidth]{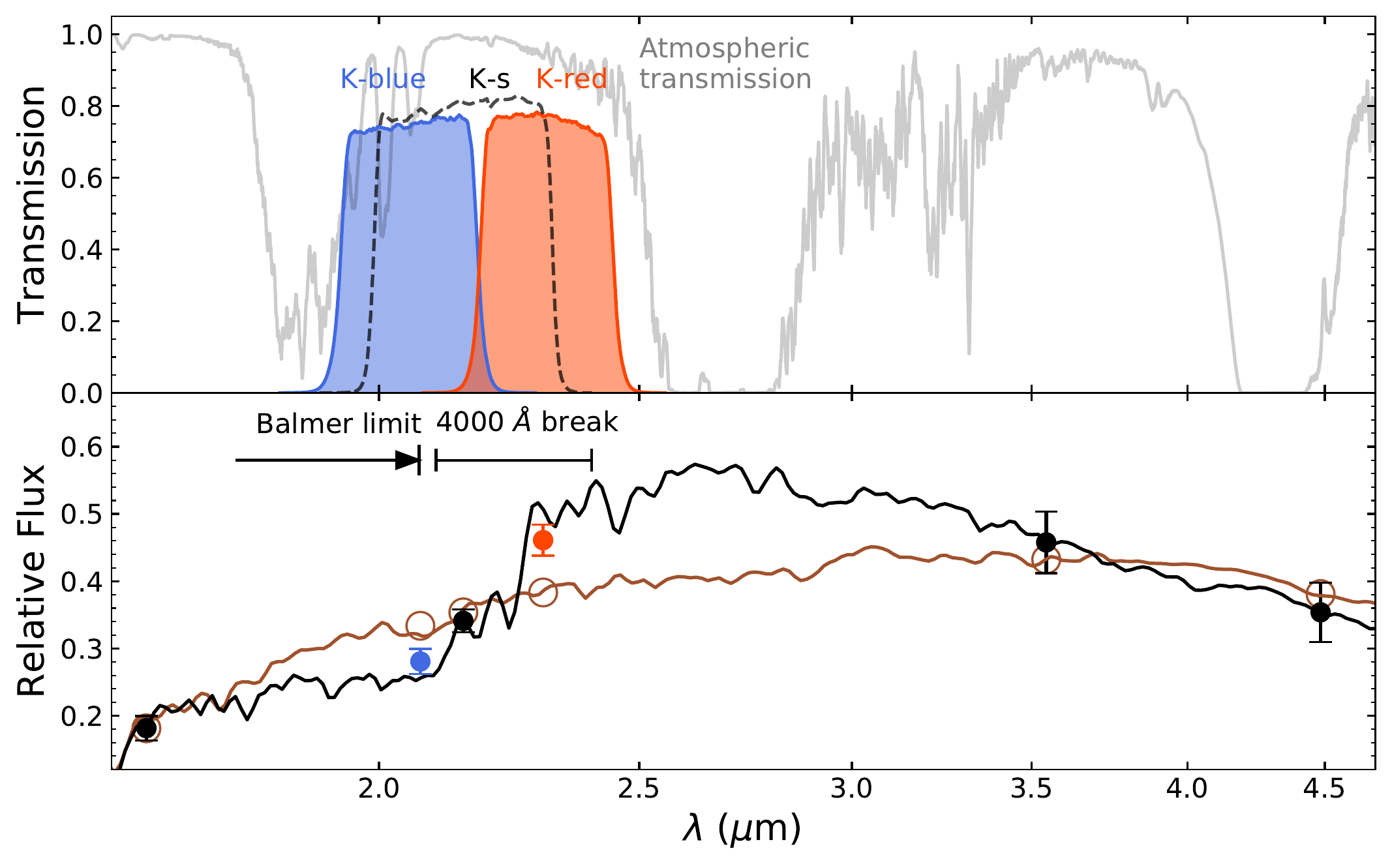}
    \caption{ 
    The new \texttt{FLAMINGOS-2} medium band $\kblue$ and $\kred$ filters sample the blue and red ends of the 1.9 - 2.5 micron $K$-band window, both overlapping the $K_s$ filter.  Full optical path transmission of the telescope+instrument is applied to these the filters.  Atmospheric transmission is shown in grey for a typical 2.3 mm water vapour column for a Cerro Pach\'on altitude.
    Bottom panel: Galaxy SEDs of a dusty galaxy at $z = 3$ (brown line) and a quiescent galaxy at $z = 4.7$ (black line) are shown with brown circles and black points showing the respective synthetic photometry of filters ($H$, $K_s$, IRAC at 3.6 and 4.5 micron) in this wavelength range.  The synthetic photometry of $\kblue$ and $\kred$ filters is shown for the quiescent galaxy in blue and red respectively.  The $K$-split filters demonstrate their ability to discriminate between these two different galaxy SEDs.  The Balmer limit and 4000~\AA\ break are shown to illustrate the extent of the spectral features traced by the $K$-split filters.  For the galaxy SED at $z = 4.7$, the $K$-split filters sample portions of both the Balmer and 4000~\AA\ breaks.
    }
    \label{fig:ks_kb_kr_bands}
\end{minipage}
\end{tabular}
\end{figure*}

Recent observational searches for massive quiescent galaxies beyond $z \sim 4$ suggest the number density declines compared to that of the $3 < z < 4$ population.  The number density of massive quiescent galaxies at $z>4$ is between $1.7-8.2\times 10^{-6}$ Mpc$^{-3}$, after correcting for completeness \citep{Pampliega2019,Merlin2019,Shahidi2020,Marsan2020}. The range in these values are indicative of the large uncertainties both due to the limited number statistics and the potential for systematic errors inherent in determining photometric redshifts for quiescent galaxies at $z>4$.

There is a clear need for a large and robust sample of quiescent galaxies at $z > 4$ with precise photometric redshifts to see how early in the Universe galaxies can shut down their star-formation.  Traditional NIR-selected surveys are well-suited for studying quiescent galaxies from $1 < z < 4$, due to the detection band sampling red-ward of the age-sensitive Balmer and 4000~\AA\ break. These breaks develop in young post-starburst galaxies at the Balmer limit at 3650~\AA, and then evolve to a 4000~\AA\ break \citep[e.g. ][]{Kauffmann2003}, as exhibited by most modern ETGs after $\sim$ 800 Myr. Beyond $z > 4$, these breaks enters the $K$-band, which makes detection more difficult and photometric redshift uncertainty worse due to the large gap between the $H$ (1.6 micron) and IRAC 3.6 micron bands.  Additionally, emission line galaxies and old, dusty galaxies at lower redshifts can mimic the spectral energy distribution (SED) shape of $z > 4$ quiescent galaxies.  These lower redshift galaxies represent a much larger sample, meaning that photometric measurement errors can potentially cause up-scattering of lower redshift galaxies and significantly contaminate the massive quiescent galaxy sample \citep{Straatman2014, Schreiber2018}.  Deriving robust redshifts for rare massive quiescent galaxies is therefore a significant challenge that must be overcome to understand $z > 4$ galaxy populations.

Previous NIR surveys have been successful in improving photometric redshift precision at lower redshifts by adding medium band filters to increase the sampling of prominent spectral features.  Surveys such as the Newfirm Medium Band Survey \citep[NMBS]{Whitaker2011} and the FourStar Galaxy Evolution survey \citep[ZFOURGE]{Straatman2016} used medium band filters (R=$\lambda/\Delta\lambda$ $\sim$ 7) that split the J and H broadband filters (R$\sim$3) into three and two respectively.  The addition of these medium band filters improved photometric redshift precision (the scatter of $|z_\mathrm{spec} - z_\mathrm{phot}|$ / $( 1 + z_\mathrm{spec} )$ by up to a factor of 2, achieving photometric redshift precision of $\approx$ 1--2$\%$.  Outliers associated with catastrophic photometric redshift solutions are also reduced by a factor of two compared to where only broad-band photometry is used.  Similar results were found by \citet{Merlin2021} using deep broad-band and medium-band filters achieving photometric redshift precision of 1.5$\%$ and outlier fractions of 3$\%$.  Medium band NIR imaging enabled the robust selection of massive quiescent galaxies to $3 < z < 4$ as shown in \citet{Straatman2014} in which 14 were identified.  \citet{Schreiber2018} used an extended data set to spectroscopically confirm 12 massive quiescent galaxies selected with NIR medium-band data with only a 20$\%$ contamination rate by lower redshift objects.  Extending this increased photometric sampling approach to longer wavelengths such as the $K$-band could produce similar results for higher redshifts.

Increased sampling of the $K$-band also offers the opportunity to identify extreme emission line galaxies (EELG) where emission lines are indicated by the boosting of flux in one medium band filter but not the other.  Adding medium band filters to the $K$-band probes emission line features such as H$\alpha$ and the [NII] doublet (6548\AA, 6584\AA) at $2 < z < 2.8$, the [OIII] doublet (4959\AA, 5007\AA) at $3 < z < 4$ and the [OII] doublet (3726\AA, 3729\AA) at $z > 4$.  EELGs are often associated with Lyman-$\alpha$ emitters, and display similar characteristics to higher redshift star-forming galaxies,  thought to be responsible for reoinizing the Universe \citep{Forrest2017}.  Spectroscopic follow-up of these galaxies can enable studies of the ionizing photon production efficiencies using H$\alpha$ \citep[as in][]{Nanayakkara2020} and provide a better understanding of the likely photo-ionisation budgets during the epoch of reionization (EoR).  Additionally, some of these starburst galaxies likely have similar characteristics to the progenitors of massive quiescent galaxies at high-redshift, as the SFHs of massive quiescent galaxies suggest they exhibited extreme starbursts with maximum star-formation rates (SFR) in excess of 1000 M$_{\odot}$ yr$^{-1}$.  Studying them could offer greater insights into the formation scenarios of higher redshift massive quiescent galaxies.

Finally, large ground-based surveys that aim to find massive quiescent galaxies at $z>4$ are especially important in the era of the James Webb Space Telescope (JWST).  Due to the relatively small field-of-view of JWST, photometrically pre-selecting these rare galaxies would be inefficient and involve multiple JWST cycles. Ground-based photometric pre-selection of $z > 4$ massive quiescent galaxy candidates is therefore optimal to identify the best candidates to follow-up with JWST.

Increased photometric sampling of the $K$-band is a logical progression to next generation deep NIR surveys and has influenced the development of several instruments that include medium-band filters in the $K$-band region \citep[including ][]{Motohara2014}.  This has now been achieved with the design and commissioning of two new medium $K$-band filters (hereafter $K$-split filters) that have been added to the \texttt{FLAMINGOS-2} instrument \citep[][hereafter \texttt{F2}]{Eikenberry2008, Gomez2012}, on the 8.1\ m Gemini South Telescope in Chile. \texttt{F2} is an ideal instrument choice due to its NIR optimized design and its location on top of Cerro Pach\'on, which features low NIR sky backgrounds.  The design of the filters was undertaken by our research collaboration (described in Section \ref{sec:filters}) which was selected by the Gemini Observatory, as part of its Instrument Upgrade Program \citep{Diaz2016}.  The $K$-split filters were manufactured by the Asahi Spectra USA inc and added to \texttt{F2} in Semester 2017A. Shortly after the FLAMINGOS-2 Extra-galactic Near-Infrared $K$-band Split (FENIKS) pilot survey (PI Papovich) was undertaken (which is described in Section \ref{sec:survey}).  A larger area follow-up survey, the FENIKS Large and Long-term Program is currently underway and will be introduced in an upcoming survey paper (Papovich et al. in prep).

In this paper we outline the filter design and selection for the $K$-split filters in section \ref{sec:filters}.  We show the theoretical performance of these filters using photometric redshift measurements of simulated galaxy catalogues in section \ref{sec:simulations}.  Section \ref{sec:survey} summarises the FENIKS pilot survey which uses the new $K$-split filters.  Section \ref{sec:data} outlines the data reduction and cataloguing methodology.  Section \ref{sec:results} shows preliminary results of the FENIKS pilot survey where we identify candidates for massive galaxies at $z > 4$, emission line galaxy candidates at $z > 2$, and show the impact of the $K$-split filters on the photometric redshifts of these galaxies.  Finally, a summary is given in Section \ref{sec:conclusions}.  Throughout, we assume a $\Lambda$CDM cosmology with $\Omega_{M}$ = 0.3, $\Omega_{\Lambda}$ = 0.7 and $H_0$ = 70 km s$^{-1}$ Mpc$^{-1}$. The photometric system is AB \citep{oke1983}.

\section{Design of the K-split Filters}
\label{sec:filters}

The $K$-band medium filters used in the FENIKS pilot survey are shown in Fig.~\ref{fig:ks_kb_kr_bands}.  The newly added filters split the $K$-band atmospheric window into two with a blue and a red component (which we name: $\kblue$ and $\kred$).  The lower panel of Fig.~\ref{fig:ks_kb_kr_bands} illustrates a key use-case for the $\kblue$ and $\kred$ filters which distinguish between a dusty and old galaxy at $z = 3$ and a massive quiescent galaxy at $z = 5$ that are otherwise degenerate with standard photometric filters at limited depths.  This is highlighted by the $H$, $K_s$ and IRAC 3.6 micron and 4.5 micron bands that are insufficient to distinguish the Balmer/4000\AA\ break, which is the most prominent spectral feature in the $z\sim5$ SED.

To translate the science goal of identifying $z>4$ quiescent galaxies into the specific filter designs, we need to identify a strong color signature from the Balmer and 4000~\AA\ breaks over the redshift range these spectral features are present ($4 \lesssim z \lesssim 6$).  To explore the optimal filter characteristics, we computed a figure of merit (FOM) that optimizes the filter wavelength position and filter widths by maximising the redshift range sensitivity and hence increase the cosmic volume probed.  In the FOM, we include the redshift range probed by the additional filters, the sky background (including the thermal emission from the telescope), atmospheric transmission window, telescope optical path, and the source brightness as a function of wavelength and redshift.  The detailed procedure for the FOM, and description of these parameters are discussed in Appendix \ref{sec:appendixA}.

\begin{figure}
    \includegraphics[width=\linewidth]{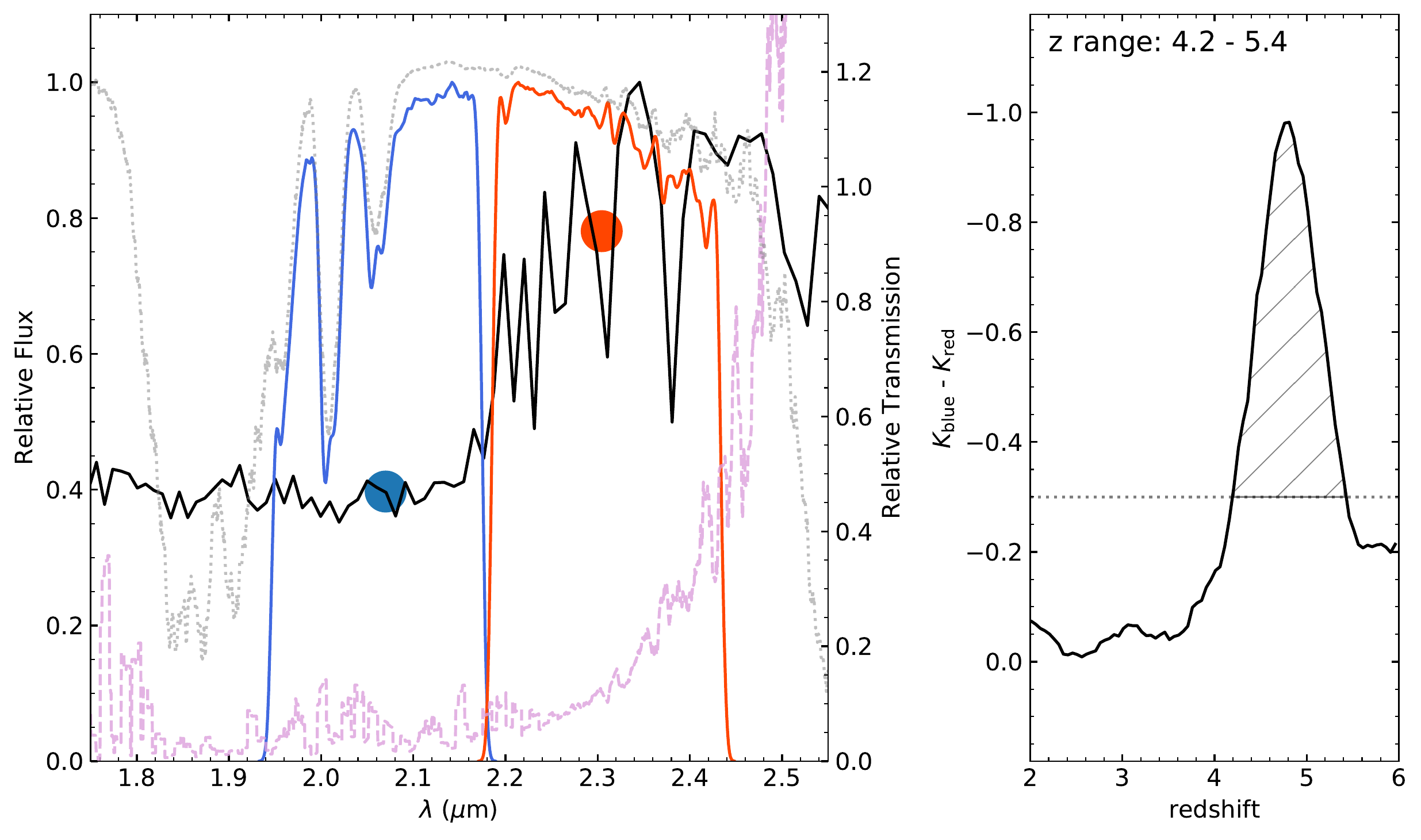}
    \caption{ 
    Filter design simulation for the dual filter set-up.  Left panel: $\kblue$ and $\kred$ filters, including atmospheric absorption (shown in blue and red respectively).  Black solid line is the fiducial 0.4 Gyr SSP model SED at $z = 4.6$.  The mauve dashed line is the Cerro Pach\'on background + telescope emission spectrum and the gray dotted line is the atmospheric transmission for a 2.3 mm water vapour column.  Red and blue dots are the synthetic photometry for $\kblue$ and $\kred$ respectively.  Right panel: $\kblue$ - $\kred$ color as a function of redshift.  The solid line is the predicted color as our model SED is redshifted from 2 to 6.  The horizontal dashed line shows the 3$\sigma$ color threshold used to approximate the redshift range probed by the filter configuration.
    }
    \label{fig:fom_ex}
\end{figure}

An illustration of the FOM procedure is shown in Fig.~\ref{fig:fom_ex} for the selected dual filter set-up with the left panel showing the simulation set-up with the many components considered, and the right panel showing the color measurement for this filter set sampling the SED in the left panel, over a specific redshift range.  To measure the redshift range probed by this filter configuration, a fiducial 0.4 Gyr single stellar population (SSP) galaxy model is redshifted from $2 < z < 6$ and the redshifts for which the $\kblue$-$\kred$ color is greater than some threshold represents the redshift range.  Here the age of the SSP galaxy is set to 0.4 Gyr so that the stellar population is sufficiently old to produce a sharp detectable Balmer break, which develops within 100-200 Myr in post starburst galaxies.  The upper redshift limit is where the Balmer and 4000~\AA\ break has redshifted beyond the $K$-band window.  The 0.3 magnitude threshold is based on achieving a 3$\sigma$ color measurement with typical $K_s$ magnitude depths achieved in the ZFOURGE catalogue \footnote{The typical 3-$\sigma$ $K$-band depth of ZFOURGE is 25.7 in a D=0.6” aperture (scaled to total)}.  The results are not sensitive to the exact choice of the threshold.  A simplification in this approach is that the apparent magnitudes of the galaxies are assumed to be constant over the $4 < z < 6$ redshift range, which is approximately correct for an instantaneous formation redshift of $6.0-7.0$.  A caveat to this methodology is that it only considers two filters and that it is not optimised for SSPs of different ages.

The main outcome of the FOM analysis is that the central wavelengths of 2 additional filters are pushed to either side of the $K$-band window with relative broad widths (R$\sim$7).  While it might seem counter-intuitive to have the $\kblue$ filter extend into the ``blue'' edge of the $K$-band atmospheric transmission window, or the $\kred$ filter overlap with exponentially increasing sky+telescope emission at wavelengths greater than 2.4 micron, these solutions are optimal when considering the fact that quiescent galaxy SEDs increase in flux with increasing wavelength. The other key factor is the fact that exposure time increases for narrower filter widths.  

Our final adopted design included slight modifications of the edges of $\kblue$ and $\kred$ filters to reduce overlap with the existing $K_s$ filter and improve photometric stability by reducing sensitivity to variations in atmospheric transmission due to water vapor and airmass. The final filter central wavelengths and widths are $\kblue$  = (2.06 $\mu$m, 0.25 $\mu$m) and $\kred$  = (2.31 $\mu$m, 0.27 $\mu$m), complementary to the existing $K_s$ (2.16 $\mu$m, 0.32 $\mu$m) filter.  This enables one to take advantage of existing deep $K_s$ photometric data with the 3 photometric filters together fully covering the $K$-band transmission window from $1.9 - 2.5$ micron.

\section{Simulations Using K-split Filters}
\label{sec:simulations}

\subsection{Photometric redshift and outliers}
\label{sec:photzsim}


To understand the impact of adding the $K$-split photometry to an existing multi-wavelength catalogue in determining photometric redshifts, a mock galaxy catalogue was simulated.  The synthetic photometry of the catalogue was generated to have similar depths and broad-band filter sets as the UltraVISTA catalogues in the COSMOS field ($u$, $B$, $V$, $g$, $r$, $i$, $z$, $J$, $H$, $K_s$, and the four SPITZER/IRAC channels at 3.6, 4.5, 5.8 and 8.0 micron), because these are representative of the current state of the art in deep fields. The depths for the $K$-split filters are $\kblue$ 5$\sigma$ depth = 25 mag, set to approximately match the $K_s$ depth in the UltraVISTA catalogue (based on the total flux), and the $\kred$ depth set to 5$\sigma$ depth = 24.3 mag, to reproduce similar observational constraints associated with this filter such as the comparatively lower observational efficiencies achieved on the \texttt{F2} instrument and brighter sky backgrounds.  These $K$-split depths are similar to the observed depths of the pilot survey shown in Section \ref{sec:survey}. 

\begin{figure*}
\begin{tabular}{cc}
\begin{minipage}[t]{0.9\linewidth}
\centering
\includegraphics[width=\linewidth]{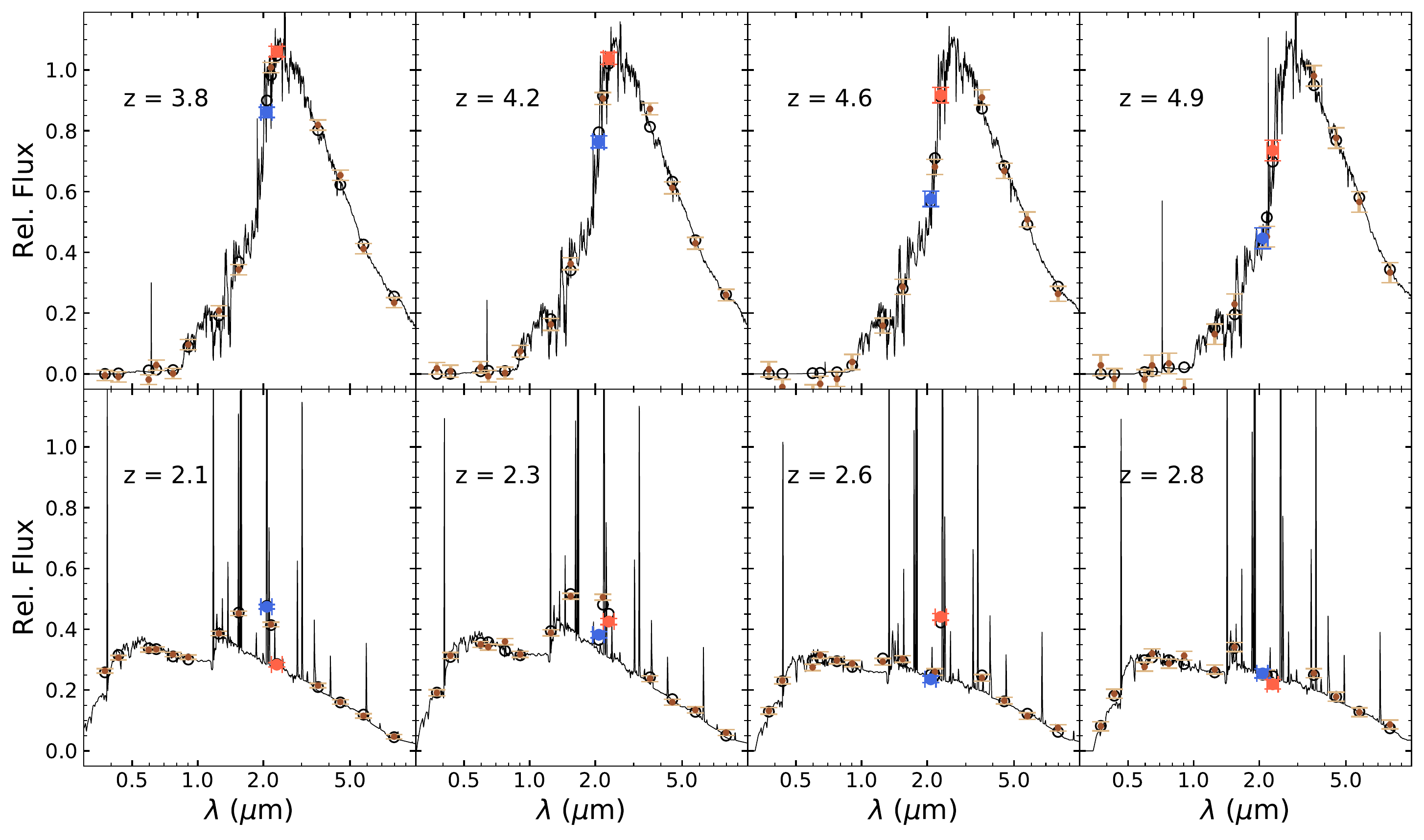}
\end{minipage}
\end{tabular}
\caption{Two example galaxy models with synthetic photometry for a range of redshifts used in the EAzY simulations to test the impact on photometric redshift precision when adding the $K$-split photometry.  In each panel, the black solid line shows galaxy SED, brown scatter points are the selection of photometry used in the simulations and the blue and red scatter points are the $\kblue$ and $\kred$ photometry respectively.  The top row shows the 0.9 Gyr SSP model, and the bottom row a dusty star-forming galaxy.}
\label{fig:sim_eg_sed}
\end{figure*}

The mock galaxy catalogues are comprised of several different aged SSPs, spanning a range of SNR to produce the expected range of galaxies one might observe in deep photometric surveys.  A total of five galaxy types were simulated: four SSPs with 0.1, 0.3, 0.6 and 0.9 Gyr ages (assuming solar metallicity and no dust attenuation) giving a range of Balmer/4000~\AA~ break amplitudes, and a dusty constant star-forming galaxy with strong emission lines.  The emission line galaxy has an ionisation potential, log(U)$= -2$, for strong [OIII] emission and a A$_V$ = 2 mag attenuation, \citep[using the][dust law]{Calzetti2001} to flatten the UV spectral slope and thus highlight the effect that the $K$-split filters can have on tracing emission line color signatures.  To establish which types of galaxies, and at what redshifts, are impacted most by the $K$-split filters, the mock catalogues are created with galaxies randomly drawn from a uniform distribution, over a large redshift range.

These galaxy models were generated using Bayesian galaxy modelling software: Bayesian Analysis of Galaxies for Physical Inference and Parameter EStimation \citep[BAGPIPES][]{Carnall2017} which uses \citet{Bruzual2003} stellar population models and emission line modelling with \texttt{Cloudy} \citep{Ferland2017}.  Examples of two galaxy models are shown in Fig.~\ref{fig:sim_eg_sed} at a range of redshifts to illustrate how the $K$-split filters move through the SEDs as they are redshifted.

To simulate a range of SNR, each galaxy has its flux scaled such that the $\kred$ band has a specific SNR (which was chosen as it has the shallowest depth in the $K$-band for these simulations).  Scaling the flux by the $\kred$ filter essentially produces a similar SNR profile across the multi-wavelength filters irrespective of redshift.  A consequence of the scaling is that it produces slightly different mass ranges at different redshifts (i.e. lower masses at low redshifts and higher masses at high redshifts). As a point of reference, a 0.9 Gyr SSP at $z\sim5$ with a SNR=5 in the $K$-split filters has a log(M*/M$_{\odot}$)=11.

To test the $K$-split filters impact on recovering the redshifts of the simulated galaxies, we used the photometric redshift fitting software EAzY \citep{Brammer2008}.  EAzY uses linear combinations of multiple template spectra to find the minimum chi-squared description of the observed SED, in order to construct a redshift probability distribution, P(z).  EAzY has been widely used to determine photometric redshift catalogues for multi-wavelength surveys including NMBS \citep{Whitaker2011}, UltraVISTA \citep{Muzzin2013b}, 3D-HST \citep{Momcheva2016}, and ZFOURGE \citep{Straatman2016}.  The template set is from ZFOURGE (see Section \ref{sec:software}) and includes standard EAzY templates plus a strong emission line galaxy template and an old and dusty galaxy template.  These two additional templates help demonstrate how the $K$-split filters can distinguish between galaxy SEDs that can resemble massive quiescent galaxies.  EAzY was run both with and without the new $K$-split medium bands to compare the impact on the recovered photometric redshifts.

\begin{figure*}
    \begin{tabular}{cc}
    \begin{minipage}[t]{0.9\linewidth}
    \centering
    \includegraphics[width=\linewidth]{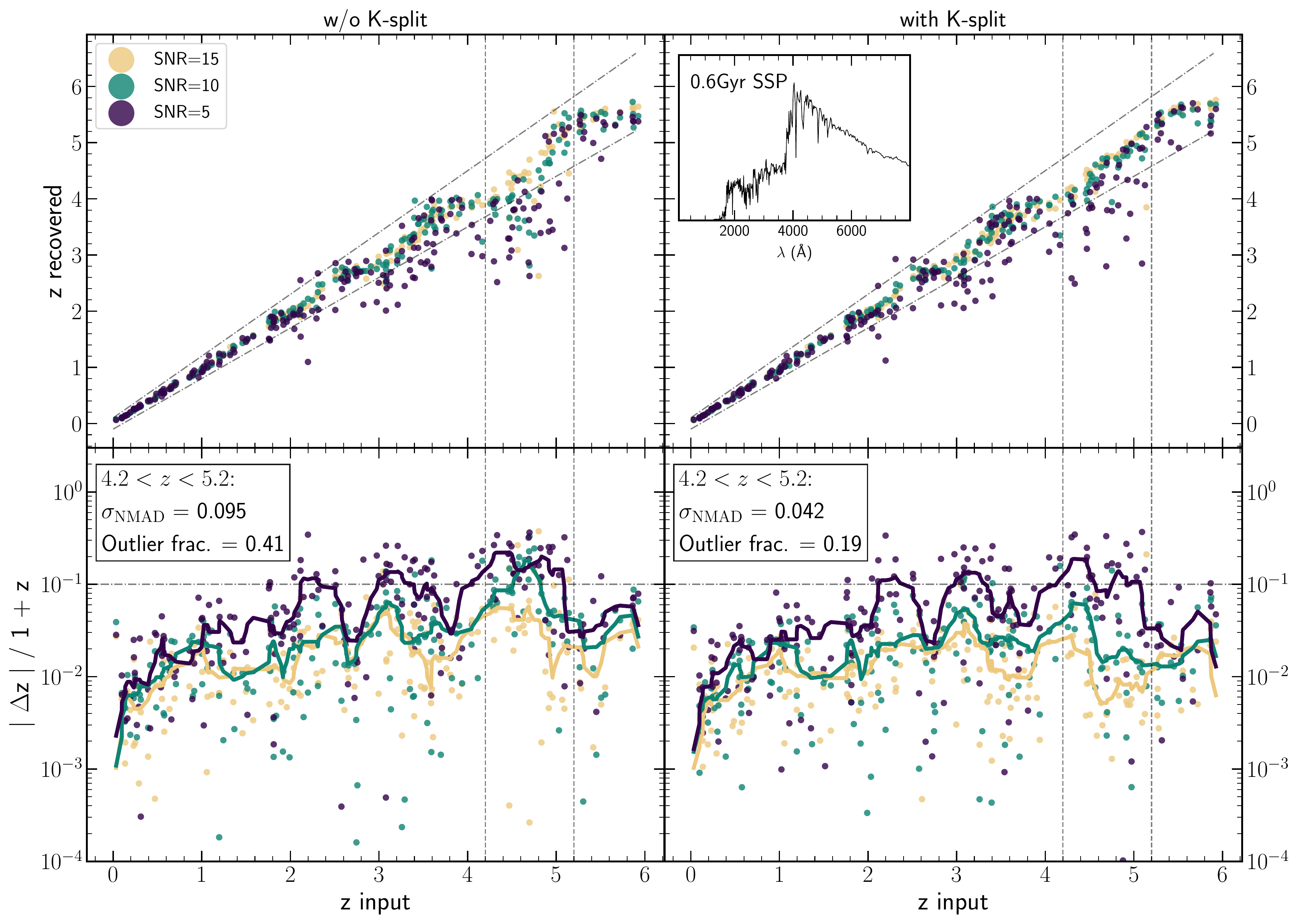}
    \end{minipage}
    \end{tabular}
    \caption{ 
    Testing the impact of the $K$-split filters on photometric redshift recovery using simulated photometry for a 0.6 Gyr SSP model.  The upper panel shows recovered redshift for a given redshift input without (left panel) and with $K$-split filters (right panel), with each scatter point colored by SNR of the $K$-split filters.  The lower panels show the corresponding residuals from the simulations, $|\Delta z|/( 1 + z )$, with and without $K$-split filters.  The solid colored lines indicate the running median of the residuals.  The gray dashed-dot lines show the $|\Delta z|/( 1 + z )$ = 0.1 outlier limit and the vertical grey dashed lines help outline the redshift ranges that are most impacted by the $K$-split filters.  A factor of 2.3 improvement in the NMAD scatter in $|\Delta z|/( 1 + z )$ is seen from $\signmad$ = 0.095 to $\signmad$ = 0.042 in the redshift range $4.2 < z < 5.2$ and outlier fraction is reduced by a factor of 2.1.
    }
    \label{fig:ez_sim_mass_gal}
\end{figure*}

\begin{figure}
    \includegraphics[width=\linewidth]{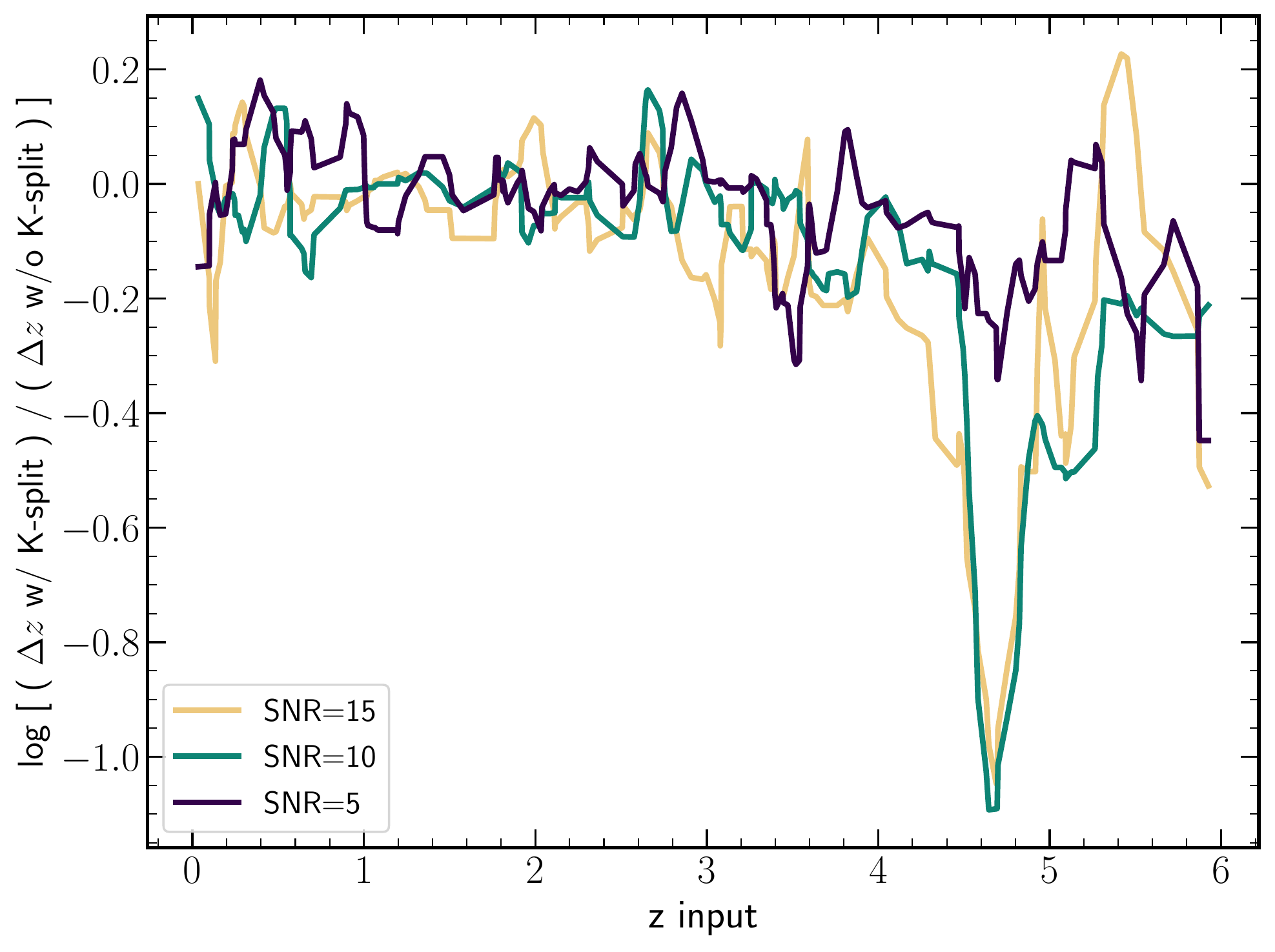}
    \caption{ 
    Ratio of the $\Delta z$ running median lines from the lower panel of Fig.~\ref{fig:ez_sim_mass_gal} for with and without $K$-split filters.  The ratio of the running median $\Delta z$ lines show the redshift range for which the $K$-split filters have an impact and the increase in photometric redshift accuracy.  The negative peak at $z\sim4.6$ shows the increase in photometric redshift accuracy by up to a factor of 10 using the $K$-split filters at $SNR\gtrsim10$.
    }
    \label{fig:deltz_ratio}
\end{figure}

\begin{figure*}
    \begin{tabular}{cc}
    \begin{minipage}[t]{0.9\linewidth}
    \centering
    \includegraphics[width=\linewidth]{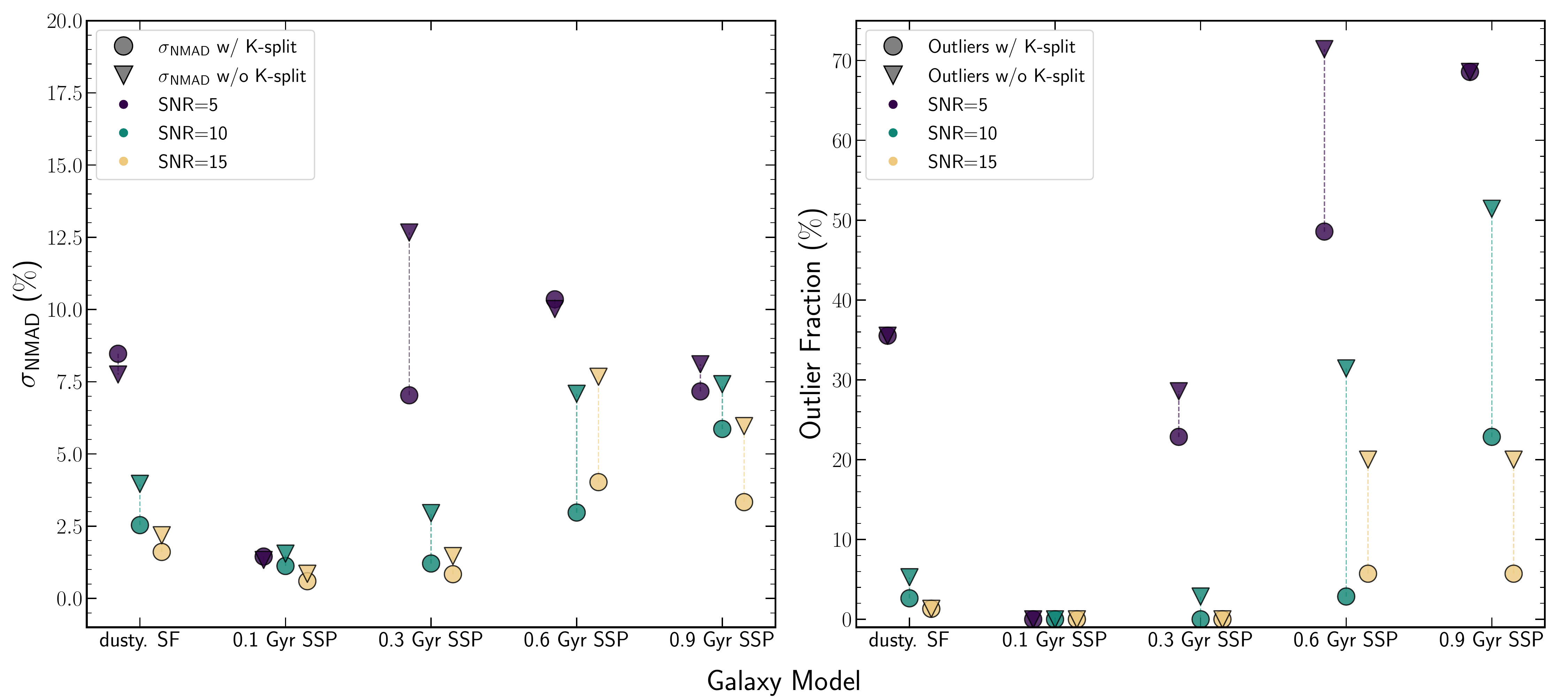}
    \end{minipage}
    \end{tabular}
    \caption{ 
    Summary of the $K$-split filters impacts on determining photometric redshifts for various galaxy models as described in the text.  $\sigma$\textsubscript{NMAD} is shown in the left panel and outlier fraction in the right panel with the circles and triangles showing these quantities for the simulations with and without the $K$-split filters respectively.  $\sigma$\textsubscript{NMAD} and outlier fractions are measured for the redshift ranges $2<z<4$ for the dusty star-forming galaxy and  $4.2<z<5.2$ for the SSPs, where the impact from the $K$-split filters are most prevalent (as seen in Figure \ref{fig:deltz_ratio} and Appendix \ref{sec:appendixB}).  Scatter points are colored similarly to Figure \ref{fig:ez_sim_mass_gal} and \ref{fig:deltz_ratio} with purple, green and yellow representing SNR=5, 10, 15 respectively.
    }
    \label{fig:ez_sim_sum}
\end{figure*}

There are two key metrics that we consider in our analysis of the EAzY results: photometric redshift precision and outlier fractions, both of which relate to the photometric redshift accuracy: $|z_\mathrm{input}$ - $z_\mathrm{recovered}|$ / ( 1 + $z_\mathrm{input}$ ) (hereafter $\Delta z / ( 1 + z )$).  The photometric precision metric is measured by the spread in the $\Delta z / ( 1 + z )$ for which we use the normalised median absolute deviation (NMAD), normalised by a factor of 1.48 (hereafter $\signmad$). The outlier fraction is defined as the number with  $\Delta z / ( 1 + z )>$ 0.1 relative to the number of galaxies in the sample.  These two metrics show the overall improvement in photometric precision and accuracy achieved by including the $K$-split filters.  The results for the 0.6 Gyr SSP mock galaxy catalogue are shown in Fig.~\ref{fig:ez_sim_mass_gal} where the upper panels show the input versus recovered redshift without and with the $K$-split filters (left and right respectively).  The lower panels show the $\Delta z / ( 1 + z )$ versus redshift and include running median curves shown for each SNR bin.  The outlier fraction and $\signmad$ are displayed for the full SNR population and for the redshift range $4.2<z<5.2$ where we see the most improvement from the $K$-split filters.  With the $K$-split filters, the outlier fraction is reduced by a factor of 2.4 and $\signmad$ improves by a factor of 2.3. The full set of diagnostic plots for the 4 other galaxy models are shown in Appendix \ref{sec:appendixB}.

To investigate the redshift ranges impacted by $K$-split filters, one can take the ratio of the running median $\Delta z$ lines with and without the $K$-split filters.  This is shown for the 0.6 Gyr SSP population in Fig.~\ref{fig:deltz_ratio}.  We can see here that the improvement in $\Delta z / ( 1 + z )$ for SNR$>$10 can be as high as a factor of 10 at a $z\sim4.6$, whereas this improvement is less at $z\lesssim4.2$ and $z\gtrsim5.2$.  Finally, one can see that there are minimum SNR requirements for the photometry to achieve photometric precision and accuracy improvement.  In the 0.6 Gyr SSP, SNR$\lesssim$5 produces minimal impact on the photometric precision.  It should be emphasised that for each increment in SNR, the SNR of all the filters are improved and so do not describe the minimum SNR required in the $K$-split filters.  An investigation varying the $K$-split SNR whilst keeping the other photometric errors constant is shown in the Section \ref{sec:snrsim}.

Figure \ref{fig:ez_sim_sum} summarises the $\signmad$ and outlier fractions for the simulations for each of the SNR bins and the five galaxy types. Note that these metrics are quoted for the redshift range $2<z<4$ for the dusty star-forming galaxy (labelled dusty SF in Figure \ref{fig:ez_sim_sum}) and the remaining 4 SSPs for the redshift range $4.2<z<5.2$, highlighting the areas where the most improvement is seen from the $K$-split filters.  All galaxy types see at least some improvement in the outlier fraction and $\signmad$, with the exception of the 0.1 Gyr SSP population, which has no strong Balmer/4000~\AA~ break and whose photometric redshift performance is dominated by the strong Lyman break.  In general, the most improvement is seen for the 0.6 Gyr and 0.9 Gyr SSP populations.  The $\signmad$ for the 0.6 Gyr and 0.9 Gyr SSP galaxies at SNR=10 are both reduced a factor of 2.4 and the outlier fraction is reduced by a factor of 11 and 2.3, respectively.  While there are similar relative improvements for the 0.3 Gyr SSPs at SNR=10 (factor of 2.4 improvement in $\signmad$ and 100$\%$ reduction in outliers), the outlier fraction and $\signmad$ values are significantly lower even without the $K$-split filters.  Interestingly, the impact of the $K$-split filters on the 0.3 Gyr SSP is most prominent at SNR=5, with large improvements to $\Delta z / (1+z)$ impacting the $\signmad$ despite relatively unchanging the outlier fraction.  Finally, for the dusty star-forming galaxy we use the redshift range $2<z<4$, where the emission lines for [OIII] and H$\alpha$ enter the $K$-split filters.  Here the outlier fraction is reduced by a factor of 2 and the $\signmad$ is improved by a factor of 1.5, although these quantities are aggregated over a broad redshift range and individual improvement in $\Delta z / ( 1 + z )$ can be as high as a factor of 4.

The results of these simulations indicate that at sufficient SNR, the most impact from the $K$-split filters is apparent for the older SSP populations, where there is very little rest-frame UV flux and the photometric redshift is strongly influenced by identifying the Balmer/4000~\AA\ break.  It should be noted here that the outlier fraction and $\signmad$ quoted in this section is dependent on the photometric filter sets and the depths used.  Therefore, while quantitative comparisons are still relevant, their absolute value may change accordingly.

\subsection{Impact of $K$-split signal-to-noise on photometric redshift precision}
\label{sec:snrsim}

The simulations in the previous section included varying the SNR for all the photometric filters, however it also useful to assess the sensitivity of the SNR of the $K$-split filters given an existing photometric catalogue at fixed depths.  As an additional outcome, we can determine the ideal depth of the $K$-split photometry given existing photometric catalogues.
To investigate the impact of the $K$-split photometry SNR on the photometric redshift probability distribution function, P(z), we conducted the following simulations. We simulate photometry for two different galaxy SEDs, fixed at $z = 4.6$, where the $K$-split filters sample the largest color signature for an evolved galaxy with a Balmer/4000\AA\ break, and incrementally increase the SNR of the $K$-split photometry while maintaining the SNR of the other bands at the same depths as those used in Section \ref{sec:photzsim}.

The two galaxy SEDs are simulated using BAGPIPES with (1) a massive post starburst galaxy with residual star-formation and strong dust attenuation and (2) a massive quiescent galaxy based on the same 0.9 Gyr SSP model used in Section \ref{sec:photzsim}. These two galaxy SEDs are chosen to be representative of the type of quiescent and quenching galaxies one might expect at $z>4$.  The filter set and depths used are the same as Section \ref{sec:photzsim}.  Galaxies are modelled with various stellar masses such that their $K_s$ band luminosities are the same.  Photometry is generated from the model galaxy with a scatter consistent with the photometric errors. In the simulation, the $K$-split photometry SNR is incrementally increased and EAzY is used to derive a P(z) using the full template set described in \ref{sec:photzsim}.

\begin{figure*}
\begin{tabular}{cc}
\begin{minipage}[t]{\linewidth}
\centering
\includegraphics[width=\linewidth]{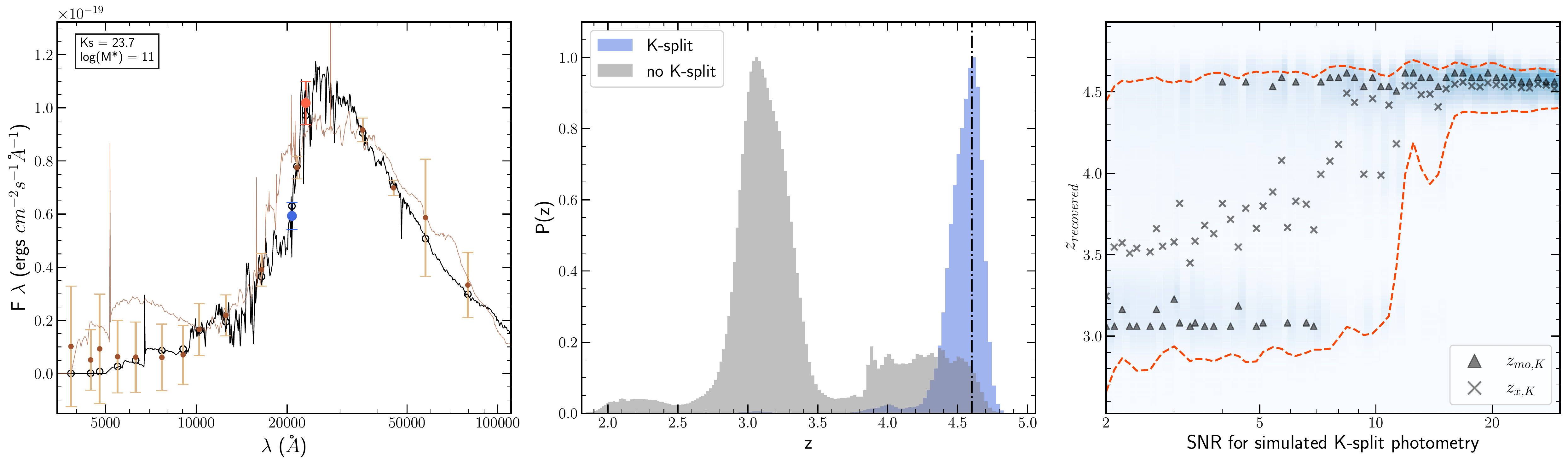}
\end{minipage}
\\
\begin{minipage}[b]{\linewidth}
\centering
\includegraphics[width=\linewidth]{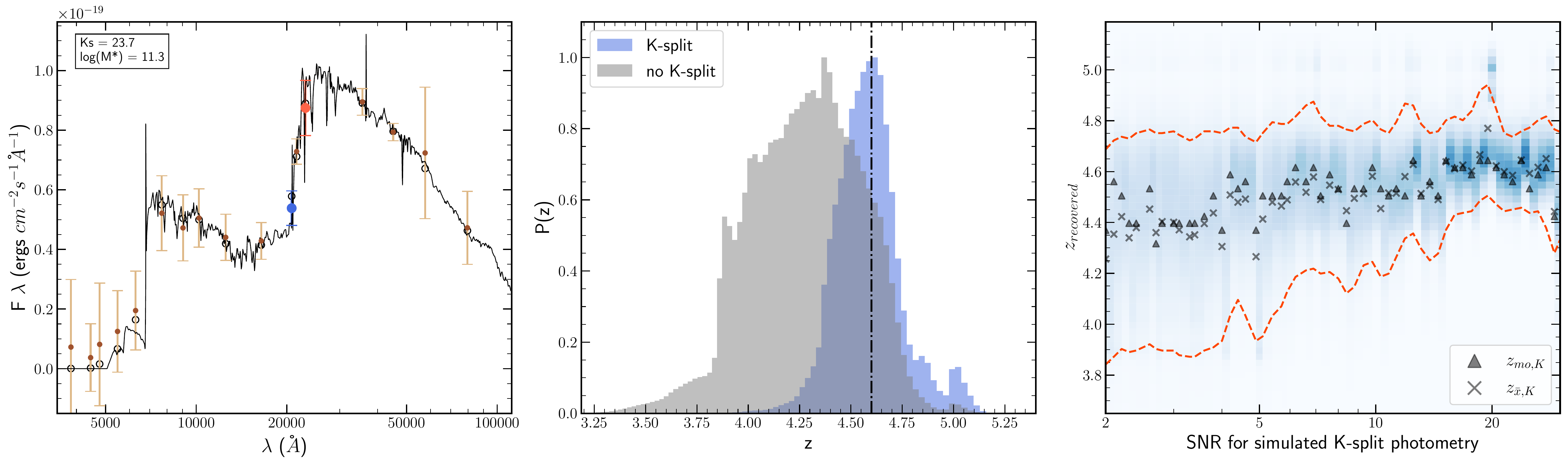}
\end{minipage}
\end{tabular}
\caption{Impact of the signal-to-noise ratio and SED shape on the photometric redshift improvement for galaxies at $z > 4$ with substantial Balmer and 4000~\AA\ breaks.  Each row shows a model galaxy at z=4.6 with simulated photometry matching the fiducial filter set used in Section \ref{sec:simulations}.  The top row shows a 0.9 Gyr SSP model and the bottom row is for a post starburst galaxy with significant dust extinction (A$_V$=1) and residual star-formation.  Left panel: same layout as for the SED panel in Fig.~\ref{fig:sim_eg_sed}, the extra brown line in the top left panel shows the best-fit SED without $K$-split photometry (a dusty galaxy at $z = 3$).  Middle panel: P(z) for respective galaxy SEDs, blue histogram is with simulated $K$-split photometry for SNR=10, gray histogram shows P(z) without $K$-split photometry, vertical dot-dashed black line shows the true redshift of the simulated galaxies.  Right panel: EAzY recovered redshift vs $K$-split SNR.  The blue shaded area represents the P(z) density.  The red dashed lines represent the upper and lower 95th percentile for each P(z).  The crosses are the redshift based on the weighted mean of the P(z) with a K luminosity prior ($z_{\bar{x},K}$) and the triangles are the redshift corresponding to the mode of the P(z), also with the luminosity prior ($z_{mo,K}$).
}
\label{fig:sim_SNR_mod_gals}
\end{figure*}

Fig.~\ref{fig:sim_SNR_mod_gals} shows the results of the test, for which there are two key takeaways.  The first is that the galaxy SED shape is important in determining the contribution of $K$-split photometry to P(z).  The narrowing of the P(z) (seen from the 95th percentile lines shown as red dashed lines in the right panel of Fig.~\ref{fig:sim_SNR_mod_gals}) for the massive quiescent galaxy SED in the top row is more significant than for the post starburst galaxy in the bottom row, with a decrease in the width of the P(z) by a factor of $\sim$ 6 and $\sim$ 2 respectively.  This is in part due to the substantial Lyman-break which is sampled by observer-frame optical bands, which already constrains the redshift of the massive Lyman-break galaxy well.  Secondly, from the steep rise in the 95th percentile line in the top-right panel of Fig~\ref{fig:sim_SNR_mod_gals} at a $K$-split SNR$\sim$10, it is evident that there is a critical SNR where the impact of the $K$-split filters starts to emerge.  Conversely, consistent with Section \ref{sec:photzsim}, it appears that there is no substantial improvement to the photometric redshift with a $K$-split SNR$\lesssim$5 in either galaxy model.

The marked improvement in photometric redshift precision of the 0.9 Gyr SSP model at SNR $\sim$ 10 can be understood by looking at the massive quiescent galaxy SED in the top row of Fig.~\ref{fig:sim_SNR_mod_gals}.  This suggests constraining the steep Balmer and 4000~\AA\ break with the $K$-split filters requires a minimum SNR of $\sim10$. At this SNR the simulation results indicate we can likely distinguish between the two different SEDs shown in the top-left panel (a quiescent galaxy at $z = 4.6$ in black and secondary solution of a dusty galaxy at $z = 3$).  Once this threshold SNR is achieved in the $K$-split photometry, the bi-modal P(z) collapses to the correct redshift of the simulated galaxy. In this way, the gain in photometric precision can also be considered synonymous with discriminating between two different galaxy SED types. For the 0.9 Gyr SSP case, the redshift probability without $K$-split filters is not well-constrained and cannot distinguish between a $z\sim3$ dusty, old galaxy and a $z\sim5$ quiescent galaxy. The inclusion of the $K$-split filters with sufficient SNR effectively excludes the $z\sim3$ solution.

A final point to take from this simulation is with respect to the interpretation of the P(z) and the chosen redshift estimate.  While it is often standard practice to choose a redshift weighted mean, with a luminosity prior ($z_{\bar{x},K}$) to avoid selecting a lower probability density peak, the upper-right panel of Fig.~\ref{fig:sim_SNR_mod_gals} shows the prior may not be appropriate when there are potential SED degeneracies.  In this case $z_{\bar{x},K}$ value falls between the two bi-modal peaks suggesting a solution that has a low likelihood.  Therefore, in this instance, the mode of the P(z), $z_{mo,K}$, is more robust.  The lesson is that large disagreements between $z_{mo,K}$ and $z_{\bar{x},K}$ should be treated with caution.

\subsection{Recovering Stellar Mass and Age with the K-Split Filters}
\label{sec:sim_mass+age}

Using the same mock catalogues from Section \ref{sec:photzsim}, we conducted stellar population modelling using FAST \citep{Kriek2009} to investigate the impact of the $K$-split filters on determining stellar mass and age, two key characteristics of massive quiescent galaxies.  FAST was run using the 0.6 Gyr and 0.9 Gyr mock galaxy catalogues both with and without the $K$-split filters and fixing the redshift to the simulated values.  Comparisons were made to the recovered masses and ages at the redshift ranges where the expected impact from the $K$-split filters is the most i.e. $4.2<z<5.2$.  The results are that the standard deviation in the residuals with and without the $K$-split filters are mostly unchanged for the stellar masses and ages in both the 0.6 Gyr and 0.9 Gyr SSP models.  There is a low overall scatter in the residuals for both age and mass without the $K$-split filters ($\lesssim$0.1 dex for all SNR).  This shows that, given the correct redshift, the overall SED shape, and hence age and mass-to-light ratio, is well constrained by the broad-band filters.  This highlights that the main impact the $K$-split filters have in selecting quiescent galaxies is from correctly determining photometric redshifts, without which the masses and ages cannot be constrained meaningfully.


%
\subsection{Predictions for Constraining Quiescent Galaxy Stellar Mass Function at $z>4$}
\label{sec:llp_pred}

\begin{figure*}
\begin{tabular}{cc}
\begin{minipage}[t]{0.45\linewidth}
\centering
\includegraphics[width=\linewidth]{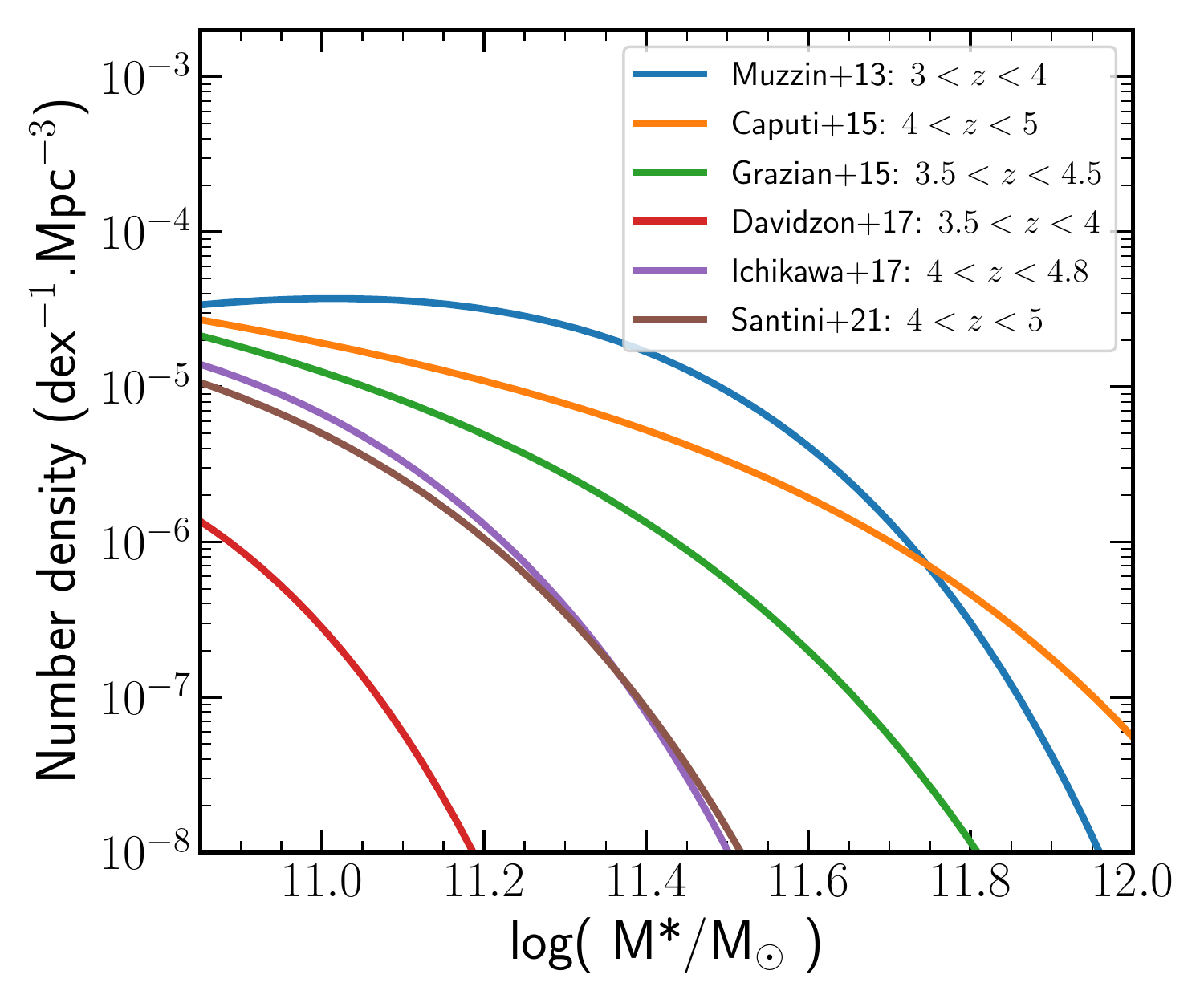}
\end{minipage}
&
\begin{minipage}[t]{0.45\linewidth}
\centering
\includegraphics[width=\linewidth]{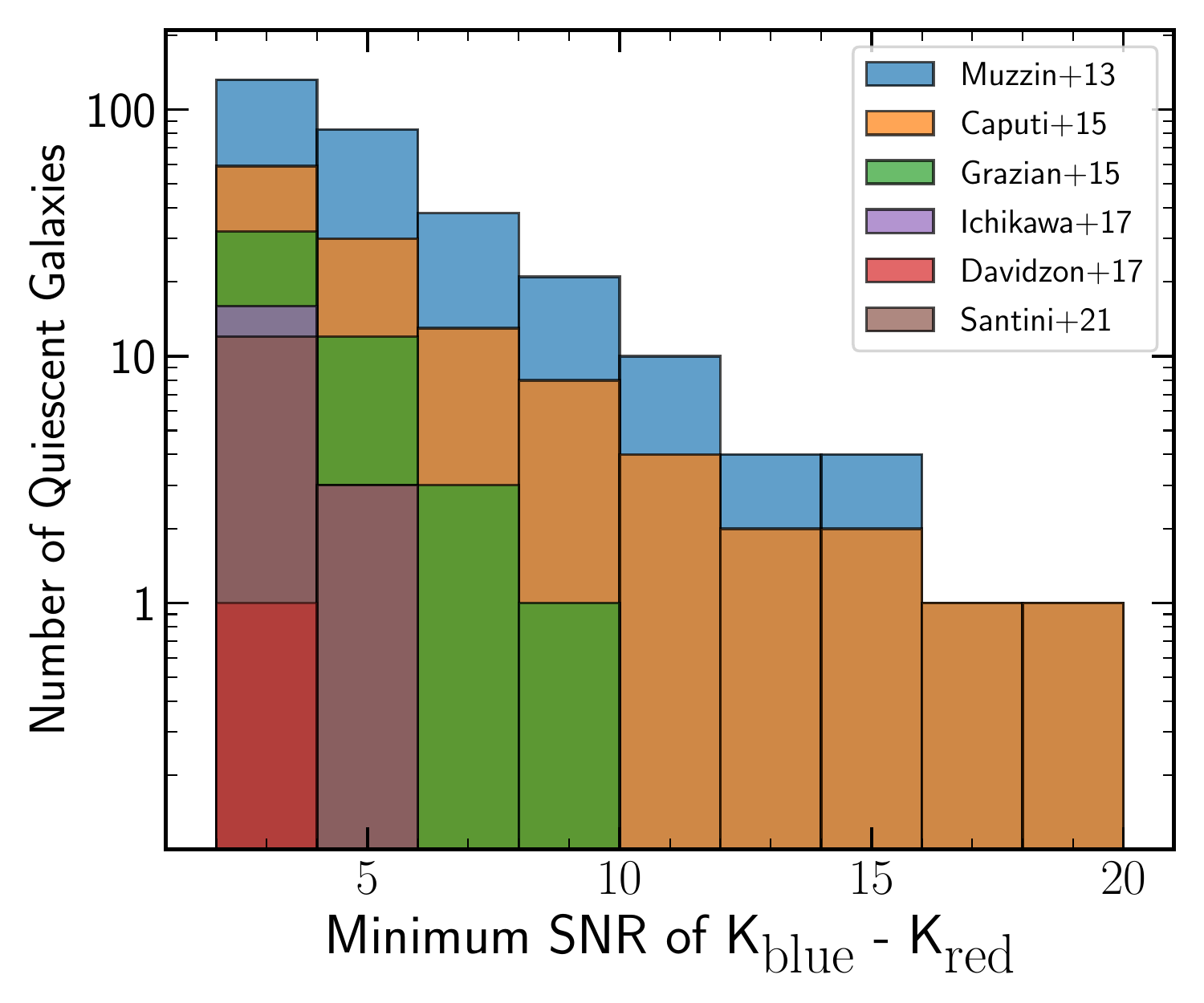}
\end{minipage}
\end{tabular}
\caption{ 
        Predictions for the cumulative numbers of quiescent galaxies at $z>4$ for a wide area survey utilising $K$-split filters.  Left: selection of quiescent GSMFs from literature including \citet{Muzzin2013b}, \citet{Caputi2015}, \citet{Grazian2015}, \citet{Ichikawa2017}, \citet{Davidzon2017} and \citet{Santini2021}.  \citet{Caputi2015} and \citet{Grazian2015} quiescent GSMF are based on the total GSMF assuming a quiescent fraction of 15$\%$.  Right: Histogram of the expected cumulative numbers of quiescent galaxies at $z>4$ with a minimum $\kblue$-$\kred$ SNR, using the corresponding quiescent GSMF from the left panels.
    }
\label{fig:llp_pred}
\end{figure*}

For a wide area survey, there is the opportunity to better constrain the quiescent galaxy stellar mass function (GSMF) at $z>4$, with more robust photometric redshifts using the $K$-split filters.  To date, various measurements of the GSMF have been made at $z\gtrsim4$, however given the challenges in photometrically selecting quiescent galaxies at high redshift, there are a wide range of values.  The left panel of Fig~\ref{fig:llp_pred} shows a selection of quiescent GSMF published in recent years at the highest redshifts.  It is noted that while these GSMFs do not all span the same redshift ranges and have different selection methods, the shape, normalization and evolution of the quiescent GSMF at high-redshift is currently not well constrained and therefore provides the range of possible values.  Given the large range in the GSMFs, spanning 2 orders of magnitude at log(M*/M$_{\odot}$)=11, there is clearly substantial motivation to further constrain them.

A wide area survey using the $K$-split filters is already underway with the FENIKS Large and Long program (PI Papovich).  This survey will cover a large area, $\sim$0.6 deg$^2$, and 5$\sigma$ depths of 24.3 mag in $\kblue$ and 23.3 mag in $\kred$, aimed at maximising the number of candidates (see Papovich in prep. for more details).

To predict the numbers of quiescent galaxies identified in such a survey, we ran the following analysis.  To determine the likely numbers of quiescent galaxies, we integrated the GSMFs in the left panel of Fig~\ref{fig:llp_pred} at log(M*/M$_{\odot}$)$>$10.9 in numerous mass bins and converted them to a distribution of masses by using the survey area and converting to a volume assuming a redshift range of $4.2<z<5.2$.  We then converted these to $K$-band luminosities by modelling a passively evolving galaxy with $z\textsubscript{form}\sim7$, and then calculated their SNR based on the $K$-split photometry using the proposed large area survey depths.  The right panel of Fig~\ref{fig:llp_pred} shows the expected number of quiescent galaxies for which we would be able to make $\kblue$-$\kred$ color measurements at a certain SNR.

At the redshift range of $4.2<z<5.2$, the passively evolving galaxy has a $\kblue$-$\kred >$ 0.4 mag.  Given the effective selection of quiescent galaxies with a SNR$\sim$10 in the $K$-split photometry (as demonstrated in Section \ref{sec:photzsim} and \ref{sec:snrsim}) one can infer a $\kblue$-$\kred>0.4$ at 3$\sigma$ significance should be sufficient to identify quiescent galaxies.  Using this selection criteria the wide area survey could robustly identify anywhere between 1 and 120 quiescent galaxies, based on the variation in quiescent GSMF present in the literature.  This survey would therefore provide strong constraints for the quiescent GSMF at $z>4$.

To test the capabilities of the $K$-split filters on a smaller pilot survey we consider a smaller survey area of 100 arcmin$^2$ and target depths motivated by the analysis in Section \ref{sec:snrsim}.  Based on the SNR analysis, a galaxy with log(M*/M$_{\odot}$)=11 and SNR$\sim$10 in the $K$-split filters translates into a target 5$\sigma$ depth for a survey of 24.0 AB magnitude in $\kred$ and 24.8 AB magnitude in $\kblue$.

Scaling the numbers of identified quiescent galaxies from the GSMF analysis above, down to match the area of the pilot survey would mean identifying between 0 and 6 quiescent galaxies.  Here we also consider direct measurements of number densities from literature.  Using the number densities from \citet{Pampliega2019}, which probe a $4<z<5$ sample with stellar mass of log(M*/$M_{\odot}$)$>$11, we expect to encounter 0.9$\pm$0.5 massive quiescent galaxies.  As a comparison, the number of quiescent galaxies identified based on existing literature values could be as high as 2.9$\pm$1.4 \citep[based on][]{Santini2021} or as low as 0.5$\pm$0.2 \citep[based on ][]{Marsan2020}, both of which include consideration for cosmic variance.  In contrast if there is no evolution of the massive quiescent galaxy number densities from $3<z<4$, using the value from \citet{Straatman2014}, one might expect 4.6$\pm$1.8 massive quiescent galaxies per pilot survey area.  While a larger survey area is ideal to identify many candidates that are not subject to field-to-field variance, a small pilot survey area such as this could be considered the minimum area to be covered to in order to identify quiescent galaxies at $z>4$.

\section{The FENIKS Pilot Survey}
\label{sec:survey}

\subsection{Survey Description}
After the design and commissioning of the $K$-split filters, a pilot survey was carried out to test the impact of the $K$-split filters using real data.  The FENIKS pilot survey (Co-PIs: C. Papovich, E. Taylor, C. Marsan) is an eight night survey with the \texttt{F2} instrument on the 8.1\ m Gemini South Telescope at Cerro Pach\'on observatory, Chile.  The observations were undertaken predominantly in queue mode starting in November 2017 and finishing April 2019 (programs GS-2017A-Q-10, GS-2017B-C-2, GS-2017B-FT-16, GS-2018A-Q-212, GS-2018A-Q-213, GS-2018A-DD-101, GS-2018B-Q-124, GS-2019A-FT-101).  Seeing conditions were generally excellent with 50th percentile seeing of 0.45" in $\kred$ (0.37" and 0.65" upper and lower quartiles) and 0.49" in $\kblue$ (0.38" and 0.74" upper and lower quartiles).

The strategy for this survey is to add the $\kblue$ and $\kred$ photometry to existing, deep photometric catalogues: UltraVISTA \citep{Muzzin2013b} DR3 (private communication) and ZFOURGE \citep{Straatman2016} catalogues, reaching depths consistent with those in section \ref{sec:llp_pred}, and in doing so, push the limits of these surveys to explorations of higher redshifts, in particular at $z > 4$.  Three fields in total were chosen across COSMOS and CDFS from pre-selected catalogues to maximize the discovery of potential $z\textsubscript{phot} > 4$ massive galaxies.  A summary of the exposure times and image depths are shown in Table~\ref{tab:obs_sum}.  The \texttt{F2} circular FOV is 6.1' diameter (29.2 arcmin$^2$ per field) making up a total survey area of 87.6 arcmin$^2$.  This survey area is $\sim$ 25$\%$ of the ZFOURGE survey area and $\sim$ 1$\%$ of the UltraVISTA deep survey area.

\begin{table}
  \begin{center}
    \caption{Observational summary}
    \label{tab:obs_sum}
    \begin{tabular}{c|c|c|c} 
    
      \hline \hline
      \textbf{Field} & \textbf{Filter} & \textbf{Total exposure} & \textbf{5$\sigma$ depth}\textsuperscript{a} \\
      (R.A., Dec.) &  & \textbf{time (hr)} & \textbf{(magnitude)}\\      
      \hline
      CDFS & $\kblue$ & 1.9 & 24.1\\
        (53.082, -27.809) & $\kred$ & 2.8 & 23.7\\
      \hline
      COSMOS 352 & $\kblue$ & 3.2 & 24.7\\
       (150.090, 1.703) & $\kred$ & 3.7 & 24.3\\
      \hline
      COSMOS 544 & $\kblue$ & 6.6 & 25.0\\
       (150.442, 2.557) & $\kred$ & 3.0 & 24.0\\
    \end{tabular}
  \end{center}
  {\raggedright
   \footnotesize{a - Depths are based on the standard deviation of random empty 0.6" ``optimal" apertures, corrected for flux outside of the aperture in the corresponding images.}\\
   }
\end{table}


\subsection{\texttt{FLAMINGOS-2} Image Reduction}
\label{sec:data}
The \texttt{F2} $\kblue$ and $\kred$ data were reduced using a custom IDL pipeline, written by one of the authors (I. Labb\'e), used in the ZFOURGE \citep{Straatman2014} and NMBS {\citep{Whitaker2011} surveys.  The pipeline utilizes a two-pass sky subtraction method based on the methodology of the \texttt{IRAF} package \citep{Tody1986} \texttt{xdimsum}, which produces sky subtracted images from sets of dithered observations.  Our final reduced images are flat fielded, point-spread-function (PSF) and pixel matched to the $K_s$ detection images (both legacy catalogue detection images have pixel scales of 0.15" per pixel) and photometrically calibrated using a nearby NIR spectro-photometric standard star (selected from the Calibration Database System \footnote{\url{https://www.stsci.edu/hst/instrumentation/reference-data-for-calibration-and-tools/astronomical-catalogs/calspec}}).  Our standard stars were observed in conjunction with our science fields to achieve $\sim$1$\%$ precision on our zero points.  A series of secondary standards were selected in each field to calibrate observations on different nights to the zeropoint of our primary calibrated science observations.  Some custom reduction methods were utilized to deal with thermal radiation from the on-instrument wave-front sensor (OIWFS) illuminating the \texttt{F2} detector.  A more detailed treatment of the data reduction can be found in an upcoming paper (Esdaile et al. in prep).

\subsection{Photometry and Catalogue}
\label{sec:catalogue}
The FENIKS pilot survey catalogues are created by adding the \texttt{F2} photometry to the existing legacy catalogues, which are based on the $K_s$ band detections.  Photometry is undertaken by placing fixed apertures at the same RA/Dec locations for each corresponding legacy catalogue object. As the legacy catalogues use large apertures for their photometry (1.2" and 2.1" diameter in ZFOURGE and UltraVISTA catalogues respectively) and the $K_s$ images for these catalogues are deeper than the \texttt{F2} images, matching aperture sizes would produce lower SNR $K$-split photometry and diminish their impact on determining photometric redshifts.  Taking advantage of the exceptional seeing conditions for the \texttt{F2} images, we can improve the SNR of the \texttt{F2} photometry by selecting smaller, `optimal’ apertures.  Optimal apertures were selected by investigating the SNR curve of growth for a star in the PSF matched \texttt{F2} images.  By limiting the aperture sizes to the PSF of the $K_s$ images (0.9” and 1.05” in ZFOURGE and UltraVISTA respectively) we were able to increase the SNR by up to 58$\%$.  Limiting the aperture to the PSF avoids introducing any systematic errors such as from centroiding or from limited knowledge of the PSF curves of growth.

Using optimal apertures however presents the problem of having different aperture sizes from the legacy catalogues, meaning color measurements would be made without the same fraction of light within an aperture.  This is particularly important to address as colors are used in determining photometric redshifts.  Therefore in order to be consistent with the aperture fluxes in the legacy catalogues, we scale our fluxes such that the ratio of the fluxes, or colors i.e. $K_s$ - $K$x, are consistent irrespective of aperture size.  This is done using Equation \ref{eq:opt_ap} below which shows the scaling as a ratio of the optimal aperture fluxes to catalogue aperture fluxes.
\begin{equation}
\label{eq:opt_ap}
    F_{K\mathrm{x,cat}} = F_{K\mathrm{x,opt}}\times\frac{F_{K\mathrm{s,cat}}}{F_{K\mathrm{s,opt}}}
\end{equation}
It is noted that these corrections assume that there is a negligible color gradient, which is reasonable for the compact sizes of massive quiescent galaxies at $z>4$.  Errors are calculated by placing random, empty, non-overlapping apertures over the field and calculating the standard deviation of these aperture fluxes.  These errors are also scaled to catalogue aperture sizes as per Equation \ref{eq:opt_ap}.  Additionally, the errors are scaled by the inverse square-root of the relative exposure time at each objects location.  This ensures objects on the periphery of the field have larger errors consistent with lower per pixel exposure time due to dithering.

The K-split fluxes are scaled to total, or left as fixed aperture fluxes, consistent with the format of the respective legacy catalogues.  We note that while the legacy catalogues have their own contamination flags, we simply inspected image cut-outs of \texttt{F2} images with galaxies of interest to ensure no contamination from nearby sources was present. Finally, the fluxes and errors are normalised to match the corresponding zero points of the catalogues which are both 25 AB magnitude for UltraVISTA and ZFOURGE catalogues (corresponding to a flux density of 3.631$\times10^{-30}$erg s$^{-1}$Hz$^{-1}$cm$^{-2}$ or 0.3631 $\mu$Jy).

\subsection{Photometric redshifts and stellar population modelling}
\label{sec:software}
Photometric redshifts are determined using EAzY.  The templates used are the same as UltraVISTA and ZFOURGE (and in the simulations of \ref{sec:photzsim}), which are both comprised of PEGASE stellar population synthesis models \citep{Fioc1999}, along with some additional SEDs to supplement the standard EAzY template set by including old and dusty galaxies and strong emission line galaxies.

Stellar population modelling was done using FAST \citep{Kriek2009}.  Stellar mass, SFR, dust extinction and age are all derived by fitting \citet{Bruzual2003} models assuming \citet{Chabrier2003} initial mass function (IMF) and exponentially declining SFH with timescale $\tau$, solar metallicity and \citet{Calzetti2001} dust attenuation law.  FAST was run using photometric redshift outputs from EAzY which is a standard technique also employed by the ZFOURGE and UltraVISTA catalogues.

\section{Results}
\label{sec:results}
\subsection{Massive Quiescent Galaxies at $z>4$}
\label{sec:mass_gal}

\begin{figure*}
\begin{tabular}{cc}
\begin{minipage}[t]{0.5\linewidth}
\centering
\includegraphics[width=\linewidth]{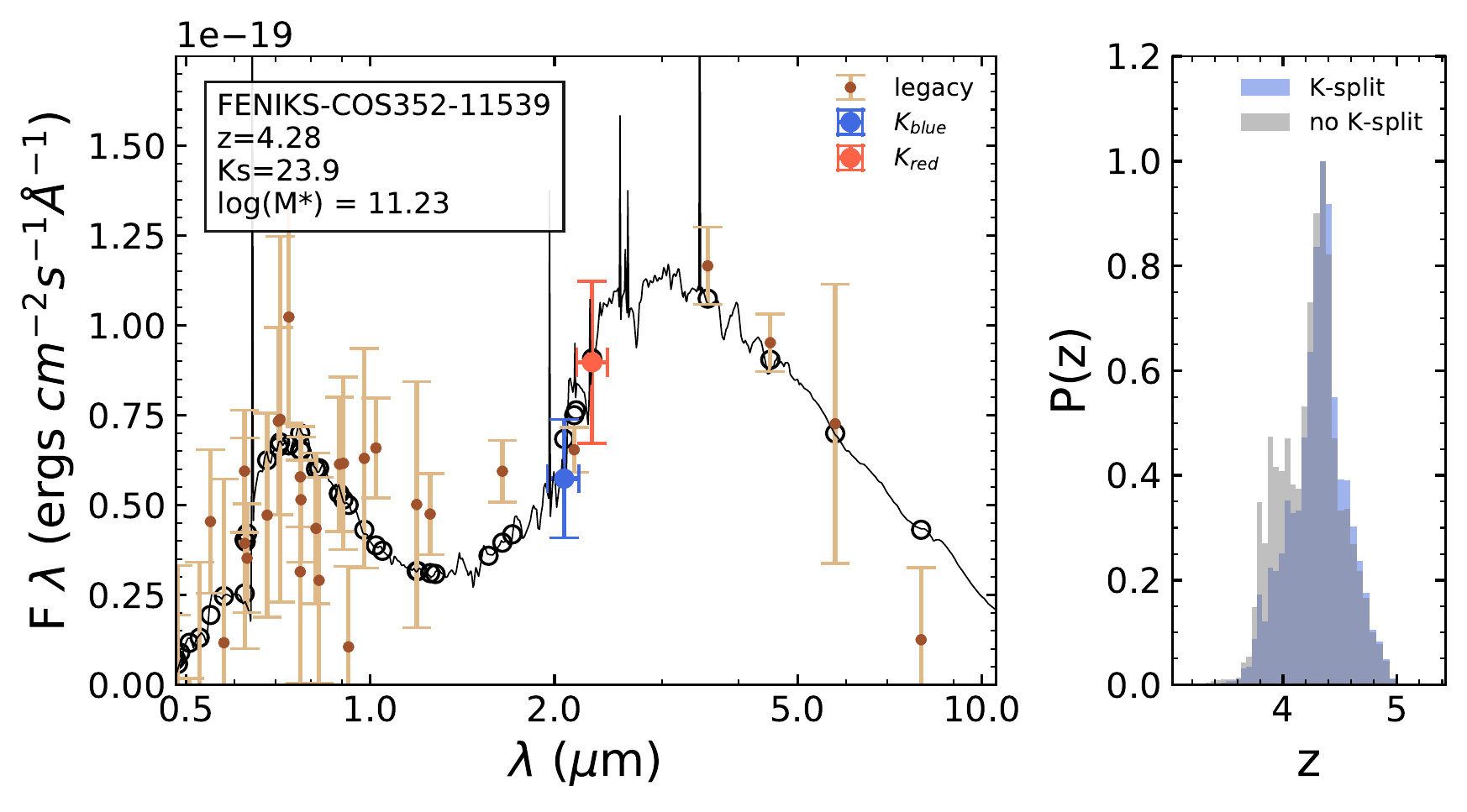}
\end{minipage}
&
\begin{minipage}[t]{0.5\linewidth}
\centering
\includegraphics[width=\linewidth]{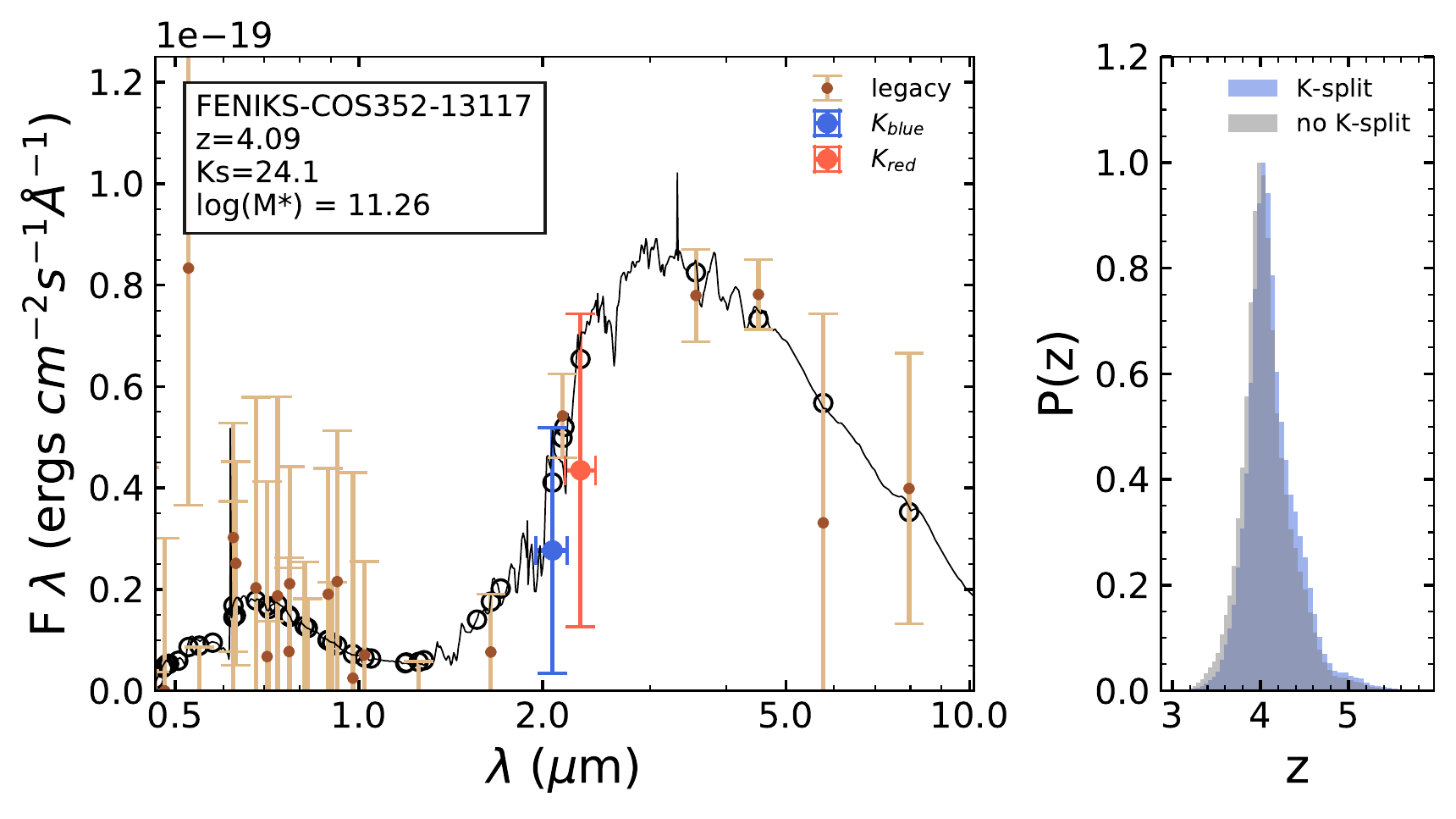}
\end{minipage} 
\\
\begin{minipage}[b]{0.5\linewidth}
\centering
\includegraphics[width=\linewidth]{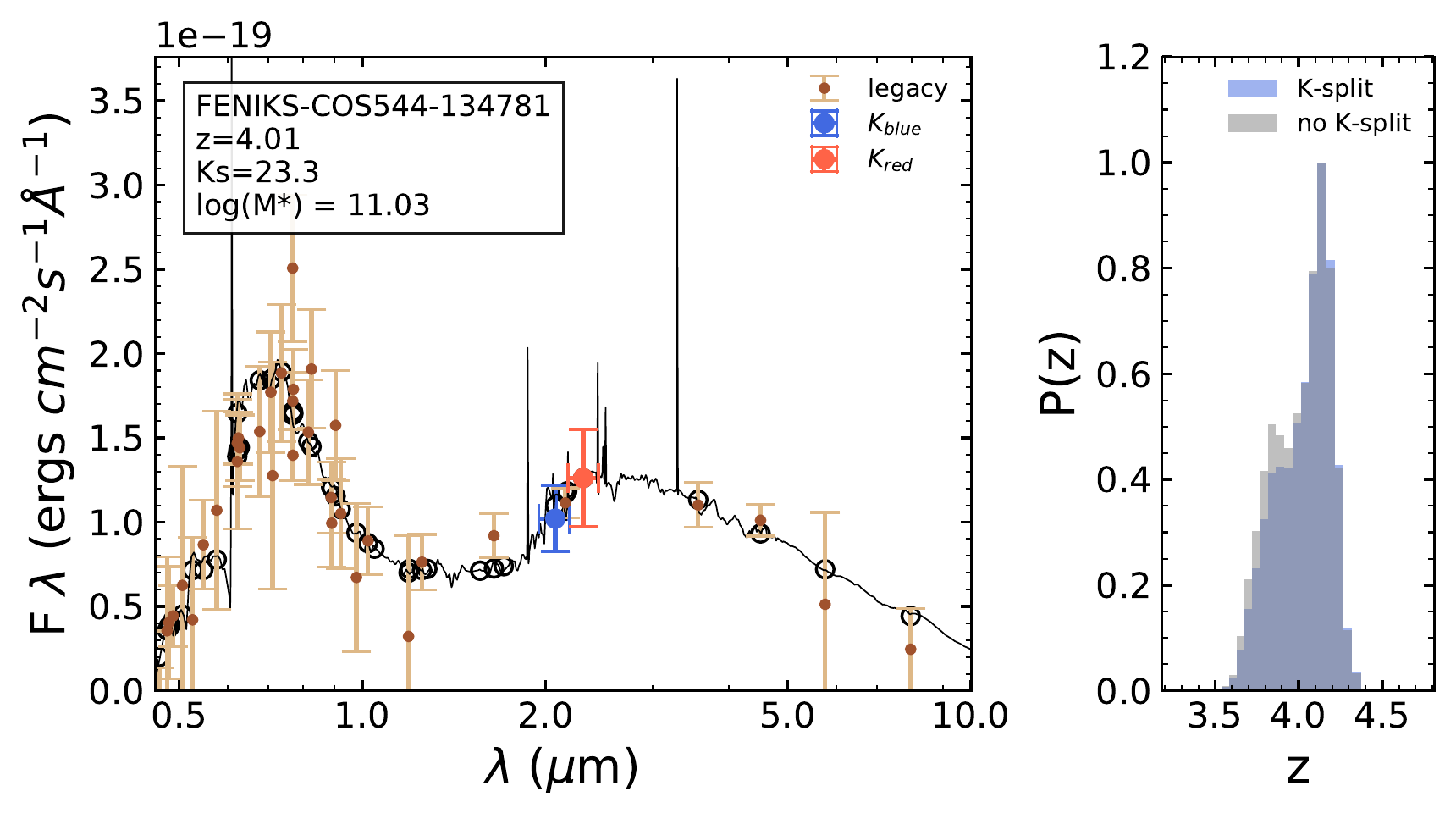}
\end{minipage}
&
\begin{minipage}[b]{0.5\linewidth}
\centering
\includegraphics[width=\linewidth]{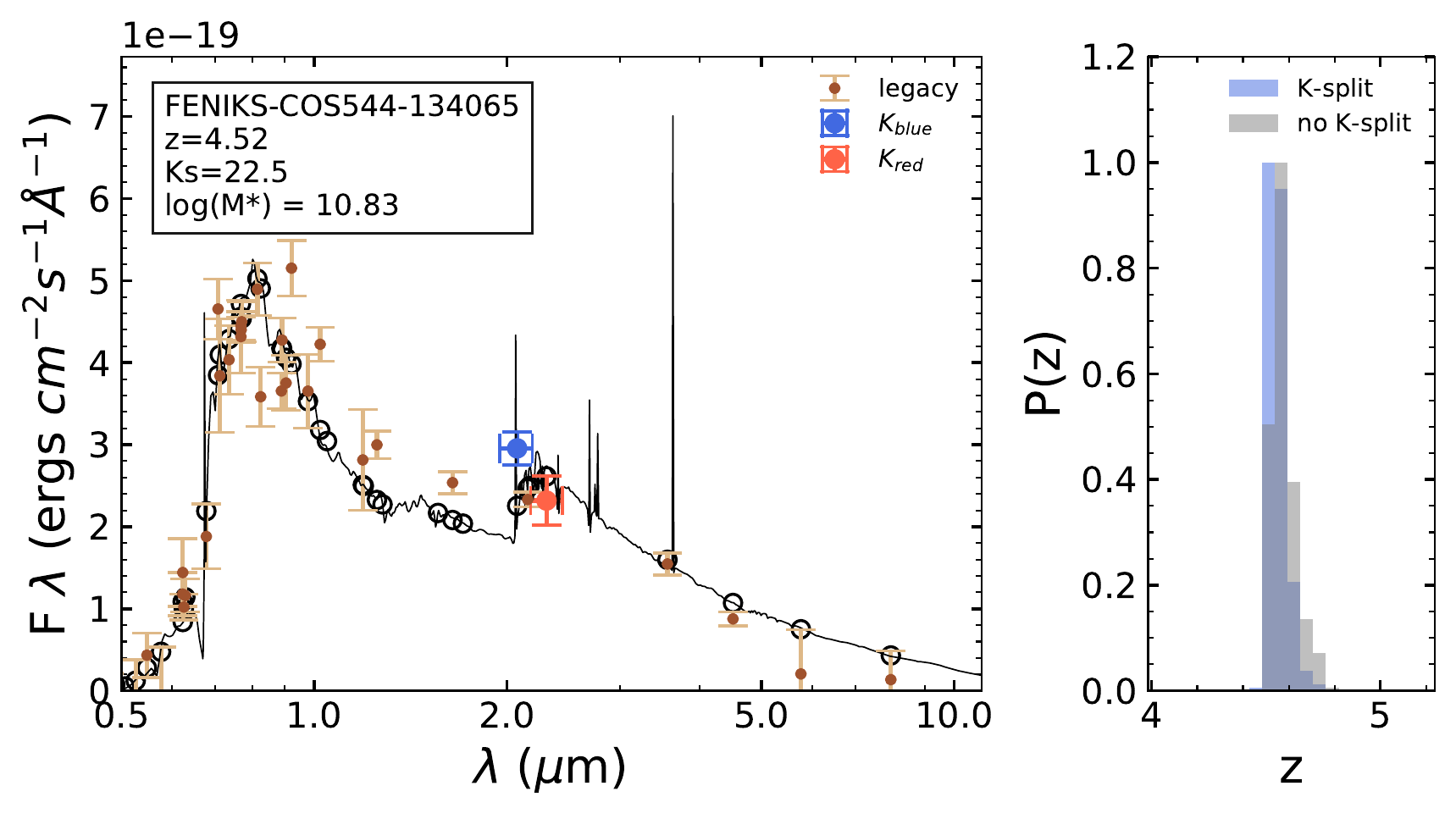}
\end{minipage} 
\\
\end{tabular}
\caption{Example of four massive galaxy candidates at $z > 4$.  For each galaxy SED, the left panel shows the best-fit EAzY SED as the black line, the tan points are legacy photometry from the UltraVISTA catalogue with corresponding 1$\sigma$ error bars, the open black circles are the synthetic photometry for best-fit SED and the blue and red points are the observed $\kblue$ and $\kred$ photometry.  For each galaxy SED, the right panel shows the photometric redshift probability distribution function, P(z), for the photometry in the left panel including the $K$-split filters in blue and without the $K$-split filters in gray.}
\label{fig:mass_gal}
\end{figure*}

In our FENIKS pilot survey we select galaxies in the redshift range $4 < z < 5.5$ where the $K$-split filters straddle the Balmer and 4000~\AA\ break.  We select galaxies with $K_s$ SNR $>$ 7, for which we estimate our catalogs are complete to log(M*/$M_{\odot}$) = 10.7 for a passively evolving stellar SED, with a formation redshift of $z\textsubscript{form}\sim7$ from the \citet{Bruzual2003} models and a \citet{Kroupa2001} IMF. Our final candidate list comprises 12 massive galaxies, from which a subset is shown in Fig.~\ref{fig:mass_gal}.

The photometric redshift probability distribution function, P(z), is shown in the right hand panels of Fig.~\ref{fig:mass_gal} for each galaxy with the $K$-split photometry (blue histogram) and without (grey histogram). The P(z)s for the 12 selected massive galaxies at $z>4$ are consistent between the original legacy catalogues and the FENIKS survey including the $K$-split photometry.  Additionally there are no substantial changes to the best-fit SEDs with and without the $K$-split photometry.  This supports the idea that the z$\textsubscript{phot}$ estimates are robust, as the $K$-split filters provide additional confidence in the SED fit.  The additional confidence is significant given the potential for emission line galaxies contaminating the selection \citep[e.g.][]{Merlin2019,Marsan2020}. Furthermore, the consistency of the photometry with the legacy catalogue best-fit SEDs reinforces that these galaxies host old stellar populations, such as the case for FENIKS-COS352-11539.

While our sample generally does not appear to show a substantial increase in P(z) precision using the $K$-split photometry, there are indications that the P(z) is changing with their inclusion.  This can be seen for FENIKS-COS352-11539, which shows a truncation of the P(z), narrowing the range of potential redshift solutions.  Considering the results of the simulations in Section \ref{sec:simulations}, the limited improvement in precision is likely driven by three factors.  Firstly, the galaxy SED shape impacts the P(z) significantly (as seen in simulations in Sections \ref{sec:photzsim} and \ref{sec:snrsim}).  Most of the massive galaxies identified have at least some rest-frame UV flux, corresponding to a detectable Lyman-break, which substantially constrains the P(z) to $z>4$, even in the absence of the $K$-split filters.  Secondly, the relatively low SNR ($\lesssim$ 5) of the $K$-split pilot survey photometry limits the constraints on the P(z) for these faint galaxies.  In particular this is seen for FENIKS-COS352-13117, which has low rest-frame UV flux but also low SNR for the $K$-split photometry.  Lastly the redshift range for the galaxies with lower rest-frame UV flux are at redshift ranges where the $K$-split filters have less of an impact ($z\lesssim4.3$), consistent with the simulations.  This highlights the serendipity of our selected galaxies and the potential field-to-field variability contributing to a lack of decisive contribution to the P(z).  For example, if a galaxy like FENIKS-COS352-13117 was present at a slightly higher redshift, the $K$-split filters would sample a brighter portion of the Balmer break, resulting in higher SNR measurements.  As an additional check, we note that there are no galaxies in our catalogues with $z\textsubscript{phot}>4$ with $\kblue$-$\kred>0.4$ at $>3\sigma$ significance, which reinforces the idea that there are no galaxies for which the $K$-split filters would have had an impact in our observed pilot fields.

Any changes to the P(z) and/or the best-fit SED will have consequences for the derived physical properties, such as stellar mass and also rest-frame colors typically used to select quiescent galaxies.  Given the limited change to both the P(z) and best-fit SED from the added $K$-split filters, these quantities do not change significantly for the selected 12 massive galaxies in the FENIKS survey catalogue.  The analysis below on quiescent galaxies in the FENIKS survey use the P(z) and physical properties derived using the $K$-split filters (as outlined in Section \ref{sec:software}).

\begin{figure}
\centering
\includegraphics[width=\linewidth]{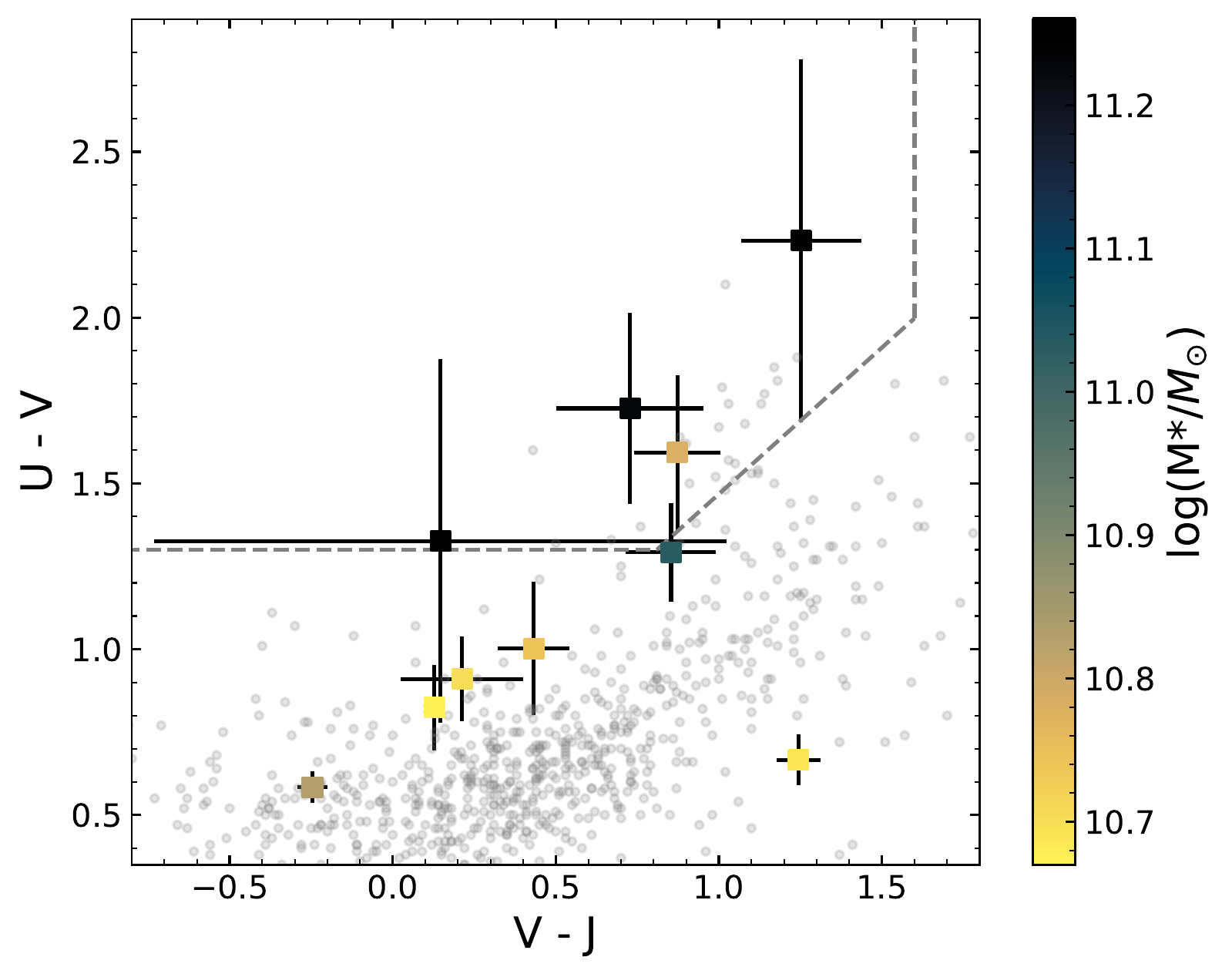}
\caption{Rest-frame V-J vs. U-V color-color plot (a ``UVJ’’ plot) for the log(M*/$M_{\odot}$) $>$ 10.7 sample from $4 < z < 5$.  UVJ errors are based on the 16th and 84th percentiles redshift probability distributions from EAzY.  The square scatter points are the $z > 4$ galaxy sample which are colored by mass.  The gray dots are for a similar mass cut of the $z \sim 2$ population from the UltraVISTA catalogue.}
\label{fig:ssfr_uvj}
\end{figure}

To determine which of our sample are quiescent we looked at both a sSFR selection and a UVJ selection\footnote{It is noted here that the rest-frame J band beyond $z \sim 4$ are based on the best-fit EAzY template from limited photometric sampling.} using rest-frame colors determined with EAzY. The UVJ selection is a rest-frame color selection which allows separation of quiescent galaxies from dust-reddened star-forming galaxies \citep{Patel2011}.  While UVJ selection has been demonstrated to be successful in identifying high-redshift quiescent galaxies, it has also been shown to exclude recently quenched galaxies with younger ages and bluer UV colors \citep[as in][]{Schreiber2018}.  sSFR selection can therefore be a useful comparison with the caveat that the SFR can be sensitive to the chosen SFH model \citep[see][]{Merlin2018, Carnall2019}.  In our analysis, we test this sensitivity by deriving SFRs based on the rest-frame UV flux of the EAzY best-fit SED.  As the EAzY templates are based on empirically derived SEDs, both UVJ colors and rest-frame UV derived SFR are independent of SFH model assumptions.  Additionally rest-frame UV SFR are more sensitive to recent SF events of order 10-30 Myr as opposed to parametric SFH models which are implicitly assessed over larger timescales $\gtrsim$ 100 Myr.  For UV SFRs, it is important to ensure the best-fit SED represents the observed photometry.  This is generally not an issue due to the flexible nature of EAzY fitting which uses linear combinations of many templates to determine best-fit SEDs.

The result of the quiescent search for our galaxy sample is that we identify four galaxies that are UVJ quiescent selected and three galaxies that have sSFR$<0.05$ Gyr$^{-1}$, two of which are not selected by the UVJ cut.  Fig.~\ref{fig:ssfr_uvj} shows the UVJ plot for our galaxy sample.  The two low sSFR galaxies that do not make the UVJ selection have bluer rest-frame U-V colors (0.6-0.9), young ages (0.1-0.2 Gyr) and are likely recently quenched galaxies similar to those identified in \citet{Schreiber2018}, and also seen in \citet{Marsan2015}.  Inspecting their SEDs show these galaxies exhibit substantial star-formation.  To further investigate their SFR, we calculate the rest-frame 1500~\AA\ flux using EAzY (with errors based on 16th and 84th percentiles the rest-frame flux) and the scaling relation for SFR\textsubscript{UV} from \citet{Madau2014}.  It is noted that these SFRs are not corrected for dust extinction and so represent lower limits.  Rest-frame SFR\textsubscript{UV} indicate that the recently quenched galaxies exhibit more substantial star-formation that would exclude them for the sSFR$<0.05$ Gyr$^{-1}$ selection and are therefore not considered as quiescent.  The quiescent sample reported here is therefore the UVJ selected sample, which represents 33$\%$ of the total massive galaxy sample.

Inspecting the SEDs of the UVJ selected quiescent galaxies show that all of them have some indication of residual star-formation (albeit at lower levels compared to the sSFR selected galaxies).  Again calculating their rest-frame UV SFR, the galaxies indicate low SFR with an average of 5$\pm$1 M$_{\odot}$yr$^{-1}$.  UV slopes, $\beta$, are fitted using the $F_{\lambda} \propto \lambda^{\beta}$ using a similar approach with rest-frame filters sampling 1500~\AA-2500~\AA\ wavelength range.  The average UV slope for the UVJ selected galaxies is $\beta$=$-1\pm$0.3, consistent with moderate far-UV attenuation ($A_{1500}$ $\sim$ 2.5 or a factor 10) based on the A$_{FUV}-\beta$ relation \citep{Noll2005}, corresponding to A$_V$ = 1.0 for a \citet{Calzetti2001} attenuation law. Therefore the ongoing SFRs could be an order of magnitude higher.

\begin{figure*}
\begin{tabular}{cc}
\begin{minipage}[t]{0.5\linewidth}
\centering
\includegraphics[width=\linewidth]{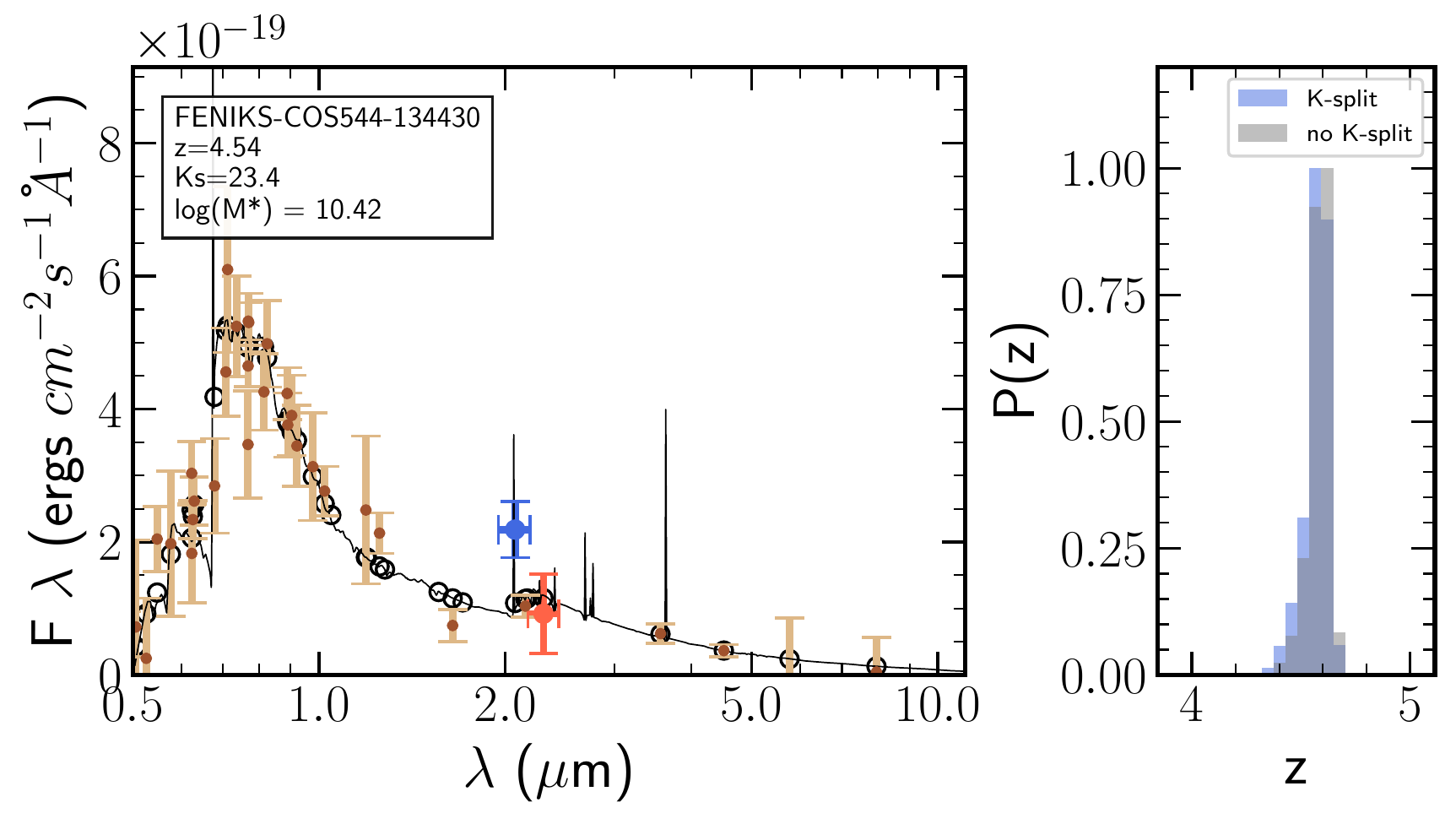}
\end{minipage}
&
\begin{minipage}[b]{0.5\linewidth}
\centering
\includegraphics[width=\linewidth]{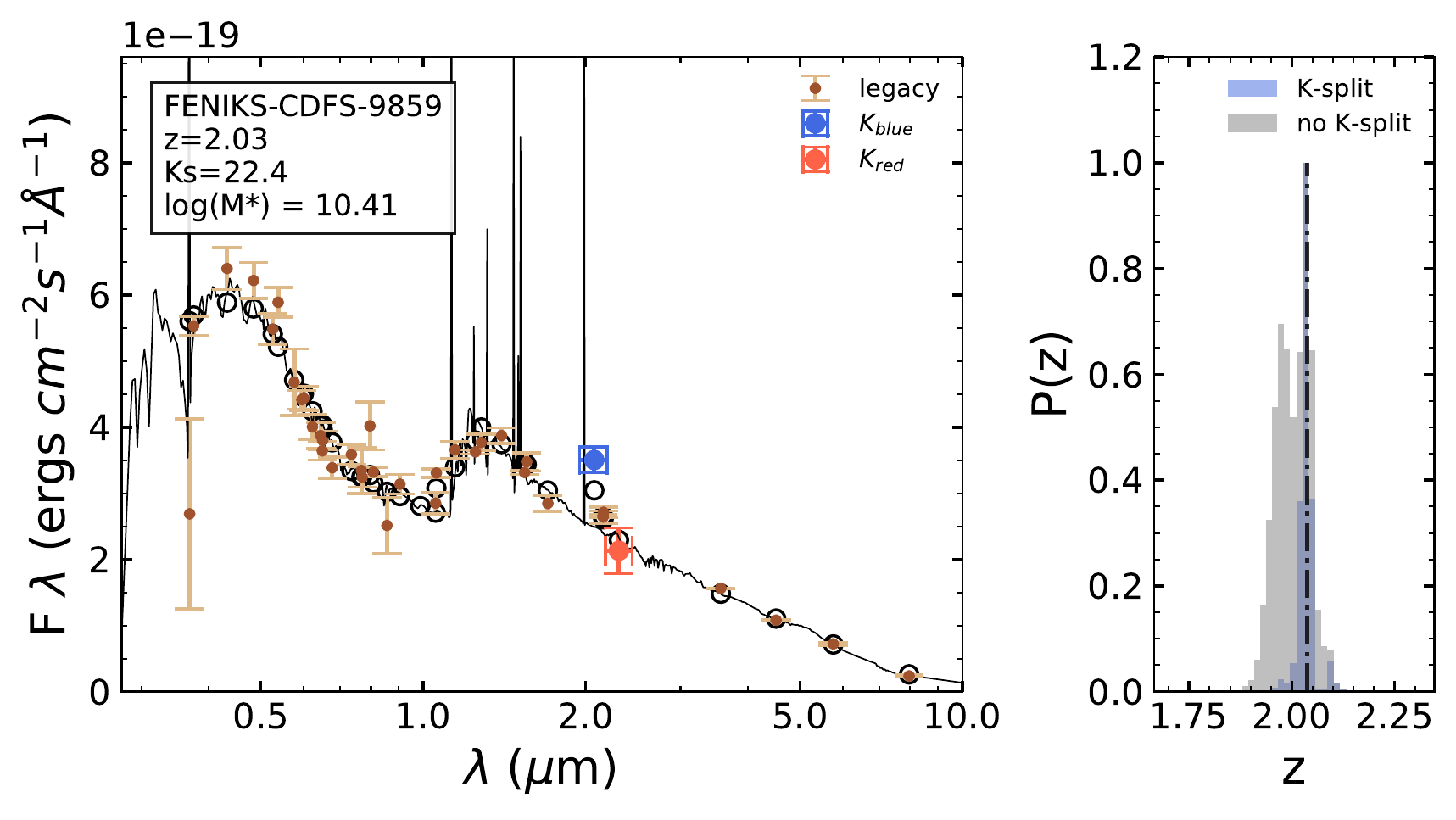}
\end{minipage} 
\\
\begin{minipage}[b]{0.5\linewidth}
\centering
\includegraphics[width=\linewidth]{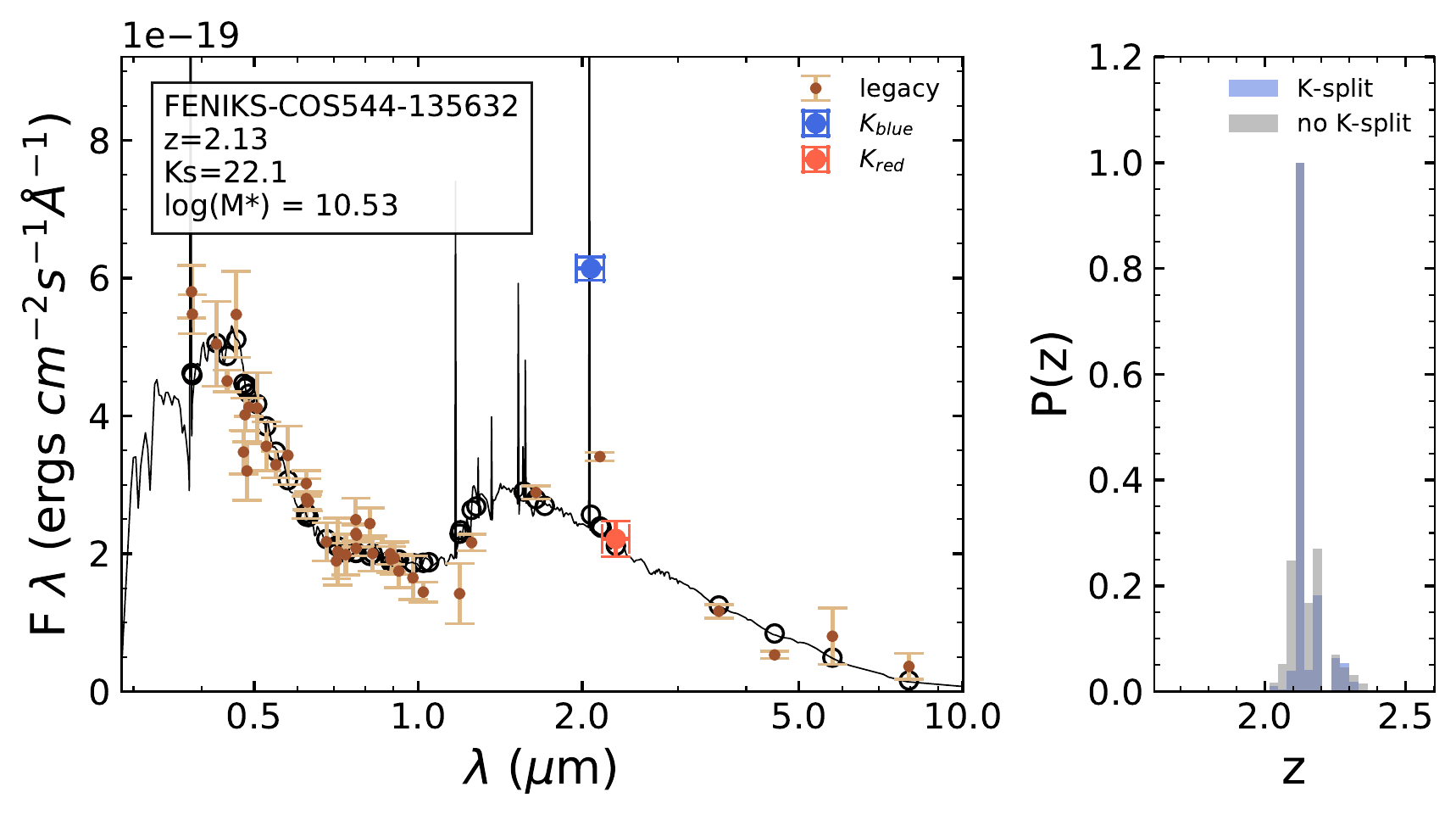}
\end{minipage}
&
\begin{minipage}[b]{0.5\linewidth}
\centering
\includegraphics[width=\linewidth]{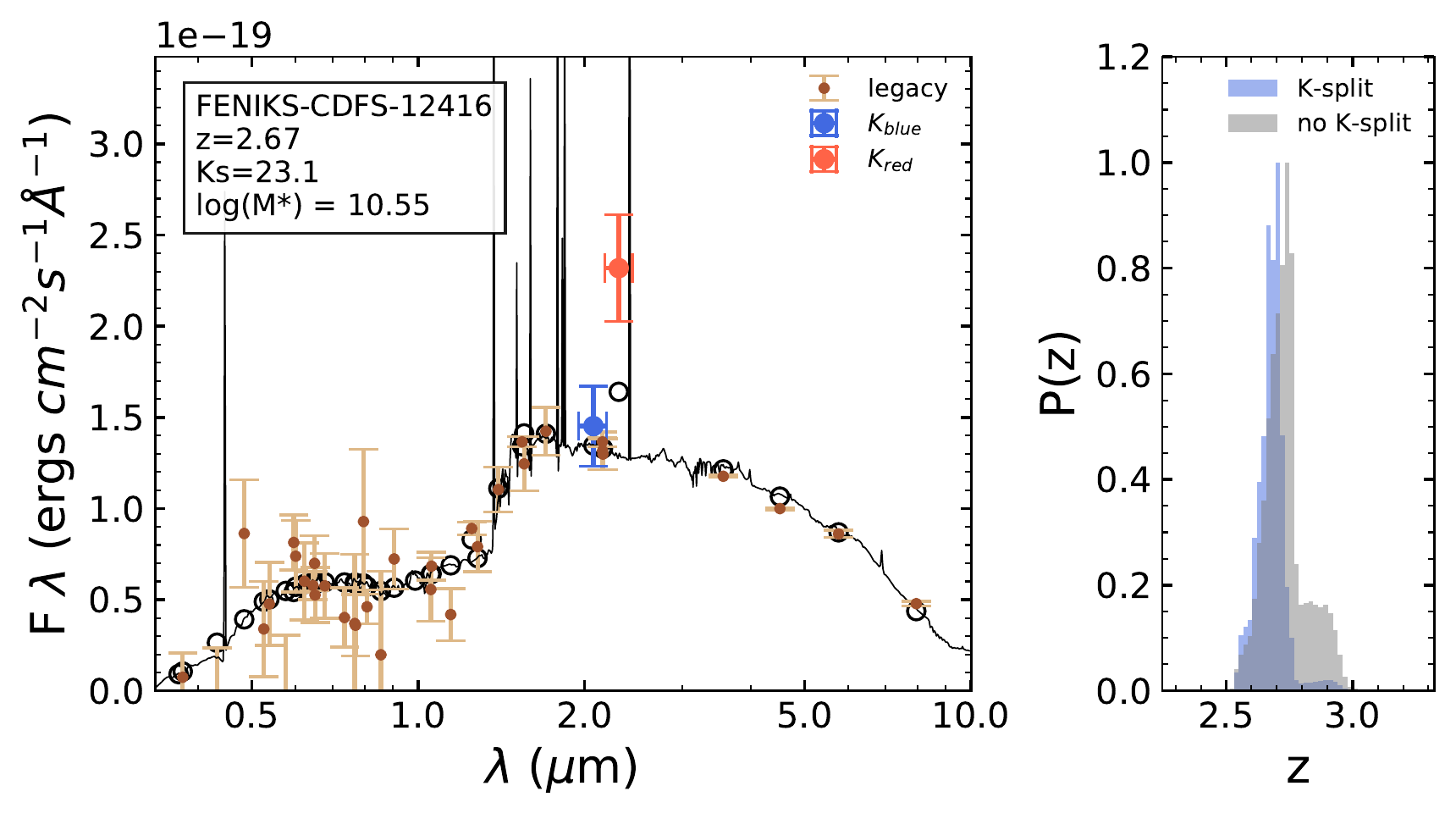}
\end{minipage} 
\\
\end{tabular}
\caption{Example of four emission line galaxy candidates at  $z > 2$ showcasing a range of strong $\kblue$-$\kred$ colors by `boosted' fluxes in either $\kblue$ or $\kred$ compared to the best-fit SED flux continuum.  The top right and both bottom panel galaxy SEDs show indications of strong H$\alpha$ emission and the top left panel galaxy SED shows some indication of [OII] emission.  The same layout as Fig.~\ref{fig:mass_gal} with the addition of spectroscopic redshifts shown in the top right P(z) plots as vertical dashed black line.}
\label{fig:em_line_gal}
\end{figure*}

The combination of the physical properties of these galaxies, i.e. high stellar masses, log($M*/M_{\odot}$) $= 11.2$, and old stellar populations of 1 Gyr, low dust attenuation and relatively low SFR indicate these galaxies are moving off the star-formation main sequence. Taken at face value, the very low $<$sSFR$>$ $\simeq$ 0.03-0.3 Gyr$^{-1}$ values are well below ($5-50\times$) the star-forming main sequence at $z = 4$ of sSFR = 1.5 Gyr $^{-1}$ \citep[see][]{Schreiber2018}}, consistent with being a post starburst galaxy. It is possible that significant star-formation could be dust obscured. Sub-millimeter data, such as those from ALMA are capable of making direct constraints on the dust obscured SFRs \citep[e.g., ][]{Schreiber2018,Santini2019}, however an archival search for our quiescent galaxy sample showed no coverage was available. Further study of them could potentially provide clues to the causes for quiescence in massive galaxies and will be addressed in an upcoming paper (Esdaile in prep).

Finally, the number of massive quiescent galaxies encountered in our pilot survey are consistent with the range of GSMFs shown in Fig~\ref{fig:llp_pred}, including a no redshift evolution of the $3<z<4$ passive GSMF in \citet{Muzzin2013b} and the \citet{Caputi2015} and \citet{Grazian2015} GSMF assuming a 15$\%$ quiescent fraction.  The significance of these agreements however are low considering small number statistics, field-to-field variation, and the potential dust obscured star-formation present in the UVJ selected galaxies.

\subsection{Emission Line Galaxies at $z > 2$}
\label{sec:em_line_gal}

The \texttt{F2} medium $K$-split filters also have the ability to select emission line galaxies at $z > 2$ (H$\alpha$ and [NII] emission $2 < z < 2.8$ and [OIII] at $3 < z < 4$).  The $K$-split filters are spaced such that they are able to measure the boosted flux of emission lines in one filter but not the other.  A $|\kblue$-$\kred|$ = 0.5 mag corresponds to very strong line equivalent widths of 350-400~\AA\ \citep{Marmol-Queralto2015}.  We made a selection of $|\kblue$-$\kred| > 0.3$ mag and identified potential emission line galaxies, a sample of which are shown in Fig.~\ref{fig:em_line_gal}.  These galaxies include examples of [OII] and H$\alpha$ and [NII] emission lines boosting either $\kblue$ or $\kred$.  The bottom left panel of Fig.~\ref{fig:em_line_gal} shows a galaxy with the $\kblue$ filter 1 magnitude above the best-fit SED continuum flux, which corresponds to an H$\alpha$ equivalent width (EW) of $\sim$ 1000~\AA\ (or 600 $M_{\odot}$ yr$^{-1}$).  Based on the H$\alpha$ flux, this galaxy has six times the star formation rate compared to the star formation main-sequence at this epoch and stellar mass \citep[similar galaxies were also identified by][]{Santini2017,Tran2020}.

To see the impact of the $K$-split filters on spectroscopically confirmed galaxies, we use spectroscopic redshifts from the existing legacy catalogues and supplement with more recent spectroscopic surveys from VANDELS \citep{McLure2018} and 3D-HST \citep{Momcheva2016,Brammer2012}.  While there is no overlap with our COSMOS fields, there is good coverage in our CDFS field with over 100, $z > 2$ spectroscopic redshifts (using only the highest quality flags).  The agreement of spectroscopic redshift and photometric redshift is good with an NMAD scatter in $|z_\mathrm{spec} - z_\mathrm{phot}|$ / $( 1 + z_\mathrm{spec} )$ of $\signmad$ = 0.022 at $z > 2$, consistent with the ZFOURGE results.  There are also a handful of strong $\kblue$ - $\kred$ color signatures that are consistent with spectroscopic redshifts as in the upper right panel of Fig.~\ref{fig:em_line_gal}, which suggests the H$\alpha$ boosted $\kblue$ is physical and not just from measurement scatter.  While the photometric redshift is generally consistent with the legacy catalogues, strong color signatures can have decisive impacts on the photometric redshift precision.  For example, there is boosting of both $\kblue$ and $K_s$ but not $\kred$, in the SED shown in the lower left panel of Fig.~\ref{fig:em_line_gal}.  There are also instances where the $\kblue$ - $\kred$ color signature highlights one of the only distinct spectral features, such as in the lower right panel of Fig.~\ref{fig:em_line_gal}.  This galaxy in particular has a dust extinction of A$_V$ = 2.0 mag and therefore rest-frame UV features are significantly suppressed, making photometric redshifts difficult to determine.  The $K$-split filters have an impact in this case increasing the photometric redshift precision $\signmad$ from 0.08 to 0.05 (similar improvement to what was seen in the simulations shown in section \ref{sec:photzsim}).

\section{Conclusions}
\label{sec:conclusions}
In this paper, we presented two new medium band filters $\kblue$ ($\lambda \textsubscript{cen}$=2.06 $\mu$m, $\Delta \lambda$=0.25 $\mu$m) and $\kred$ ($\lambda \textsubscript{cen}$=2.31 $\mu$m, $\Delta \lambda$=0.27 $\mu$m), we outlined their design and ran simulations to test their performance and presented the first results for the new filters in a pilot of the \text{F2} Extra-galactic Near-Infrared $K$-band Split (FENIKS) survey.

We showed the procedure of designing the optimal central wavelength position and filter widths of the additional medium band $K$-split filters.  The goal of the filter design procedure was to maximize the redshift range of the Universe for which a strong color signature across the $K$-band window is measurable.  The filter design took into consideration the sky background + telescope emission, atmospheric transmission, telescope throughput, and was optimized for a source with a strong 3650~\AA\ Balmer break, to produce two new filters on the blue and red side of the traditional $K_s$ ($\lambda \textsubscript{cen}$=2.16 $\mu$m, $\Delta \lambda$=0.32 $\mu$m) filter. Together, the three filters cover the entire $1.9-2.5$ $\mu$m $K$-band transmission window, improving sampling of the Balmer and 4000~\AA\ break at $4 \lesssim z \lesssim 5.5$.

We used simulations with the photometric redshift software EAzY to test the potential benefit of the $K$-split filters.  We simulated a large range of galaxy models, with a range of SNRs, including various SSP ages and a dusty star-forming galaxy to test the different redshift ranges for which the $K$-split filters can show improvement. Using these models we produced multi-wavelength mock catalogues with representative photometric depths and filter sets from existing deep photometric surveys to recover input redshifts both with and without the $K$-split filters.  From these simulations, we find that the impact of $K$-split filters on photometric redshift depends on the age, SNR and redshift of the galaxies.  Given sufficient SNR, the addition of the $K$-split filters can substantially improve both the photometric redshift precision and the outlier fraction (by a factor of $\gtrsim$2) for the older SSP models ($\gtrsim$ 0.6 Gyr) in the redshift range $4.2 < z < 5.2$.  Younger SSP models show less improvement likely due to the presence of the Lyman-break, which provides an additional break to constrain the photometric redshift. The dusty star forming galaxies see improvements in photometric redshift precision by a factor of 1.5 at $2 < z < 4$.

To investigate the SNR requirements of the $K$-split filters given an existing photometric catalogue, and their impact on different SED types, we conducted additional simulations.  These simulations included two SEDs: a 0.9 Gyr SSP and a post starburst galaxy with residual star-formation, and considered a range of SNR in the $K$-split filters.  Generally, photometric redshift constraints improve most between SNR$\sim5-10$ and the relative improvement is largest for quiescent SEDs with no rest-frame UV. Galaxies with detectable Lyman breaks, such as the post starburst SED, tend to already have well-constrained redshift so the impact of the $K$-split filters is smaller.

Using the newly commissioned \texttt{F2} filters we conducted a pilot of the FENIKS survey which covered 87.6 arcmin$^2$ across three fields in the COSMOS and CDFS fields with 5$\sigma$ depths ranging from 23.7-24.3 mag in $\kred$ and 24.1-25.0 mag in $\kblue$.  We created catalogues by adding the \texttt{F2} photometry to existing multi-wavelength photometric catalogues from ZFOURGE \citep{Straatman2016} and UltraVISTA \citep{Muzzin2013b} DR3 (private communication).  We used these catalogues to determine redshifts using EAzY and physical properties such as stellar mass and SFR using FAST with additional checks using rest-frame UV flux based on the best-fits from EAzY.

To a stellar mass limit of log(M*/$M_{\odot}$) $>$ 10.7, we identified 12 massive galaxies at $4 < z < 5$ in our pilot survey.  Applying both UVJ and sSFR$<$0.05 Gyr$^{-1}$ selections to identify quiescent galaxies we found four UVJ selected galaxies and two additional galaxies from the sSFR selection not in the UVJ selection.  The two sSFR galaxies have young ages, bluer UV colors and are consistent with the recently quenched galaxies observed in \citet{Schreiber2018}.  Their rest-frame UV determined SFR showed them to have more substantial SFR than permitted by the sSFR$<$0.05 Gyr$^{-1}$ and were therefore omitted from the quiescent sample, producing a final sample of four galaxies (representing 33$\%$ of the total massive galaxy sample).

The SEDs of the UVJ selected galaxies show evidence of large Balmer and 4000~\AA\ breaks, consistent with the $\kblue$ and $\kred$ photometry. In contrast to UVJ quiescent galaxies at $3 < z < 4$ \citep[e.g.,][]{Straatman2014}, most galaxies show low but detectable levels of rest-frame UV emission with a mean blue UV slope $<\beta>\sim-1$, indicating mildly obscured residual star formation. The galaxies are thus not fully quenched. Nevertheless, the range of inferred mean sSFR = 0.03-0.3 Gyr$^{-1}$ is about an order of magnitude below the star-forming main sequence at $z = 4$ \citep{Schreiber2018}.  We are likely identifying galaxies that are transitioning from the star-forming main sequence.

For the selected quiescent galaxies found in our sample we found modest improvements in the P(z) after inclusion of the $K$-split filters. This result, combined with the simulation work presented here, indicates that the current pilot observations and sample (with $K$-split SNR=1-5) do not yet fully exploit the discriminatory power of the new $K$-split filters. Deeper observations, brighter targets, and galaxies with lower rest-frame UV emission would likely see more impact from the $K$-split filters.

We found good agreement of photometric redshifts and spectroscopic redshifts for a selection of 100+ galaxies at $2<z<4$ for which we have spectra.  Several of these galaxies exhibit strong color signatures in the new medium-band filters that indicated extreme nebular emission lines. The new filters are sensitive to emission lines due to relatively narrow widths ($R\sim7$) and the ability to identify and account for emission line contribution results in reduced photometric redshift uncertainty for these galaxies.

The full FENIKS Survey (Gemini Large Program) is now underway with a total of 170 hours of Gemini time allocated to the project. With a sky area of up to 0.6 deg$^2$ the larger survey volume will likely yield a high-quality sample of the most massive galaxies at $z > 4$. We will be able to identify massive quiescent galaxies and accurately measure their number densities, their contribution to the galaxy mass function and the overall quiescent fraction at these redshifts.  These objects are high priority science targets for spectroscopic follow-up with JWST.  Given JWST's relatively small FOV compared to the number density of these massive quiescent galaxies, identifying them from ground based surveys such as FENIKS is of paramount importance, and will lead the way for future studies of high redshift massive galaxies.

\section*{Data Availability}
The raw images corresponding to the pilot survey (for the program IDs listed in Section \ref{sec:survey}) are available from the Gemini archive\footnote{\url{https://archive.gemini.edu/}}.  A public release of the catalogue is planned to coincide with a future paper outlining the data reduction and catalogue procedure for the FENIKS pilot survey (Esdaile et al. in prep).  Catalogues can be provided upon reasonable request to the author.

\section*{Acknowledgements}

Based on observations obtained at the international Gemini Observatory, a program of NSF’s OIR Lab, which is managed by the Association of Universities for Research in Astronomy (AURA) under a cooperative agreement with the National Science Foundation on behalf of the Gemini Observatory partnership: the National Science Foundation (United States), National Research Council (Canada), Agencia Nacional de Investigación y Desarrollo(Chile), Ministerio de Ciencia, Tecnología e Innovación (Argentina), Ministério da Ciência, Tecnologia, Inovações e Comunicações (Brazil), and Korea Astronomy and Space Science Institute (Republic of Korea).

Based on data products from observations made with ESO Telescopes at the La Silla Paranal Observatory under ESO programme ID 179.A-2005 and on data prod- ucts produced by TERAPIX and the Cambridge Astronomy Survey Unit on behalf of the UltraVISTA consortium.

The authors thank the referee for their contributions and detailed comments that markedly improved the manuscript.  JE acknowledges an Astro3D scholarship from CE17010013.  KG acknowledges support from CE17010013 and Laureate Fellowship FL180100060.  JAD and CP acknowledge the generous support from George P. and Cynthia Woods Mitchell Institute for Fundamental Physics and Astronomy at Texas A$\&$M University.  This material is based upon work supported by the National Science Foundation under Cooperative Agreement No. AST0525280.  Z.C.M. gratefully acknowledges support from the Faculty of Science at York University as a York Science Fellow. This material is based upon work supported by the National Science Foundation under Grant No. AST-2009442.

\software{SEP \citep{Barbary2016}, Source Extractor \citep{Bertin1996}, Astropy\footnote{\url{http://www.astropy.org}} \citep{Astropy2013, Astropy2018}, IRAF \citep{Tody1986}, BAGPIPES \citep{Carnall2017}, EAzY \citep{Brammer2008}, FAST \citep{Kriek2009}}




\bibliographystyle{aasjournal}
\bibliography{library.bib}

\begin{thebibliography}{}
\expandafter\ifx\csname natexlab\endcsname\relax\def\natexlab#1{#1}\fi
\providecommand{\url}[1]{\href{#1}{#1}}
\providecommand{\dodoi}[1]{doi:~\href{http://doi.org/#1}{\nolinkurl{#1}}}
\providecommand{\doeprint}[1]{\href{http://ascl.net/#1}{\nolinkurl{http://ascl.net/#1}}}
\providecommand{\doarXiv}[1]{\href{https://arxiv.org/abs/#1}{\nolinkurl{https://arxiv.org/abs/#1}}}

\bibitem[{{Alcalde Pampliega} {et~al.}(2019){Alcalde Pampliega},
  P{\'{e}}rez-Gonz{\'{a}}lez, Barro, S{\'{a}}nchez, Eliche-Moral, Cardiel,
  Hern{\'{a}}n-Caballero, Rodriguez-Mu{\~{n}}oz, Bl{\'{a}}zquez, \&
  Esquej}]{Pampliega2019}
{Alcalde Pampliega}, B., P{\'{e}}rez-Gonz{\'{a}}lez, P.~G., Barro, G., {et~al.}
  2019, The Astrophysical Journal, 876, 135, \dodoi{10.3847/1538-4357/ab14f2}

\bibitem[{Barbary(2016)}]{Barbary2016}
Barbary, K. 2016, The Journal of Open Source Software, 1, 58,
  \dodoi{10.21105/joss.00058}

\bibitem[{Bertin \& Arnouts(1996)}]{Bertin1996}
Bertin, E., \& Arnouts, S. 1996, Astronomy and Astrophysics Supplement Series,
  117, 393, \dodoi{10.1051/aas:1996164}

\bibitem[{Brammer {et~al.}(2012)Brammer, van Dokkum, Franx, Fumagalli, Patel,
  Rix, Skelton, Kriek, Nelson, Schmidt, Bezanson, da~Cunha, Erb, Fan,
  Schreiber, Illingworth, Labb{\'{e}}, Leja, Lundgren, Magee, Marchesini,
  McCarthy, Momcheva, Muzzin, Quadri, Steidel, Tal, Wake, Whitaker, \&
  Williams}]{Brammer2012}
Brammer, G., van Dokkum, P., Franx, M., {et~al.} 2012, Astrophysical Journal,
  Supplement Series, 200, \dodoi{10.1088/0067-0049/200/2/13}

\bibitem[{Brammer {et~al.}(2008)Brammer, van Dokkum, \& Coppi}]{Brammer2008}
Brammer, G.~B., van Dokkum, P.~G., \& Coppi, P. 2008, The Astrophysical
  Journal, 686, 1503, \dodoi{10.1086/591786}

\bibitem[{Bruzual \& Charlot(2003)}]{Bruzual2003}
Bruzual, G., \& Charlot, S. 2003, Monthly Notices of the Royal Astronomical
  Society, 344, 1000, \dodoi{10.1046/j.1365-8711.2003.06897.x}

\bibitem[{Calzetti(2001)}]{Calzetti2001}
Calzetti, D. 2001, Publications of the Astronomical Society of the Pacific,
  113, 1449, \dodoi{10.1086/324269}

\bibitem[{Caputi {et~al.}(2015)Caputi, Ilbert, Laigle, McCracken, {Le
  F{\`{e}}vre}, Fynbo, Milvang-Jensen, Capak, Salvato, \&
  Taniguchi}]{Caputi2015}
Caputi, K.~I., Ilbert, O., Laigle, C., {et~al.} 2015, Astrophysical Journal,
  810, \dodoi{10.1088/0004-637X/810/1/73}

\bibitem[{Carnall {et~al.}(2019)Carnall, Leja, Johnson, McLure, Dunlop, \&
  Conroy}]{Carnall2019}
Carnall, A.~C., Leja, J., Johnson, B.~D., {et~al.} 2019, The Astrophysical
  Journal, 873, 44, \dodoi{10.3847/1538-4357/ab04a2}

\bibitem[{Carnall {et~al.}(2017)Carnall, McLure, Dunlop, \&
  Dav{\'{e}}}]{Carnall2017}
Carnall, A.~C., McLure, R.~J., Dunlop, J.~S., \& Dav{\'{e}}, R. 2017, Monthly
  Notices of the Royal Astronomical Society, 480, 4379,
  \dodoi{10.1093/mnras/sty2169}

\bibitem[{Chabrier(2003)}]{Chabrier2003}
Chabrier, G. 2003, Publications of the Astronomical Society of the Pacific,
  115, 763, \dodoi{10.1086/376392}

\bibitem[{Conselice {et~al.}(2016)Conselice, Wilkinson, Duncan, \&
  Mortlock}]{Conselice2016}
Conselice, C.~J., Wilkinson, A., Duncan, K., \& Mortlock, A. 2016, The
  Astrophysical Journal, 830, 83, \dodoi{10.3847/0004-637x/830/2/83}

\bibitem[{Davidzon {et~al.}(2017)Davidzon, Ilbert, Laigle, Coupon, McCracken,
  Delvecchio, Masters, Capak, Hsieh, {Le F{\`{e}}vre}, Tresse, Bethermin,
  Chang, Faisst, {Le Floc'h}, Steinhardt, Toft, Aussel, Dubois, Hasinger,
  Salvato, Sanders, Scoville, \& Silverman}]{Davidzon2017}
Davidzon, I., Ilbert, O., Laigle, C., {et~al.} 2017, Astronomy {\&}
  Astrophysics, 605, A70, \dodoi{10.1051/0004-6361/201730419}

\bibitem[{Diaz {et~al.}(2016)Diaz, Goodsell, \& Kleinman}]{Diaz2016}
Diaz, R., Goodsell, S., \& Kleinman, S. 2016, in Proc.SPIE, Vol. 9908,
  \dodoi{10.1117/12.2232106}

\bibitem[{Eikenberry {et~al.}(2008)Eikenberry, Elston, Raines, Julian, Hanna,
  Warner, Julian, Bandyopadhyay, Bennett, Bessoff, Branch, Corley, Dewitt,
  Eriksen, Frommeyer, Gonzalez, Herlevich, Hon, Marin-Franch, Marti, Murphey,
  Rambold, Rashkin, Leckie, Gardhouse, Fletcher, Hardy, Dunn, \&
  Wooff}]{Eikenberry2008}
Eikenberry, S., Elston, R., Raines, S.~N., {et~al.} 2008, in Proc.SPIE, ed.
  I.~S. McLean \& M.~M. Casali, Vol. 7014, 70140V, \dodoi{10.1117/12.788326}

\bibitem[{Ferland {et~al.}(2017)Ferland, Chatzikos, Guzm{\'{a}}n, Lykins, van
  Hoof, Williams, Abel, Badnell, Keenan, Porter, \& Stancil}]{Ferland2017}
Ferland, G.~J., Chatzikos, M., Guzm{\'{a}}n, F., {et~al.} 2017, Revista
  Mexicana de Astronomia y Astrofisica, 53, 385.
\newblock \doarXiv{1705.10877}

\bibitem[{Fioc \& Rocca-Volmerange(1999)}]{Fioc1999}
Fioc, M., \& Rocca-Volmerange, B. 1999.
\newblock \doarXiv{9912179}

\bibitem[{Forrest {et~al.}(2017)Forrest, Tran, Broussard, Allen, Apfel, Cowley,
  Glazebrook, Kacprzak, Labb{\'{e}}, Nanayakkara, Papovich, Quadri, Spitler,
  Straatman, \& Tomczak}]{Forrest2017}
Forrest, B., Tran, K.-V.~H., Broussard, A., {et~al.} 2017, The Astrophysical
  Journal, 838, L12, \dodoi{10.3847/2041-8213/aa653b}

\bibitem[{Forrest {et~al.}(2020)Forrest, Annunziatella, Wilson, Marchesini,
  Muzzin, Cooper, Marsan, McConachie, Chan, Gomez, Kado-Fong, Barbera,
  Labb{\'{e}}, Lange-Vagle, Nantais, Nonino, Pe{\~{n}}a, Saracco, Stefanon, \&
  van~der Burg}]{Forrest2020}
Forrest, B., Annunziatella, M., Wilson, G., {et~al.} 2020, The Astrophysical
  Journal, 890, L1, \dodoi{10.3847/2041-8213/ab5b9f}

\bibitem[{Glazebrook {et~al.}(2017)Glazebrook, Schreiber, Labb{\'{e}},
  Nanayakkara, Kacprzak, Oesch, Papovich, Spitler, Straatman, Tran, \&
  Yuan}]{Glazebrook2017}
Glazebrook, K., Schreiber, C., Labb{\'{e}}, I., {et~al.} 2017, Nature, 544, 71,
  \dodoi{10.1038/nature21680}

\bibitem[{Gomez {et~al.}(2012)Gomez, Diaz, Pessev, Prado, Candia, Hogan, Perez,
  Lazo, Luis, Rogers, Gigoux, Solis, Tollestrup, Stephens, Schirmer, Gomez,
  Diaz, Pessev, Prado, Candia, Hogan, Perez, Lazo, Luis, Rogers, Gigoux, Solis,
  Tollestrup, Stephens, \& Schirmer}]{Gomez2012}
Gomez, P.~L., Diaz, R., Pessev, P., {et~al.} 2012, American Astronomical
  Society, AAS Meeting 219, 413.07

\bibitem[{Grazian {et~al.}(2015)Grazian, Fontana, Santini, Dunlop, Ferguson,
  Castellano, Amorin, Ashby, Barro, Behroozi, Boutsia, Caputi, Chary, Dekel,
  Dickinson, Faber, Fazio, Finkelstein, Galametz, Giallongo, Giavalisco,
  Grogin, Guo, Kocevski, Koekemoer, Koo, Lee, Lu, Merlin, Mobasher, Nonino,
  Papovich, Paris, Pentericci, Reddy, Renzini, Salmon, Salvato, Sommariva,
  Song, \& Vanzella}]{Grazian2015}
Grazian, A., Fontana, A., Santini, P., {et~al.} 2015, Astronomy {\&}
  Astrophysics, 575, A96, \dodoi{10.1051/0004-6361/201424750}

\bibitem[{Henriques {et~al.}(2015)Henriques, White, Thomas, Angulo, Guo,
  Lemson, Springel, \& Overzier}]{Henriques2015}
Henriques, B. M.~B., White, S. D.~M., Thomas, P.~A., {et~al.} 2015, Monthly
  Notices of the Royal Astronomical Society, 451, 2663,
  \dodoi{10.1093/mnras/stv705}

\bibitem[{Ichikawa \& Matsuoka(2017)}]{Ichikawa2017}
Ichikawa, A., \& Matsuoka, Y. 2017, The Astrophysical Journal, 843, L7,
  \dodoi{10.3847/2041-8213/aa78f8}

\bibitem[{Kauffmann {et~al.}(2003)Kauffmann, Heckman, White, Charlot, Tremonti,
  Brinchmann, Bruzual, Peng, Seibert, Bernardi, Blanton, Brinkmann, Castander,
  Cs{\'{a}}bai, Fukugita, Ivezic, Munn, Nichol, Padmanabhan, Thakar, Weinberg,
  \& York}]{Kauffmann2003}
Kauffmann, G., Heckman, T.~M., White, S.~D., {et~al.} 2003, Monthly Notices of
  the Royal Astronomical Society, 341, 33,
  \dodoi{10.1046/j.1365-8711.2003.06291.x}

\bibitem[{Kriek {et~al.}(2009)Kriek, {Van Dokkum}, Labb{\'{e}}, Franx,
  Illingworth, Marchesini, \& Quadri}]{Kriek2009}
Kriek, M., {Van Dokkum}, P.~G., Labb{\'{e}}, I., {et~al.} 2009, Astrophysical
  Journal, 700, 221, \dodoi{10.1088/0004-637X/700/1/221}

\bibitem[{Kroupa(2001)}]{Kroupa2001}
Kroupa, P. 2001, Monthly Notices of the Royal Astronomical Society, 322, 231,
  \dodoi{10.1046/j.1365-8711.2001.04022.x}

\bibitem[{Laigle {et~al.}(2016)Laigle, McCracken, Ilbert, Hsieh, Davidzon,
  Capak, Hasinger, Silverman, Pichon, Coupon, Aussel, {Le Borgne}, Caputi,
  Cassata, Chang, Civano, Dunlop, Fynbo, Kartaltepe, Koekemoer, {Le
  F{\`{e}}vre}, {Le Floc'h}, Leauthaud, Lilly, Lin, Marchesi, Milvang-Jensen,
  Salvato, Sanders, Scoville, Smolcic, Stockmann, Taniguchi, Tasca, Toft,
  Vaccari, \& Zabl}]{Laigle2016}
Laigle, C., McCracken, H.~J., Ilbert, O., {et~al.} 2016, The Astrophysical
  Journal Supplement Series, 224, 24, \dodoi{10.3847/0067-0049/224/2/24}

\bibitem[{Madau \& Dickinson(2014)}]{Madau2014}
Madau, P., \& Dickinson, M. 2014, Annual Review of Astronomy and Astrophysics,
  52, 415, \dodoi{10.1146/annurev-astro-081811-125615}

\bibitem[{Marchesini {et~al.}(2010)Marchesini, Whitaker, Brammer, van Dokkum,
  Labb{\'{e}}, Muzzin, Quadri, Kriek, Lee, Rudnick, Franx, Illingworth, \&
  Wake}]{Marchesini2010}
Marchesini, D., Whitaker, K.~E., Brammer, G., {et~al.} 2010, The Astrophysical
  Journal, 725, 1277, \dodoi{10.1088/0004-637X/725/1/1277}

\bibitem[{Marmol-Queralto {et~al.}(2015)Marmol-Queralto, McLure, Cullen,
  Dunlop, Fontana, \& McLeod}]{Marmol-Queralto2015}
Marmol-Queralto, E., McLure, R.~J., Cullen, F., {et~al.} 2015, Monthly Notices
  of the Royal Astronomical Society, 460, 3587, \dodoi{10.1093/mnras/stw1212}

\bibitem[{Marsan {et~al.}(2017)Marsan, Marchesini, Brammer, Geier, Kado-Fong,
  Labb{\'{e}}, Muzzin, \& Stefanon}]{Marsan2017}
Marsan, Z.~C., Marchesini, D., Brammer, G.~B., {et~al.} 2017, The Astrophysical
  Journal, 842, 21, \dodoi{10.3847/1538-4357/aa7206}

\bibitem[{Marsan {et~al.}(2015)Marsan, Marchesini, Brammer, Stefanon, Muzzin,
  Fern{\'{a}}ndez-Soto, Geier, Hainline, Intema, Karim, Labb{\'{e}}, Toft, \&
  van Dokkum}]{Marsan2015}
---. 2015, The Astrophysical Journal, 801, 133,
  \dodoi{10.1088/0004-637X/801/2/133}

\bibitem[{Marsan {et~al.}(2020)Marsan, Muzzin, Marchesini, Stefanon, Martis,
  Annunziatella, Chan, Cooper, Forrest, Gomez, McConachie, \&
  Wilson}]{Marsan2020}
Marsan, Z.~C., Muzzin, A., Marchesini, D., {et~al.} 2020, arXiv.
\newblock \doarXiv{2010.04725}

\bibitem[{Martis {et~al.}(2016)Martis, Marchesini, Brammer, Muzzin,
  Labb{\'{e}}, Momcheva, Skelton, Stefanon, van Dokkum, \&
  Whitaker}]{Martis2016}
Martis, N.~S., Marchesini, D., Brammer, G.~B., {et~al.} 2016, The Astrophysical
  Journal, 827, L25, \dodoi{10.3847/2041-8205/827/2/L25}

\bibitem[{McLure {et~al.}(2018)McLure, Pentericci, Cimatti, Dunlop, Elbaz,
  Fontana, Nandra, Amorin, Bolzonella, Bongiorno, Carnall, Castellano,
  Cirasuolo, Cucciati, Cullen, {De Barros}, Finkelstein, Fontanot, Franzetti,
  Fumana, Gargiulo, Garilli, Guaita, Hartley, Iovino, Jarvis, Juneau, Karman,
  Maccagni, Marchi, M{\'{a}}rmol-Queralt{\'{o}}, Pompei, Pozzetti, Scodeggio,
  Sommariva, Talia, Almaini, Balestra, Bardelli, Bell, Bourne, Bowler, Brusa,
  Buitrag, Caputi, Cassata, Charlot, Citro, Cresci, Cristiani, Curtis-Lake,
  Dickinson, Fazio, Ferguson, Fiore, Franco, Fynbo, Galametz, Georgakakis,
  Giavalisco, Grazian, Hathi, Jung, Kim, Koekemoer, Khusanova, {Le F'evre},
  Lotz, Mannucci, Maltby, Matsuoka, {Mc Leod}, Mendez, Mendez-Abreu, Mignoli,
  Moresco, Mortlock, Nonino, Pannella, Papovich, Popesso, Rosario, Salvato,
  Santini, Schaerer, Schreiber, Stark, Tasca, Thomas, Treu, Vanzella, Wild,
  Williams, Zamorani, \& Zucca}]{McLure2018}
McLure, R.~J., Pentericci, L., Cimatti, A., {et~al.} 2018, Monthly Notices of
  the Royal Astronomical Society, 479, 25, \dodoi{10.1093/mnras/sty1213}

\bibitem[{Merlin {et~al.}(2018)Merlin, Fontana, Castellano, Santini, Torelli,
  Boutsia, Wang, Grazian, Pentericci, Schreiber, Ciesla, McLure, Derriere,
  Dunlop, \& Elbaz}]{Merlin2018}
Merlin, E., Fontana, A., Castellano, M., {et~al.} 2018, Monthly Notices of the
  Royal Astronomical Society, 473, 2098, \dodoi{10.1093/mnras/stx2385}

\bibitem[{Merlin {et~al.}(2019)Merlin, Fortuni, Torelli, Santini, Castellano,
  Fontana, Grazian, Pentericci, Pilo, \& Schmidt}]{Merlin2019}
Merlin, E., Fortuni, F., Torelli, M., {et~al.} 2019, Monthly Notices of the
  Royal Astronomical Society, 490, 3309, \dodoi{10.1093/mnras/stz2615}

\bibitem[{Merlin {et~al.}(2021)Merlin, Castellano, Santini, Cipolletta,
  Boutsia, Schreiber, Buitrago, Fontana, Elbaz, Dunlop, Grazian, McLure,
  McLeod, Nonino, Milvang-Jensen, Derriere, Hathi, Pentericci, Fortuni, \&
  Calabr{\`{o}}}]{Merlin2021}
Merlin, E., Castellano, M., Santini, P., {et~al.} 2021, Astronomy and
  Astrophysics, 649, A22, \dodoi{10.1051/0004-6361/202140310}

\bibitem[{Momcheva {et~al.}(2016)Momcheva, Brammer, van Dokkum, Skelton,
  Whitaker, Nelson, Fumagalli, Maseda, Leja, Franx, Rix, Bezanson, Cunha,
  Dickey, Schreiber, Illingworth, Kriek, Labb{\'{e}}, Lange, Lundgren, Magee,
  Marchesini, Oesch, Pacifici, Patel, Price, Tal, Wake, van~der Wel, \&
  Wuyts}]{Momcheva2016}
Momcheva, I.~G., Brammer, G.~B., van Dokkum, P.~G., {et~al.} 2016, The
  Astrophysical Journal Supplement Series, 225, 27,
  \dodoi{10.3847/0067-0049/225/2/27}

\bibitem[{Motohara {et~al.}(2014)Motohara, Konishi, Takahashi, Tateuchi,
  Kitagawa, Todo, Kato, Ohsawa, Aoki, Asano, Doi, Kamizuka, Kawara, Kohno,
  Koshida, Minezaki, Miyata, Morokuma, Okada, Sako, Soyano, Tamura, Tanabe,
  Tanaka, Tarusawa, Uchiyama, \& Yoshii}]{Motohara2014}
Motohara, K., Konishi, M., Takahashi, H., {et~al.} 2014, in Ground-based and
  Airborne Instrumentation for Astronomy V, ed. S.~K. Ramsay, I.~S. McLean, \&
  H.~Takami, Vol. 9147 (SPIE), 91476K, \dodoi{10.1117/12.2054861}

\bibitem[{Muzzin {et~al.}(2013)Muzzin, Marchesini, Stefanon, Franx,
  Milvang-Jensen, Dunlop, Fynbo, Brammer, Labbe, \& van Dokkum}]{Muzzin2013b}
Muzzin, A., Marchesini, D., Stefanon, M., {et~al.} 2013, The Astrophysical
  Journal Supplement Series, 206, 8, \dodoi{10.1088/0067-0049/206/1/8}

\bibitem[{Nanayakkara {et~al.}(2020)Nanayakkara, Brinchmann, Glazebrook,
  Bouwens, Kewley, Tran, Cowley, Fisher, Kacprzak, Labbe, \&
  Straatman}]{Nanayakkara2020}
Nanayakkara, T., Brinchmann, J., Glazebrook, K., {et~al.} 2020, The
  Astrophysical Journal, 889, 180, \dodoi{10.3847/1538-4357/ab65eb}

\bibitem[{Noll \& Pierini(2005)}]{Noll2005}
Noll, S., \& Pierini, D. 2005, Astronomy {\&} Astrophysics, 444, 137,
  \dodoi{10.1051/0004-6361:20053635}

\bibitem[{Oke \& Gunn(1983)}]{oke1983}
Oke, J.~B., \& Gunn, J.~E. 1983, The Astrophysical Journal, 266, 713,
  \dodoi{10.1086/160817}

\bibitem[{Patel {et~al.}(2011)Patel, Holden, Kelson, Franx, van~der Wel, \&
  Illingworth}]{Patel2011}
Patel, S.~G., Holden, B.~P., Kelson, D.~D., {et~al.} 2011, Astrophysical
  Journal Letters, 748, \dodoi{10.1088/2041-8205/748/2/L27}

\bibitem[{Price-Whelan {et~al.}(2018)Price-Whelan, Sipőcz, G{\"{u}}nther, Lim,
  Crawford, Conseil, Shupe, Craig, Dencheva, Ginsburg, VanderPlas, Bradley,
  P{\'{e}}rez-Su{\'{a}}rez, de~Val-Borro, Aldcroft, Cruz, Robitaille, Tollerud,
  Ardelean, Babej, Bach, Bachetti, Bakanov, Bamford, Barentsen, Barmby,
  Baumbach, Berry, Biscani, Boquien, Bostroem, Bouma, Brammer, Bray,
  Breytenbach, Buddelmeijer, Burke, Calderone, Rodr{\'{i}}guez, Cara, Cardoso,
  Cheedella, Copin, Corrales, Crichton, D'Avella, Deil, Depagne, Dietrich,
  Donath, Droettboom, Earl, Erben, Fabbro, Ferreira, Finethy, Fox, Garrison,
  Gibbons, Goldstein, Gommers, Greco, Greenfield, Groener, Grollier, Hagen,
  Hirst, Homeier, Horton, Hosseinzadeh, Hu, Hunkeler, Ivezi{\'{c}}, Jain,
  Jenness, Kanarek, Kendrew, Kern, Kerzendorf, Khvalko, King, Kirkby, Kulkarni,
  Kumar, Lee, Lenz, Littlefair, Ma, Macleod, Mastropietro, McCully, Montagnac,
  Morris, Mueller, Mumford, Muna, Murphy, Nelson, Nguyen, Ninan, N{\"{o}}the,
  Ogaz, Oh, Parejko, Parley, Pascual, Patil, Patil, Plunkett, Prochaska,
  Rastogi, Janga, Sabater, Sakurikar, Seifert, Sherbert, Sherwood-Taylor, Shih,
  Sick, Silbiger, Singanamalla, Singer, Sladen, Sooley, Sornarajah, Streicher,
  Teuben, Thomas, Tremblay, Turner, Terr{\'{o}}n, van Kerkwijk, de~la Vega,
  Watkins, Weaver, Whitmore, Woillez, \& Zabalza}]{Astropy2018}
Price-Whelan, A.~M., Sipőcz, B.~M., G{\"{u}}nther, H.~M., {et~al.} 2018, The
  Astronomical Journal, 156, 123, \dodoi{10.3847/1538-3881/aabc4f}

\bibitem[{Qin {et~al.}(2017)Qin, Mutch, Duffy, Geil, Poole, Mesinger, \&
  Wyithe}]{Qin2017}
Qin, Y., Mutch, S.~J., Duffy, A.~R., {et~al.} 2017, Monthly Notices of the
  Royal Astronomical Society, 471, 4345, \dodoi{10.1093/mnras/stx1852}

\bibitem[{Robitaille {et~al.}(2013)Robitaille, Tollerud, Greenfield,
  Droettboom, Bray, Aldcroft, Davis, Ginsburg, Price-Whelan, Kerzendorf,
  Conley, Crighton, Barbary, Muna, Ferguson, Grollier, Parikh, Nair,
  G{\"{u}}nther, Deil, Woillez, Conseil, Kramer, Turner, Singer, Fox, Weaver,
  Zabalza, Edwards, {Azalee Bostroem}, Burke, Casey, Crawford, Dencheva, Ely,
  Jenness, Labrie, Lim, Pierfederici, Pontzen, Ptak, Refsdal, Servillat, \&
  Streicher}]{Astropy2013}
Robitaille, T.~P., Tollerud, E.~J., Greenfield, P., {et~al.} 2013, Astronomy
  {\&} Astrophysics, 558, A33, \dodoi{10.1051/0004-6361/201322068}

\bibitem[{{Rodr{\'{i}}guez Montero} {et~al.}(2019){Rodr{\'{i}}guez Montero},
  Dav{\'{e}}, Wild, Angl{\'{e}}s-Alc{\'{a}}zar, \& Narayanan}]{Montero2019}
{Rodr{\'{i}}guez Montero}, F., Dav{\'{e}}, R., Wild, V.,
  Angl{\'{e}}s-Alc{\'{a}}zar, D., \& Narayanan, D. 2019, Monthly Notices of the
  Royal Astronomical Society, 490, 2139, \dodoi{10.1093/mnras/stz2580}

\bibitem[{Santini {et~al.}(2017)Santini, Fontana, Castellano, Criscienzo,
  Merlin, Amorin, Cullen, Daddi, Dickinson, Dunlop, Grazian, Lamastra, McLure,
  Micha{\l}owski, Pentericci, \& Shu}]{Santini2017}
Santini, P., Fontana, A., Castellano, M., {et~al.} 2017, The Astrophysical
  Journal, 847, 76, \dodoi{10.3847/1538-4357/aa8874}

\bibitem[{Santini {et~al.}(2019)Santini, Merlin, Fontana, Magnelli, Paris,
  Castellano, Grazian, Pentericci, Pilo, \& Torelli}]{Santini2019}
Santini, P., Merlin, E., Fontana, A., {et~al.} 2019, Monthly Notices of the
  Royal Astronomical Society, 486, 560, \dodoi{10.1093/mnras/stz801}

\bibitem[{Santini {et~al.}(2021)Santini, Castellano, Merlin, Fontana, Fortuni,
  Kodra, Magnelli, Menci, Calabr{\`{o}}, Lovell, Pentericci, Testa, \&
  Wilkins}]{Santini2021}
Santini, P., Castellano, M., Merlin, E., {et~al.} 2021, Astronomy {\&}
  Astrophysics, 652, A30, \dodoi{10.1051/0004-6361/202039738}

\bibitem[{Schreiber {et~al.}(2018)Schreiber, Glazebrook, Nanayakkara, Kacprzak,
  Labb{\'{e}}, Oesch, Yuan, Tran, Papovich, Spitler, \&
  Straatman}]{Schreiber2018}
Schreiber, C., Glazebrook, K., Nanayakkara, T., {et~al.} 2018, Astronomy {\&}
  Astrophysics, 618, A85, \dodoi{10.1051/0004-6361/201833070}

\bibitem[{Shahidi {et~al.}(2020)Shahidi, Mobasher, Nayyeri, Hemmati, Wiklind,
  Chartab, Dickinson, Finkelstein, Pacifici, Papovich, Ferguson, Fontana,
  Giavalisco, Koekemoer, Newman, Sattari, \& Somerville}]{Shahidi2020}
Shahidi, A., Mobasher, B., Nayyeri, H., {et~al.} 2020, The Astrophysical
  Journal, 897, 44, \dodoi{10.3847/1538-4357/ab96c5}

\bibitem[{Spitler {et~al.}(2014)Spitler, {S Straatman}, Labb{\'{e}},
  Glazebrook, Tran, Kacprzak, Quadri, Papovich, {Eric Persson}, van Dokkum,
  Allen, Kawinwanichakij, Kelson, McCarthy, Mehrtens, Monson, Nanayakkara,
  Rees, Tilvi, Tomczak, \& Mitchell}]{Spitler2014}
Spitler, L.~R., {S Straatman}, C.~M., Labb{\'{e}}, I., {et~al.} 2014, The
  Astrophysical Journal Letters, 787, \dodoi{10.1088/2041-8205/787/2/L36}

\bibitem[{Stefanon {et~al.}(2015)Stefanon, Marchesini, Muzzin, Brammer, Dunlop,
  Franx, Fynbo, Labb{\'{e}}, Jensen, \& van Dokkum}]{Stefanon2015}
Stefanon, M., Marchesini, D., Muzzin, A., {et~al.} 2015, The Astrophysical
  Journal, 803, 11, \dodoi{10.1088/0004-637X/803/1/11}

\bibitem[{Straatman {et~al.}(2014)Straatman, Labb{\'{e}}, Spitler, Allen,
  Altieri, Brammer, Dickinson, van Dokkum, Inami, Glazebrook, Kacprzak,
  Kawinwanichakij, Kelson, McCarthy, Mehrtens, Monson, Murphy, Papovich,
  Persson, Quadri, Rees, Tomczak, Tran, \& Tilvi}]{Straatman2014}
Straatman, C. M.~S., Labb{\'{e}}, I., Spitler, L.~R., {et~al.} 2014, The
  Astrophysical Journal, 783, L14, \dodoi{10.1088/2041-8205/783/1/L14}

\bibitem[{Straatman {et~al.}(2015)Straatman, Labb{\'{e}}, Spitler, Glazebrook,
  Tomczak, Allen, Brammer, Cowley, van Dokkum, Kacprzak, Kawinwanichakij,
  Mehrtens, Nanayakkara, Papovich, Persson, Quadri, Rees, Tilvi, Tran, \&
  Whitaker}]{Straatman2015}
---. 2015, The Astrophysical Journal, 808, L29,
  \dodoi{10.1088/2041-8205/808/1/L29}

\bibitem[{Straatman {et~al.}(2016)Straatman, Spitler, Quadri, Labb{\'{e}},
  Glazebrook, Persson, Papovich, Tran, Brammer, Cowley, Tomczak, Nanayakkara,
  Alcorn, Allen, Broussard, van Dokkum, Forrest, van Houdt, Kacprzak,
  Kawinwanichakij, Kelson, Lee, McCarthy, Mehrtens, Monson, Murphy, Rees,
  Tilvi, \& Whitaker}]{Straatman2016}
Straatman, C. M.~S., Spitler, L.~R., Quadri, R.~F., {et~al.} 2016, The
  Astrophysical Journal, 830, 51, \dodoi{10.3847/0004-637X/830/1/51}

\bibitem[{Taylor {et~al.}(2010)Taylor, Franx, Glazebrook, Brinchmann, van~der
  Wel, \& van Dokkum}]{Taylor2010}
Taylor, E.~N., Franx, M., Glazebrook, K., {et~al.} 2010, The Astrophysical
  Journal, 720, 723, \dodoi{10.1088/0004-637X/720/1/723}

\bibitem[{Tody(1986)}]{Tody1986}
Tody, D. 1986, in Instrumentation in Astronomy VI, ed. D.~L. Crawford, Vol.
  0627 (SPIE), 733, \dodoi{10.1117/12.968154}

\bibitem[{Tran {et~al.}(2020)Tran, Forrest, Alcorn, Yuan, Nanayakkara, Cohn,
  Cowley, Glazebrook, Gupta, Kacprzak, Kewley, Labb{\'{e}}, Papovich, Spitler,
  Straatman, \& Tomczak}]{Tran2020}
Tran, K.-V.~H., Forrest, B., Alcorn, L.~Y., {et~al.} 2020, The Astrophysical
  Journal, 898, 45, \dodoi{10.3847/1538-4357/ab8cba}

\bibitem[{Trujillo {et~al.}(2009)Trujillo, Cenarro, de~Lorenzo-C{\'{a}}ceres,
  Vazdekis, de~la Rosa, \& Cava}]{Trujillo2009}
Trujillo, I., Cenarro, A.~J., de~Lorenzo-C{\'{a}}ceres, A., {et~al.} 2009, The
  Astrophysical Journal, 692, L118, \dodoi{10.1088/0004-637X/692/2/L118}

\bibitem[{Valentino {et~al.}(2020)Valentino, Tanaka, Davidzon, Toft,
  G{\'{o}}mez-Guijarro, Stockmann, Onodera, Brammer, Ceverino, Faisst,
  Gallazzi, Hayward, Ilbert, Kubo, Magdis, Selsing, Shimakawa, Sparre,
  Steinhardt, Yabe, \& Zabl}]{Valentino2019}
Valentino, F., Tanaka, M., Davidzon, I., {et~al.} 2020, The Astrophysical
  Journal, 889, 93, \dodoi{10.3847/1538-4357/ab64dc}

\bibitem[{Whitaker {et~al.}(2011)Whitaker, Labb{\'{e}}, van Dokkum, Brammer,
  Kriek, Marchesini, Quadri, Franx, Muzzin, Williams, Bezanson, Illingworth,
  Lee, Lundgren, Nelson, Rudnick, Tal, \& Wake}]{Whitaker2011}
Whitaker, K.~E., Labb{\'{e}}, I., van Dokkum, P.~G., {et~al.} 2011, The
  Astrophysical Journal, 735, 86, \dodoi{10.1088/0004-637X/735/2/86}

\end{thebibliography}



\appendix

\section{Design of K-split Filters}
\label{sec:appendixA}
Here we summarize the study we undertook to optimize the central wavelength and width of the $K$-split filters.  Our goal is to maximize the efficiency in identifying galaxies at $z > 4$ with strong 4000~\AA\ and Balmer breaks.  This was done by undertaking a figure of merit (FOM) analysis which compared the potential cosmological volume probed, that dual filters in the $K$-band enable, compared to the exposure time required to achieve a given SNR.  As described in Section \ref{sec:filters} the inclusion of an additional filter alongside the $K_s$ filter enables a color measurement and hence a redshift range, which can subsequently be converted to a volume.  We note here that the volume metric does not need consider more than two filters, and as such our analysis was limited to considering the addition of a single filter alongside the $K_s$ filter, and a dual set of filters without the $K_s$ filter.  Our final filter design and survey has a 3 filter solution, including the $K_s$ filter, and the impact of this $K$-band filter set on the photometric redshift precision is outlined in Section \ref{sec:simulations}.

Adding a filter to the $K$-band requires an additional exposure time cost with many components that contribute (i.e. sky brightness, source brightness and SED shape, filter width etc).  It is therefore useful to quantify this additional exposure time relative to the $K_s$ band such that:
\begin{equation}
    r_{exp} = \frac{K_x,exp}{K_s,exp}
\end{equation}
Where the required exposure time is a function of source brightness, background brightness and filter width such that:
\begin{equation}
    r_{exp} = \frac{bg_{r}}{w_{r}.src_{r}^2}
\end{equation}
Here each term is relative to the $K_s$ band, where \textit{bg} is the mean background photon flux density, \textit{w} is the width of the additional filter and \textit{src} is the photon flux density of the source.  This relative exposure time cost and the volume probed by the filter set are the two components used in a FOM to find the optimal filter configuration.  The FOM is defined as:

\begin{equation}
    FOM = \frac{Volume}{\sum_i r_{exp,i}}
\end{equation}

When considering an additional filter, we start with a top-hat filter of a given width and central wavelength and slightly taper the edges to avoid strong impacts from features at the edge of the bands.  As described in Section \ref{sec:filters}, we use a 0.4 Gyr SSP model galaxy to determine the volume measurement of a given filter set.  This is also used to determine the relative source brightness in one filter compared to $K_s$.  For measurements of the sky background a synthetic night sky spectrum in the $K$-band is used plus an additional scaled black body component to represent the telescope emission to estimate sky background contribution.  Atmospheric transmission and optical throughput for Gemini South telescope is applied to all flux measurements (including sky background and source measurements).

\begin{figure*}
\begin{tabular}{cc}
\begin{minipage}{\linewidth}
    \centering
    \includegraphics[width=\linewidth]{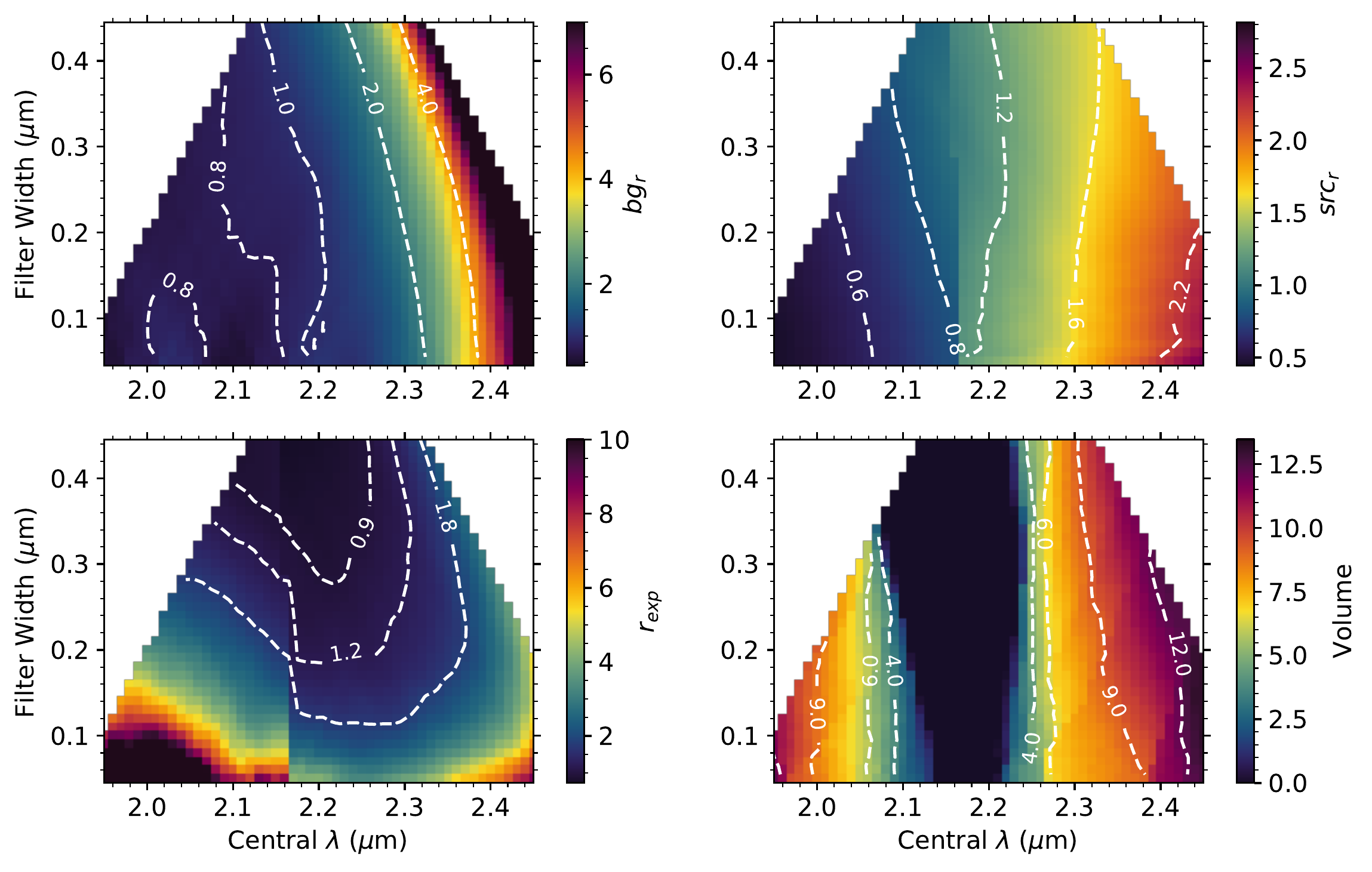}
\end{minipage}
\end{tabular}
    \caption{
    Components of the FOM analysis for a single additional filter, evaluated at a grid of central wavelength positions and filter widths.  Top left: the relative sky background flux contribution compared to $K_s$ filter.  Top right: the measured flux for an SED at a redshift (corresponding to the maximum color signature) relative to the $K_s$ filter.  Bottom left: relative exposure time to achieve the same signal-to-noise ratio in the $K_s$ filter.  Bottom right: the volume probed with a minimum $|K_s-K_x| > 0.3$ mag.  In all panels the white dashed lines are contour lines for the respective FOM components.  The white area represents where filter width and central wavelength positions extend beyond atmospheric transmission $<$ 0.1.
    }
\label{fig:fom_components}
\end{figure*}

Now using the definitions and set up above we can consider our FOM analysis with 1 and 2 filter sets.  For the single filter analysis we construct a 2D grid of filter widths and central wavelengths that cover the $K$-band window.  Here each position represents a potential single filter configuration alongside the $K_s$ filter.  The FOM components are calculated at each grid point and then the final FOM grid is calculated.  Boundaries are set such that filter solutions do not extend beyond the $K$-band atmospheric transmission windows i.e. central wavelength - 0.5 $\times$ filter width $>$ 1.9 micron and central wavelength + 0.5 $\times$ filter width $<$ 2.5 micron.  These rule out FOM solutions in the upper left and right portion of the FOM grid.  The individual components of the single filter FOM analysis evaluated over a grid of central wavelengths and filter widths are shown in Fig.~\ref{fig:fom_components}.  

\begin{figure*}
\begin{tabular}{cc}
\begin{minipage}{\linewidth}
    \centering
    \includegraphics[width=0.6\linewidth]{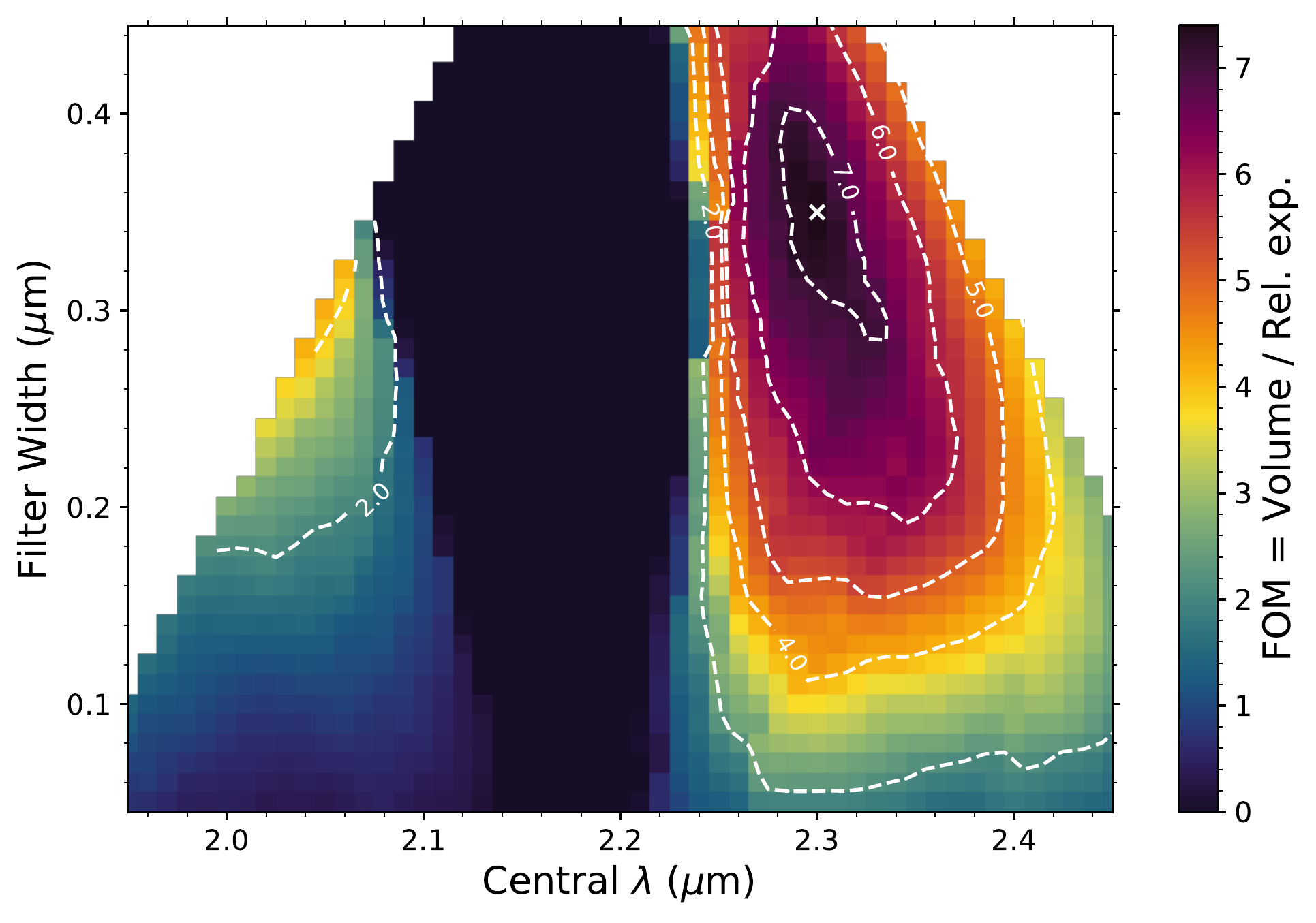}
\end{minipage}
\end{tabular}
    \caption{
     Single filter simulation from the FOM analysis.  Each grid point represents the FOM value for a given filter width and central wavelength position of an additional $K$-band filter.  White dashed lines are contour lines for the FOM values.  White x marks the optimal filter width and central wavelength position from the FOM analysis.  The white area represents where filter width and central wavelength positions extend beyond atmospheric transmission $<$ 0.1.
    }
    \label{fig:fom_single_filt}
\end{figure*}

One can develop an intuition for the driving factors in the FOM by looking at the individual components in Fig.~\ref{fig:fom_components}.  For example, one can see while the sky background increases rapidly with increasing wavelength (top left panel of Fig.~\ref{fig:fom_components}), the model galaxy with a redshift corresponding to the largest color signature, also increases with increasing wavelength (top right panel of Fig.~\ref{fig:fom_components}).  The result is that the relative exposure time cost is flatter at longer wavelengths than one might expect (bottom left panel of Fig.~\ref{fig:fom_components}).  Finally the volume probed is intuitively maximized at either ends of the $K$-band window as shown in the lower right panel of Fig.~\ref{fig:fom_components}.  The final FOM grid is shown in Fig.~\ref{fig:fom_single_filt} and shows the optimal single filter configuration at 2.3 micron with a filter width of 0.35 micron.  Note however that the peak of the FOM is relatively flat and there is substantial room to move to a different single filter configuration without significant reduction in the FOM value.

\begin{figure*}
\begin{tabular}{cc}
\begin{minipage}{\linewidth}
    \centering
    \includegraphics[width=\linewidth]{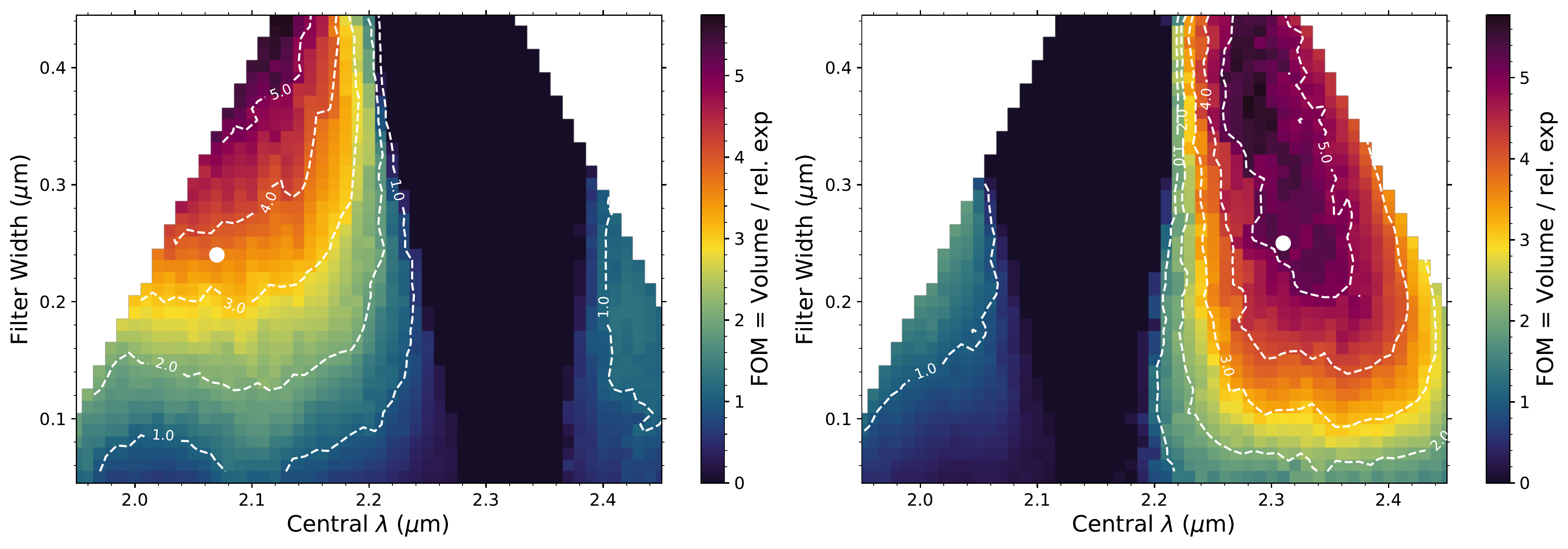}
\end{minipage}
\end{tabular}
    \caption{
     Dual filter simulation from the FOM analysis.  Left panel is the "blue" filter optimal solution and the right panel is the "red" filter.  Same figure layout as for Fig.~\ref{fig:fom_single_filt}.  The infilled white circles are the final design central wavelength and filter width for the $K$-split filters.
    }
    \label{fig:fom_dual_filt}
\end{figure*}

A similar analysis can be extended to two filters by using a 4D grid comprising two filter widths and two central wavelength positions.  In this analysis the volume probed is measured from the two additional filters while the relative exposure time is the sum of the individual filters compared to the $K_s$ filter (see equation 3).  The resulting FOM is shown in Fig.~\ref{fig:fom_dual_filt}.  These central wavelength and filter width positions peak at very similar locations to a single filter solution, which suggests that the dual filter solution is also complementary to the $K_s$ filter.  The FOM analysis indicates that a single filter solution is most effective and efficient.  A two filter solution is slightly less efficient due to the increased exposure time for the second filter, but just as effective.

While the single filter solution is optimal in the context of the FOM analysis conducted here, there are benefits to adopting a two filter strategy.  Where $K_s$ observations already exist, the two filter strategy becomes a three filter strategy, enabling full sampling of the $K$-band window, thus both maximising the discovery space and also permitting Nyquist sampling of the $K$-band window.  One can also take advantage of variable sky conditions and utilize whichever filter is optimal for observing, for example the thermal background, impacting mostly the $\kred$ filter, is usually higher at the beginning of the night.  It's for these reasons that a two filter strategy was adopted.

Qualitatively speaking, the dual filter FOM analysis suggests broad filters, pushed to either side of the $K$-band window are the most effective solution.  This has the dual filters overlapping substantially to minimize the relative exposure time term.  However, for a solution that is complementary with $K_s$, slightly narrower filters are more ideal as they further increase the volume probed and similarly increase the color signature measured across the $K$-band window, i.e. $\kblue$ - $K_s$ or $K_s$ - $\kred$ colors.  Therefore the final design trades-off filter width in this regard to incorporate the $K_s$ filter to arrive at an optimal three filter solution.

Finally, due to the overlap of the $K$-split filters and the $K$-band atmospheric transmission window, we smoothed of the filter edges to account for variation in transmission from varying water column depths and minimize the impact of this to less than 1$\%$.

\section{Photometric Redshift Recovery Simulations}
\label{sec:appendixB}
Here we present the extended sample of galaxy models and their diagnostic plots based on the analysis in Section \ref{sec:photzsim}.

\begin{figure*}
    \begin{tabular}{cc}
    \begin{minipage}[t]{0.9\linewidth}
    \centering
    \includegraphics[width=\linewidth]{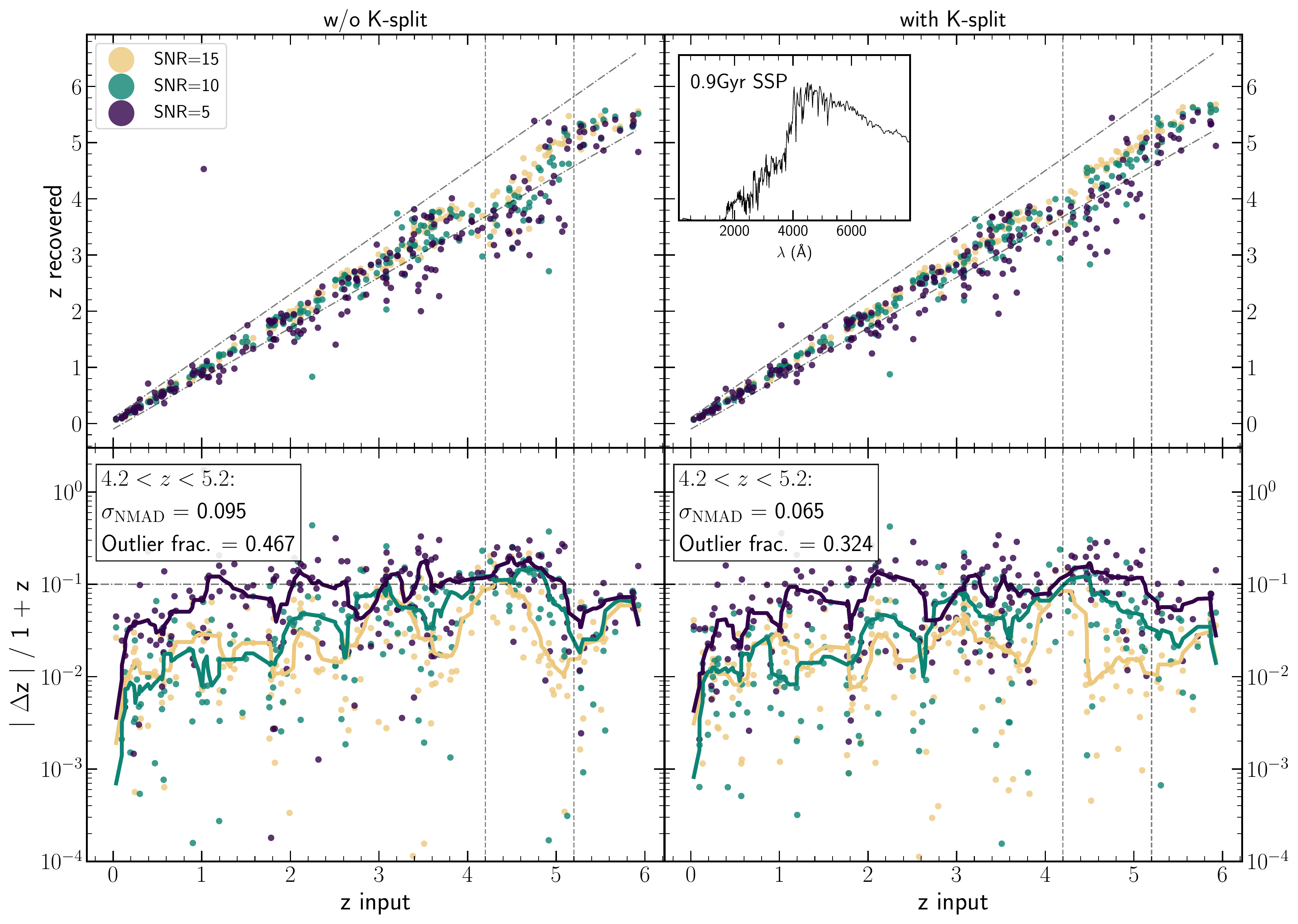}\\
    \centering\small (a)
    \end{minipage}
    \\
    \begin{minipage}[t]{0.6\linewidth}
    \centering
    \includegraphics[width=\linewidth]{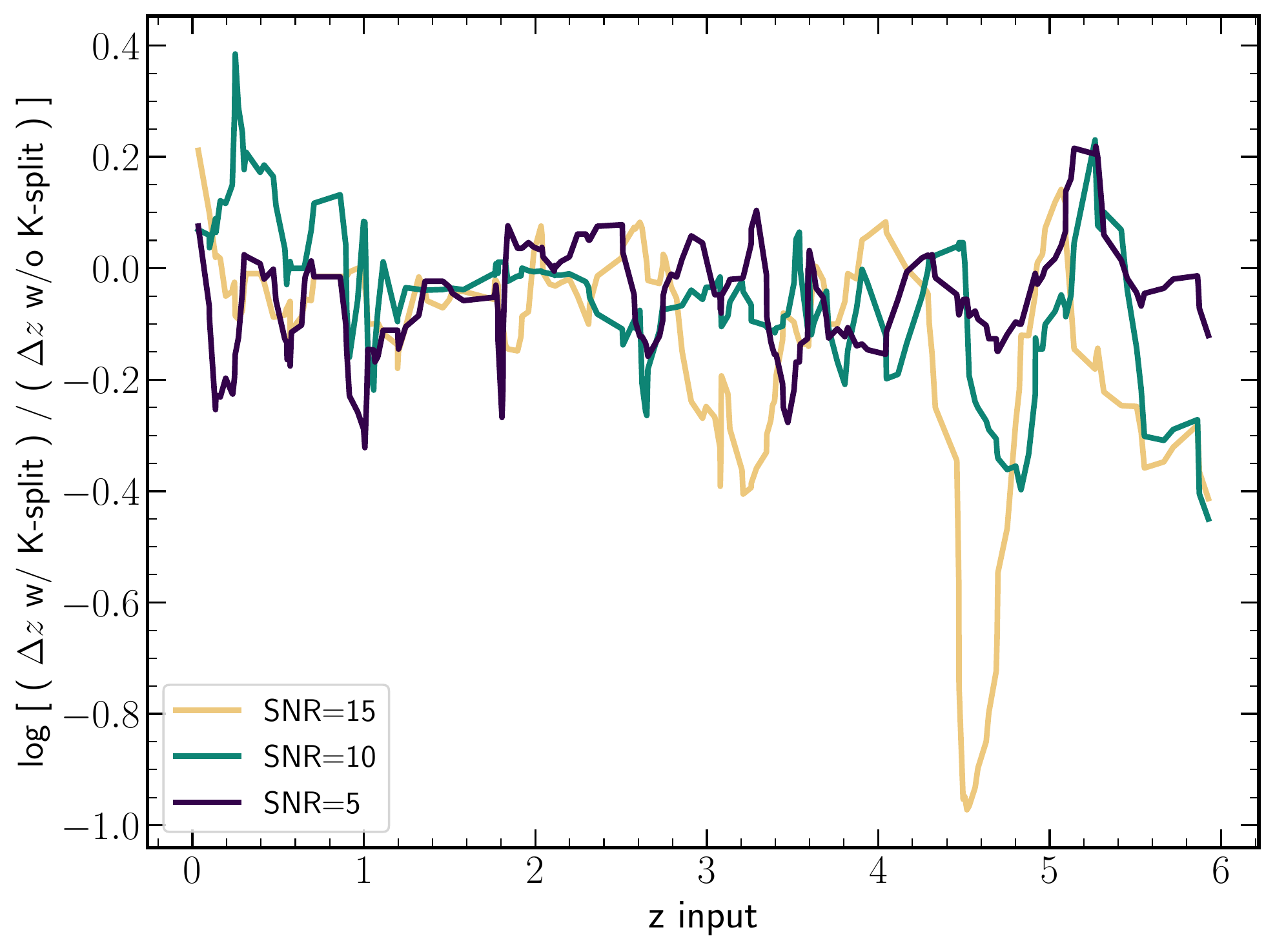}\\
    \centering\small (b)
    \end{minipage}    

    \end{tabular}
    \caption{ 
    (a) - Testing the impact of the $K$-split filters on photometric redshift recovery using simulated photometry for a 0.9 Gyr SSP model. Same layout as Fig~\ref{fig:ez_sim_mass_gal}.  (b) - Ratio of the $\Delta z$ running median lines from the lower panels from (a) for with and without $K$-split filters. Same layout as Fig~\ref{fig:deltz_ratio}.
    }
    \label{fig:0.9Gyr_photz}
\end{figure*}

\begin{figure*}
    \begin{tabular}{cc}
    \begin{minipage}[t]{0.9\linewidth}
    \centering
    \includegraphics[width=\linewidth]{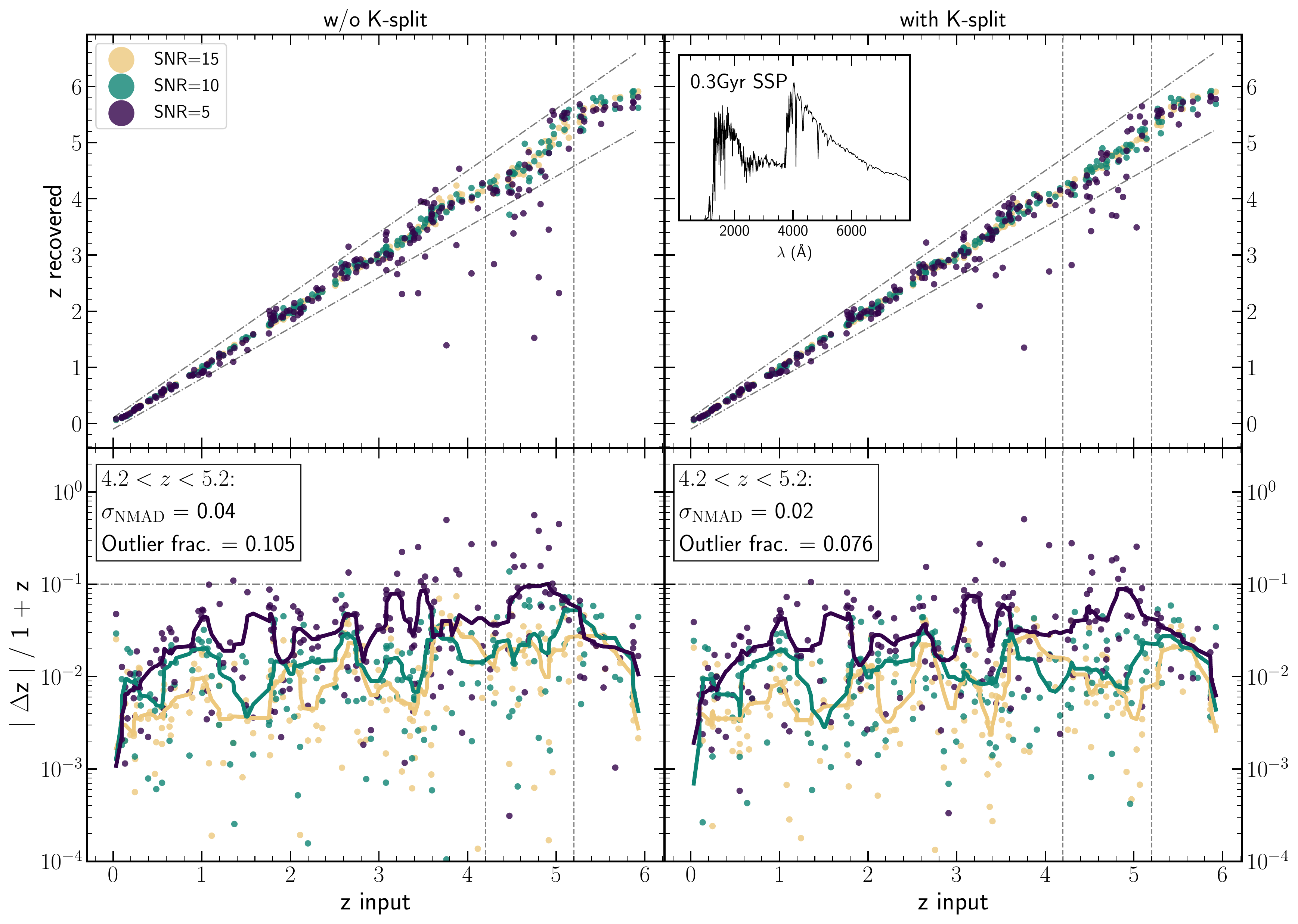}\\
    \centering\small (a)
    \end{minipage}
    \\
    \begin{minipage}[t]{0.6\linewidth}
    \centering
    \includegraphics[width=\linewidth]{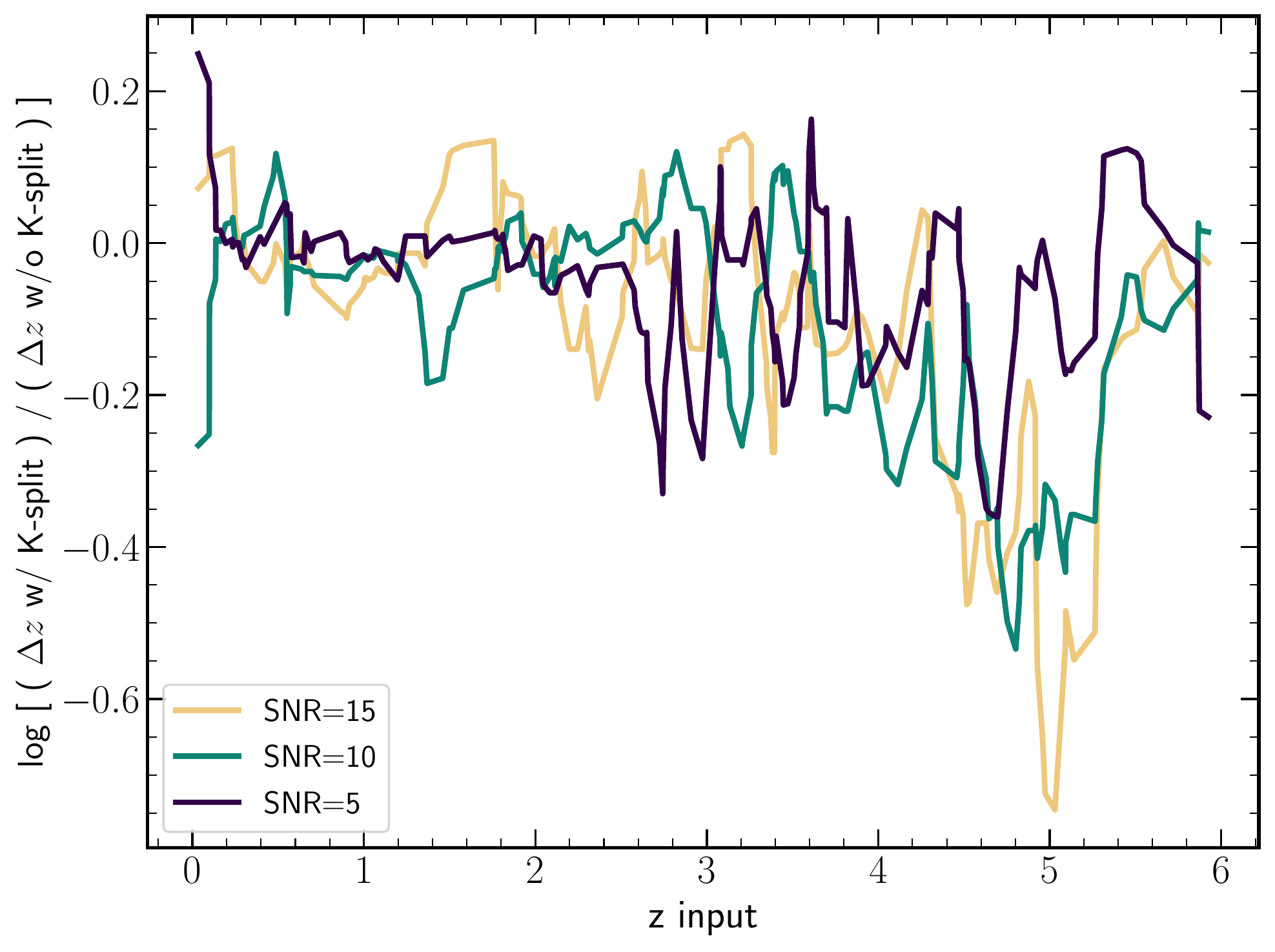}\\
    \centering\small (b)
    \end{minipage}    

    \end{tabular}
    \caption{ 
    (a) - Testing the impact of the $K$-split filters on photometric redshift recovery using simulated photometry for a 0.3 Gyr SSP model. Same layout as Fig~\ref{fig:ez_sim_mass_gal}.  (b) - Ratio of the $\Delta z$ running median lines from the lower panels from (a) for with and without $K$-split filters. Same layout as Fig~\ref{fig:deltz_ratio}.
    }
    \label{fig:0.3Gyr_photz}
\end{figure*}

\begin{figure*}
    \begin{tabular}{cc}
    \begin{minipage}[t]{0.9\linewidth}
    \centering
    \includegraphics[width=\linewidth]{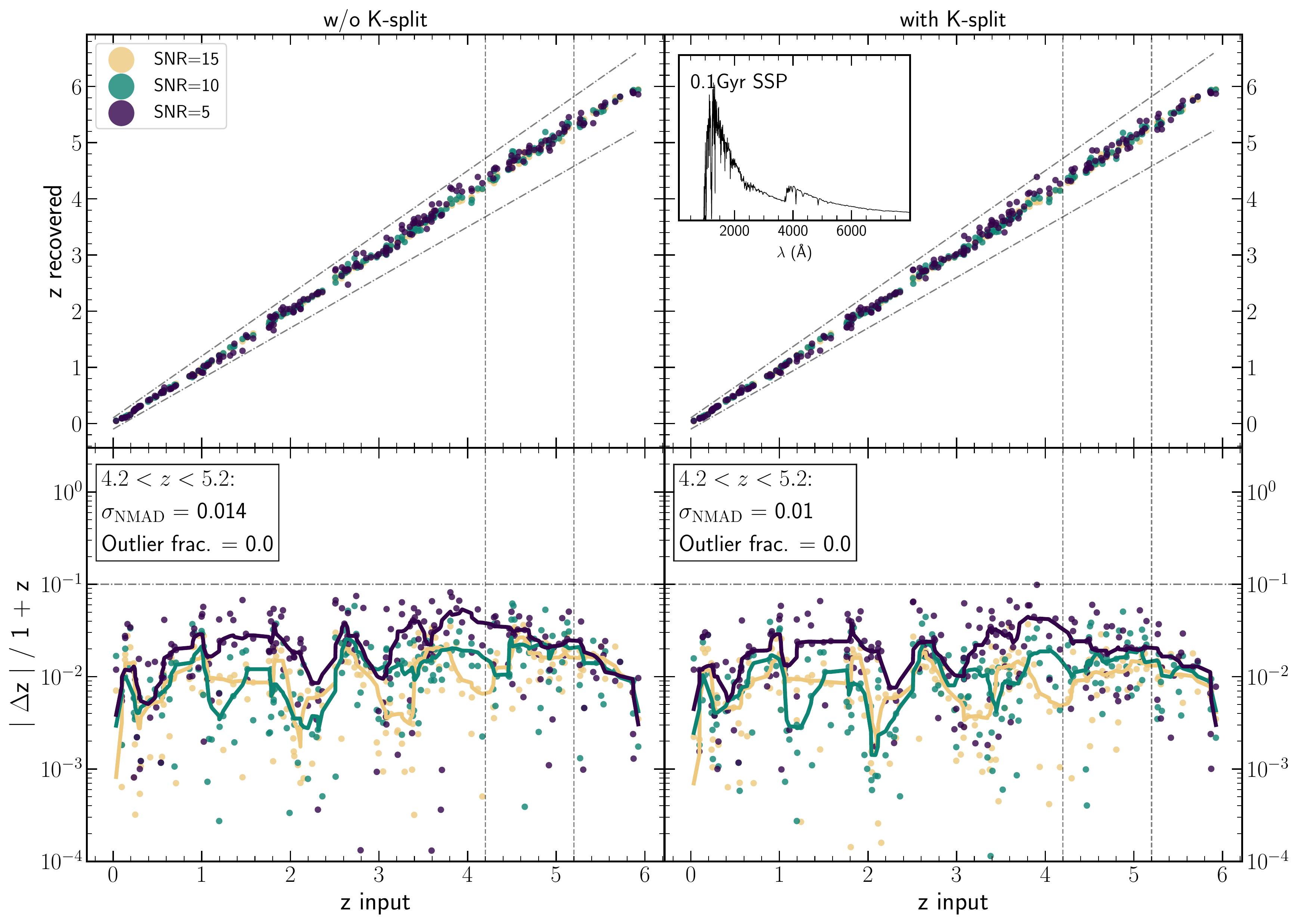}\\
    \centering\small (a)
    \end{minipage}
    \\
    \begin{minipage}[t]{0.6\linewidth}
    \centering
    \includegraphics[width=\linewidth]{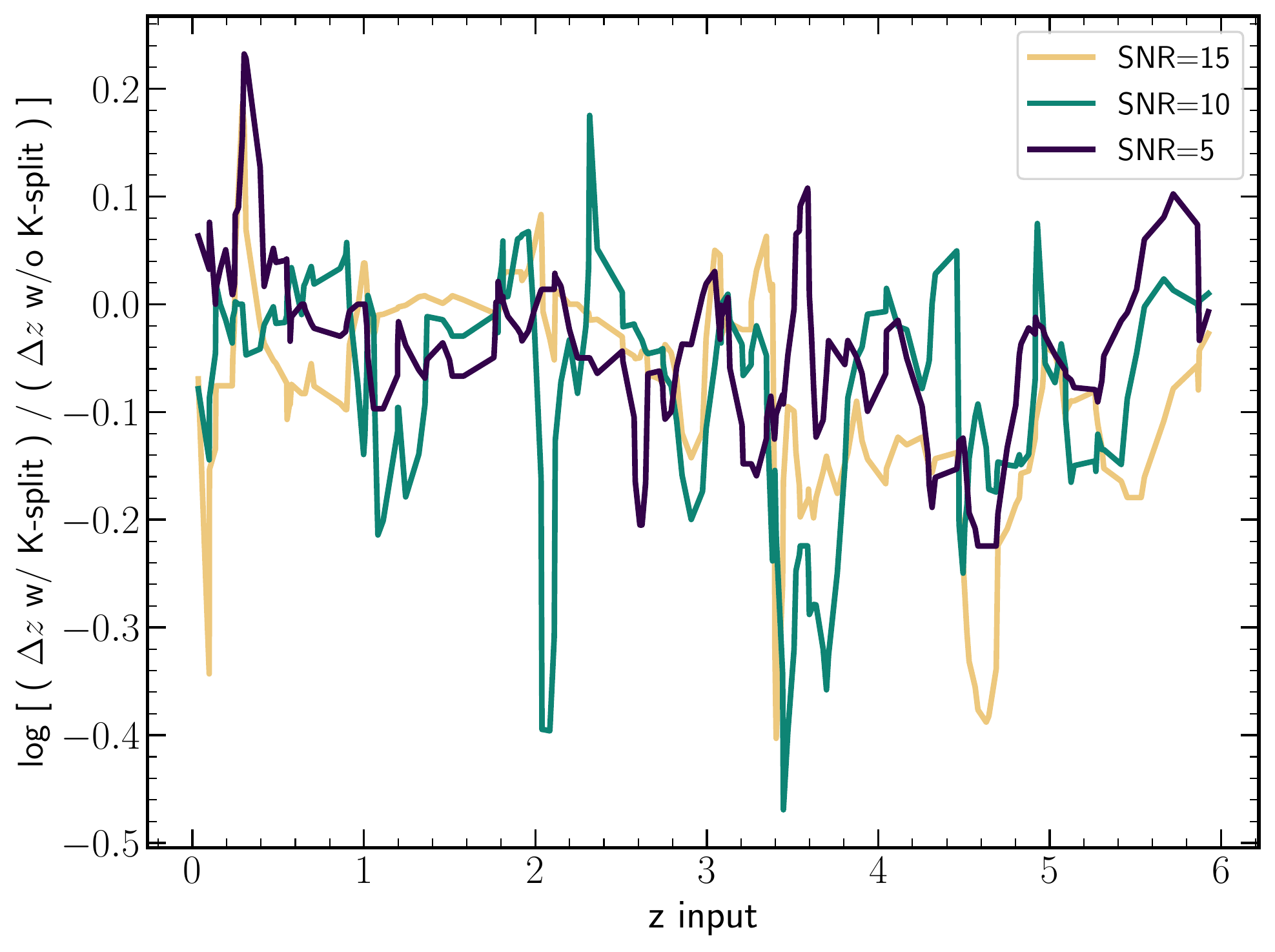}\\
    \centering\small (b)
    \end{minipage}    
    \end{tabular}
    \caption{ 
    (a) - Testing the impact of the $K$-split filters on photometric redshift recovery using simulated photometry for a 0.1 Gyr SSP model. Same layout as Fig~\ref{fig:ez_sim_mass_gal}.  (b) - Ratio of the $\Delta z$ running median lines from the lower panels from (a) for with and without $K$-split filters. Same layout as Fig~\ref{fig:deltz_ratio}.
    }
    \label{fig:0.1Gyr_photz}
\end{figure*}

\begin{figure*}
    \begin{tabular}{cc}
    \begin{minipage}[t]{0.9\linewidth}
    \centering
    \includegraphics[width=\linewidth]{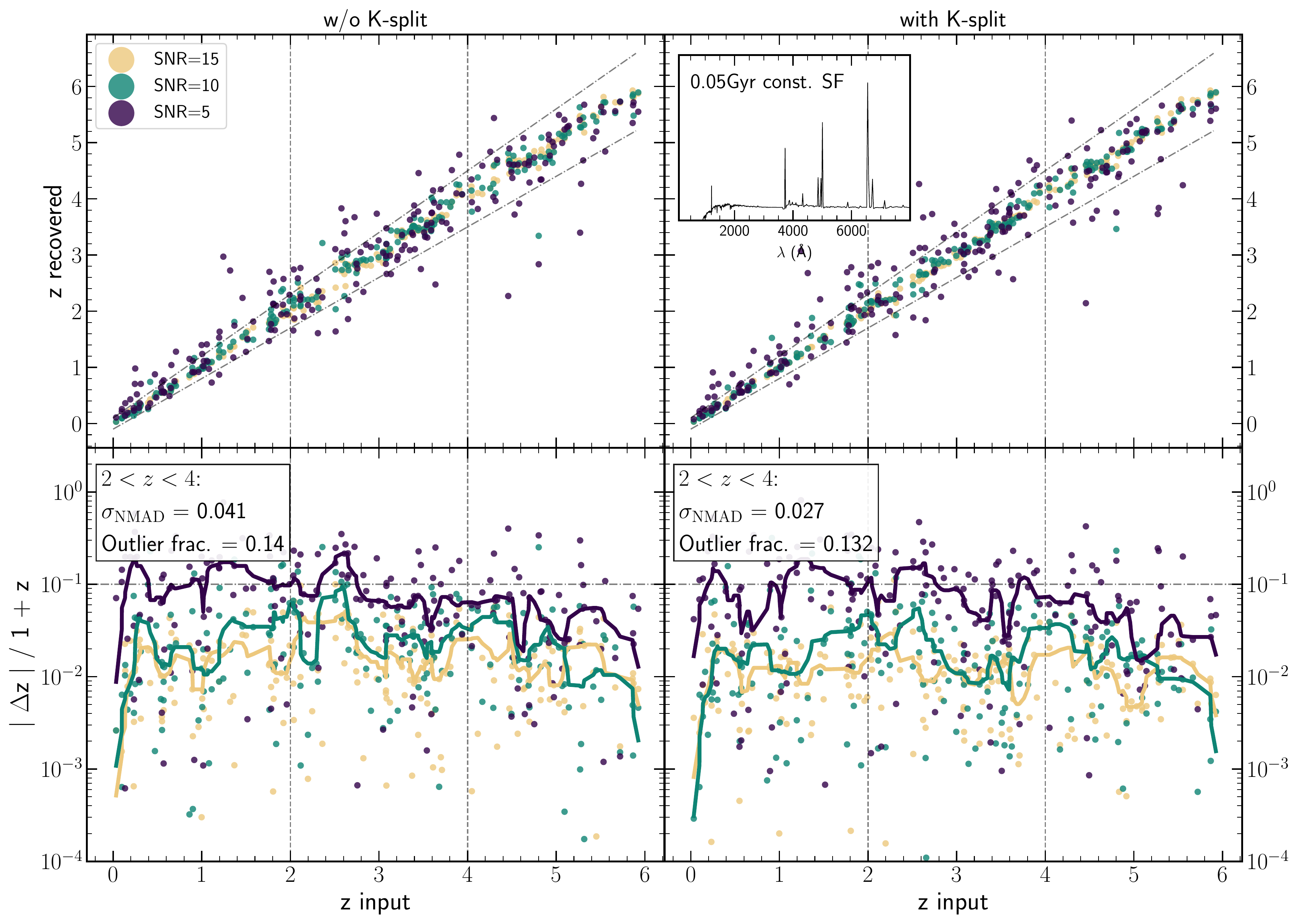}\\
    \centering\small (a)
    \end{minipage}
    \\
    \begin{minipage}[t]{0.6\linewidth}
    \centering
    \includegraphics[width=\linewidth]{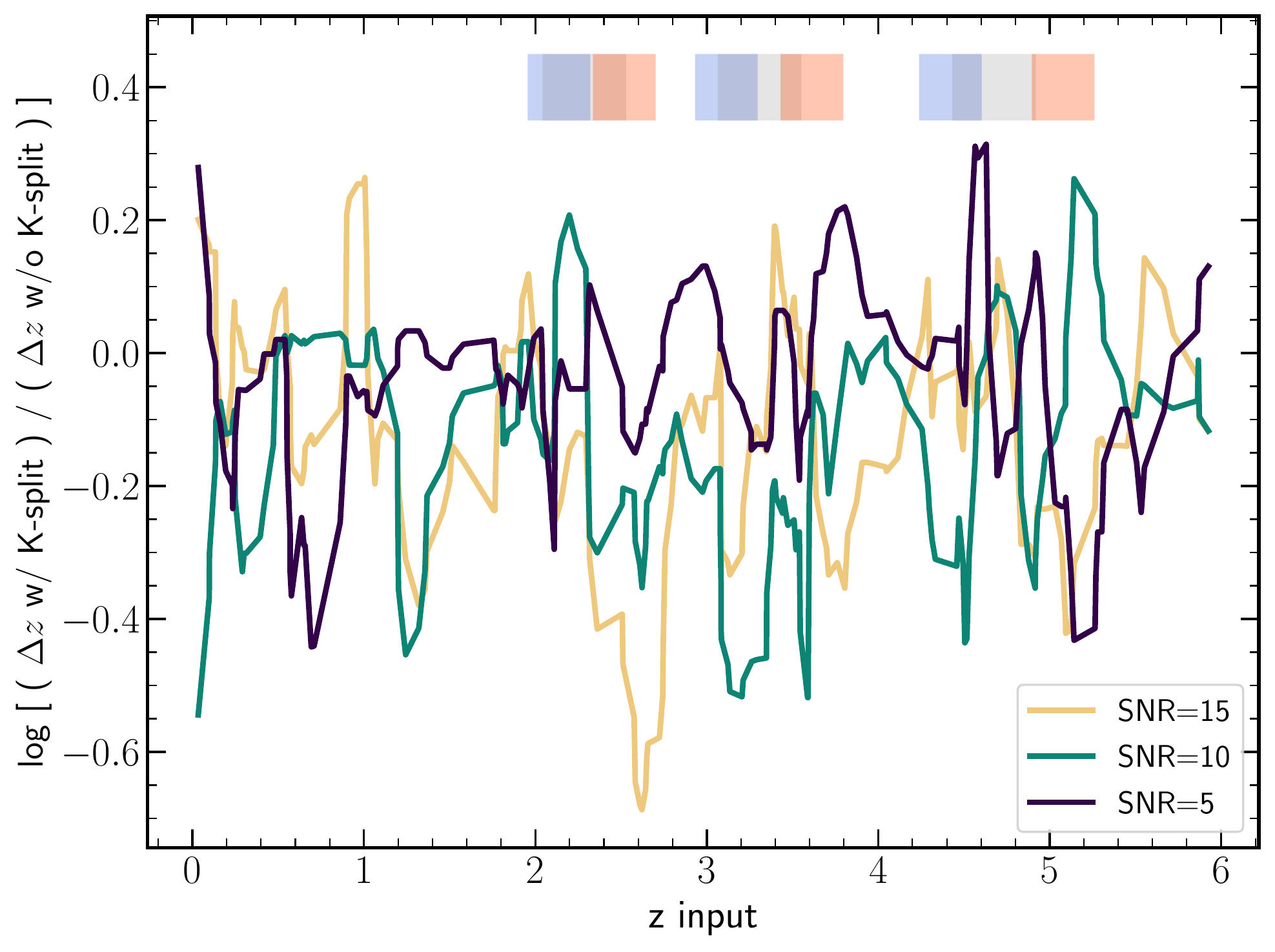}\\
    \centering\small (b)
    \end{minipage}    

    \end{tabular}
    \caption{ 
    (a) - Testing the impact of the $K$-split filters on photometric redshift recovery using simulated photometry for a dusty star-forming model. Same layout as Fig~\ref{fig:ez_sim_mass_gal}.  (b) - Ratio of the $\Delta z$ running median lines from the lower panels from (a) for with and without $K$-split filters. Same layout as Fig~\ref{fig:deltz_ratio}.  Blue, gray and red shading in lower panel represent the redshift ranges for which, from left to right, H$\alpha$, [OIII] and [OII] emission lines enter the $\kblue$, $K_s$ and $\kred$ filters respectively.
    }
    \label{fig:el_photz}
\end{figure*}




\end{document}